\documentclass{emulateapj}

\begin{document}
\title{Infrared Variability of Evolved Protoplanetary Disks: Evidence for Scale Height Variations in the Inner Disk}
\author{Flaherty, K.M. \altaffilmark{1}, Muzerolle, J. \altaffilmark{2}, Rieke, G. \altaffilmark{1}, Gutermuth, R. \altaffilmark{3}, Balog, Z.\altaffilmark{4}, Herbst, W. \altaffilmark{5}, Megeath, S.T. \altaffilmark{6}, Kun, M.\altaffilmark{7}}
\email{kflaherty@as.arizona.edu}

\altaffiltext{1}{Steward Observatory, University of Arizona, Tucson, AZ 85721}
\altaffiltext{2}{Space Telescope Science Institute, 3700 San Martin Dr., Baltimore, MD, 21218}
\altaffiltext{3}{Department of Astronomy, University of Massachusetts, Amherst, MA 01003}
\altaffiltext{4}{Max-Planck Institut f\"{u}r Astronomie, K\"{o}nigstuhl 17, D-69117 Heidelberg, Germany}
\altaffiltext{5}{Department of Astronomy, Wesleyan University, Middletown, CT 06459}
\altaffiltext{6}{Department of Physics and Astronomy, University of Toledo, Toledo, OH}
\altaffiltext{7}{Konkoly Observatory, H-1121 Budapest, Konkoly Thege \'{u}t 15-17, Hungary}

\begin{abstract}
We present the results of a multi-wavelength multi-epoch survey of five evolved protoplanetary disks in the IC 348 cluster that show significant infrared variability. Using 3-8\micron\ and 24\micron\ photometry along with 5-40\micron\ spectroscopy from the Spitzer Space Telescope, as well as ground-based 0.8-5\micron\ spectroscopy, optical spectroscopy and near-infrared photometry, covering timescales of days to years, we examine the variability in the disk, stellar and accretion flux. We find substantial variations (10-60\%) at all infrared wavelengths on timescales of weeks to months for all of these young stellar objects. This behavior is not unique when compared to other cluster members and is consistent with changes in the structure of the inner disk, most likely scale height fluctuations on a dynamical timescale. Previous observations, along with our near-infrared photometry, indicate that the stellar fluxes are relatively constant; stellar variability does not appear to drive the large changes in the infrared fluxes. Based on our near-infrared spectroscopy of the Pa$\beta$ and Br$\gamma$ lines we find that the accretion rates are variable in most of the evolved disks but the overall rates are probably too small to cause the infrared variability. We discuss other possible physical causes for the variability, including the influence of a companion, magnetic fields threading the disk, and X-ray flares.
\end{abstract}

\section{Introduction}
The general picture of the formation and evolution of low mass stars involves the complex interplay of the newly created star and the reservoir of dust and gas surrounding it, feeding it and often shielding it from out view. Early on, material from a dense envelope surrounds the protostars, falling onto the disk from where it accretes onto the protostar building up a majority of the stellar mass \citep{ter84,ada87}. After hundreds of thousands of years the envelope dissipates, the protostar has accreted most of its mass, and the system is left with a disk of gas and dust. These circumstellar disks are usually visible for millions of years before much of the dust and gas is either accumulated into planetesimals and planets, or removed by a photo-evaporative flow \citep{her07}. Left behind is a debris disk, with a small trace of optically thin dust sustained by the collisions of planetesimals \citep{wya08}. While these different evolutionary stages of low mass star formation have been fairly well established, the physical processes behind the movement from one stage to the next are not well understood.

Transition disks, and evolved disks in general, represent a stage of the disk's life that is believed to lie between fully optically thick disks and debris disks. These systems are characterized by a deviation from the typical SED of a full flared disk, often due to the presence of large, many AU wide regions of optically thin material or a severe flattening of an otherwise flared disk\footnote{Different studies apply different definitions to the term 'transition disk' in an attempt to best encapsulate the evolution of circumstellar disks. Some authors restrict this analysis to sources that contain a gap in the disk that is nearly devoid of dust \citep{muz10}. These sources are characterized by an infrared SED that has little to no excess at wavelengths less than $\sim10\micron$ with a strong excess at longer wavelengths. Some authors allow for some optically thick material at the very inner edge of the gap, close to the dust sublimation radius, which leads to a strong excess in the near-infrared \citep{esp07}.  Other studies include disks that have merely been flattened, rather than having a gap \citep{cie10}. These full disks do not show a sharp dip in the infrared SED, but instead show an overall steepening until it almost resembles the emission from an infinitesimally thin, perfectly flat disk. All of these disk geometries represent some form of dust evolution, but not necessarily the same type of evolution. Here we choose to use the term 'evolved disks' to describe any system that shows significant evolution in its dust content, while 'transition disk' only describes those systems with either a hole or gap.}. 
 These SED shapes may be indicative of the presence of planets \citep{bry99,naj07}, a photoevaporative flow combined with viscous evolution \citep{ale06,pas11} or a significant amount of small dust settling to the midplane \citep{cie10}. Resolved observations confirm the presence of gaps filled with optically thin dust spanning many AU \citep{and11}, as well as the presence of optically thick material very close to the star interior to the optically thin gap \citep{pot10} in some of these evolved disks. Gas has been observed in many of these systems \citep{pon08,sal11} at levels higher than expected given the drop in dust surface density needed to explain the SED. Understanding the gas and dust in these situations is crucial for a complete picture of planet formation and early stellar evolution.

Recently some transition disks have been found to have large amplitude infrared variability, which could potentially be related to the clearing of the disk. \citet{muz09} study one member of the IC 348 cluster in which the flux shortward of 8\micron\ increases while the flux longward of 8\micron\ decreases (and vice-versa), both in as little as one week. \citet{esp11} see this 'seesaw' behavior in a collection of 14 transition disks in Taurus and Chameleon observed with Spitzer 5-40\micron\  IRS spectroscopy. The strength, speed, and wavelength dependence of the variability was surprising given the previously known sources of variability in pre-main sequence stars, such as starspots and stellar occultations. Even large outbursts, such as seen in FU Ori or EXor objects, take place over months and years \citep{har96,her08}. Fluctuations in the 20\micron\ flux on weekly timescales are unexpected given that this wavelength traces dust at a few AU where the dynamical timescale is closer to years. Recent observations suggest that rapid infrared variability may in fact be quite common among pre-main sequence stars \citep{mor11}. Geometric models of the disk suggest that this variability may be due to changes in the scale height of the inner disk \citep{fla10,esp11} 

In \citet{fla11} we presented the results of a detailed observing campaign focused on the source LRLL 31 in IC 348. Using infrared observations from Spitzer, as well as ground-based near-infrared spectroscopy, we were able to confirm the geometric model of inner disk perturbations, as well as rule out accretion as the driving source of the variability. Here we present multi-wavelength multi-epoch observations of five additional evolved disks in IC 348 designed to determine if the behavior observed in LRLL 31 is common, as well as put further constraints on the source of the vertical perturbation. These additional sources were chosen to have strong infrared variations as well as evidence for significant evolution in their dust structure based on the SED. In Section~\ref{observations} we detail the data obtained for our sample, review the properties of LRLL 31, and discuss in detail the characteristics of the stellar, dust and gas emission for each source. We then use this information in Section~\ref{perturbe} to determine if the same process is operating in all of these stars. Assuming this process is common among normal disks, in Section~\ref{cause} we discuss what could be its physical cause.

\section{Variability Observations}\label{observations}
\subsection{Data}
We obtained multiple epochs (2007-2009) of data from 4000\AA\ out to 40\micron\ on the IC 348 evolved disks LRLL 2, 21, 58, 67, 1679 \citep{luh03,mue07} as part of the same program that observed the IC 348 transition disk LRLL 31 \citep{fla11}. We also include previous infrared Spitzer observations taken as early as 2004 to extend our temporal coverage. Table~\ref{obs_log} lists all of the observations for each source, including the times when these measurements were taken. These observations include:
\begin{itemize}
\item 10 epochs of Spitzer 24\micron\ photometry with typical uncertainties less than 2\% (Table~\ref{mips_phot}). Two epochs were taken seven months apart in 2004 as part of the c2d map of IC 348 \citep{reb07} while another epoch was taken in 2004 as part of GTO observations \citep{lad06}. We obtained five consecutive days of monitoring in Sep 2007 followed in Mar 2008 by two epochs separated by one week.
\item 7 epochs of Spitzer IRAC cold-mission 3.6,4.5,5.8 and 8.0\micron\ photometry with uncertainties less than 2\% (Table~\ref{irac_cm_phot}). One epoch was taken in 2004 as part of the c2d coverage \citep{jor06} while GTO observations were also performed in 2004 \citep{lad06}. We obtained five consecutive days of photometry in Mar 2009.
\item 4 epochs of Spitzer IRS 5-40\micron\ spectra. Two spectra were taken a week apart in Oct 2007, while two more were taken in Mar 2008, also separated by one week
\item 38 epochs of Spitzer IRAC warm-mission 3.6 and 4.5\micron\ photometry spread over 40 days in Fall 2009 with uncertainties of 1-2\% (Table~\ref{irac_wm_phot}). The cadence for this photometry ranged from four observations a day to one observation every two days throughout the visibility window.
\item One epoch of low-resolution optical spectra taken with the CAFOS instrument on the 2.2-m telescope of the Calar-Alto Observatory of LRLL 21, 58 and 1679 in Fall 2009.
\item Multiple epochs of 0.8-5\micron\ spectroscopy with Spex on IRTF in Fall 2008 (LRLL 21, 1679) and Fall 2009 (LRLL 2, 21, 58, 1679). The exact number of epochs and the wavelength coverage of the spectra varies from source to source, with each object getting at least two 0.8-2.5\micron\ spectra.
\item One epoch of high-resolution optical spectra of LRLL 21, 58, 67 and 1679 taken with Hectochelle on the MMT in Feb 2008.
\end{itemize}
Details on the data reduction can be found in \citet{fla11}. We also observed four non-accreting diskless weak line T Tauri stars (WTTS) in Taurus (K0 star LkCa 19, A2 star HD 57928, M3 star LkCa21, M1 star JH 108) with Spex to use as standards for the 0.8-2.5\micron\ spectroscopy. We have found that pre-main sequence stars provide much better templates than the dwarfs and giants in the IRTF spectral library \citep{ray09} when comparing the spectral shape and strength of the absorption lines. This is especially true for the spectral shapes of the late-type stars and the absorption lines in the early type stars. For the IRS spectra, unlike with LRLL 31, we did not obtain both Short-Low (SL) and Long-Low (LL) spectra at every epoch, due to the significant background emission that made it difficult to extract the LL spectra. For LRLL 2 and 58 we obtained four epochs of SL spectra, but only included LL spectra in the first two epochs. These two sources, especially LRLL 58, are the most susceptible to systematic uncertainties in the long wavelength end due to the large background. The background emission also leads to a systematic offset between the SL and LL spectra that is seen in LRLL 2, 58. When comparing spectra taken on different epochs we have scaled the LL spectra to match the SL spectra. For LRLL 67 the offset maybe due to calibration uncertainties causing the short-wavelength end of the LL2 spectrum ($\sim14-16\micron$) to flatten out.

We chose to focus on systems with SEDs indicative of a significant evolution in the dust distribution from that of a full, optically thick, flared disk. Recent radiative transfer modeling has shown that two of the sources we study (LRLL 21 and 31) have large gaps that extend from the inner disk out to tens of AU, another source (LRLL 67) has no optically thick material within tens of AU of the star, while another source (LRLL 2) has a flattened full disk with no gap \citep{esp12}. Given that our sources represent a wide range of geometries, we choose to use the term 'evolved disks' when describing the entire sample. Evolved disks can be further subdivided into those systems with SEDs characteristic of a flattened disk and those systems with SEDs characteristic of an optically thin gap, which we refer to as transition disks. The transition disks can then be further divided into those that only have optically thin material within a few AU, such as LRLL 67, and those that have some optically thick material at the dust sublimation radius, which we refer to as pre-transition disks. As we will discuss in section ~\ref{perturbe}, the most important piece of the disk when it comes to the variability is the optically thick dust near the sublimation radius, which we refer to as the 'inner disk'.

\subsection{LRLL 31}
Here we summarize previously published results on LRLL 31 \citep{muz09,fla11}. The data set is similar to, but more extensive than, the other evolved disks. One of the main questions we are trying to answer is whether or not the variability observed in LRLL 31 is common or is unique to this one source. We start here with the detailed observations and in section~\ref{perturbe} we try to determine if we can synthesize this information into one simple model. 

\subsubsection{Stellar Properties}
LRLL 31 is a G6 star with a luminosity of 4.3$L_{\odot}$, $R_*$=2.1$R_{\odot}$, M=1.5$M_{\odot}$ and $A_V=8.8$ \citep{fla11}. We estimate the extinction using the shape of the 0.8-2.5\micron\ spectra, ignoring the K-band whose shape may be influenced by dust emission. Contemporaneous with the near-infrared spectra, we were able to obtain J-band photometry, which is dominated by stellar emission, allowing us to calculate the stellar luminosity. Our multiple epochs of J-band photometry do not show significant fluctuations in the luminosity, although there may be long term trends in the stellar flux. The near-infrared spectra show that the extinction is constant on timescales of weeks and years. There is some evidence for variations in the radial velocity of the star, suggesting the presence of a companion, although with only three epochs this conclusion is not definitive.

\subsubsection{Infrared Variability}
Our infrared observations cover $\lambda=2-40\micron$ on timescales of days to years allowing us to look for any wavelength dependence in the fluctuations, as well as to constrain the variability timescale. In the IRS spectra, LRLL 31 displays large fluctuations ($\sim60\%$) in as little as one week with a clear wavelength dependence in which the SED appears to pivot around 8\micron. The 24\micron\ photometry displays large fluctuations (30\%) consistent with those seen in the IRS spectra. The flux of the silicate feature also changes with time, and the changes are correlated with the long-wavelength fluctuations. The 3.6 and 4.5\micron\ monitoring shows large (0.3 mag) fluctuations predominately on weekly timescales. Multiple epochs measuring the shape of the 2-5\micron\ excess find that this excess looks like a single temperature blackbody at $\sim$1700K whose strength rapidly varies with time. The [3.6]-[4.5] color of the system appears to get redder as it gets brighter. There was no evidence for periodicity in the light curve.

\begin{figure}
\includegraphics[scale=.5]{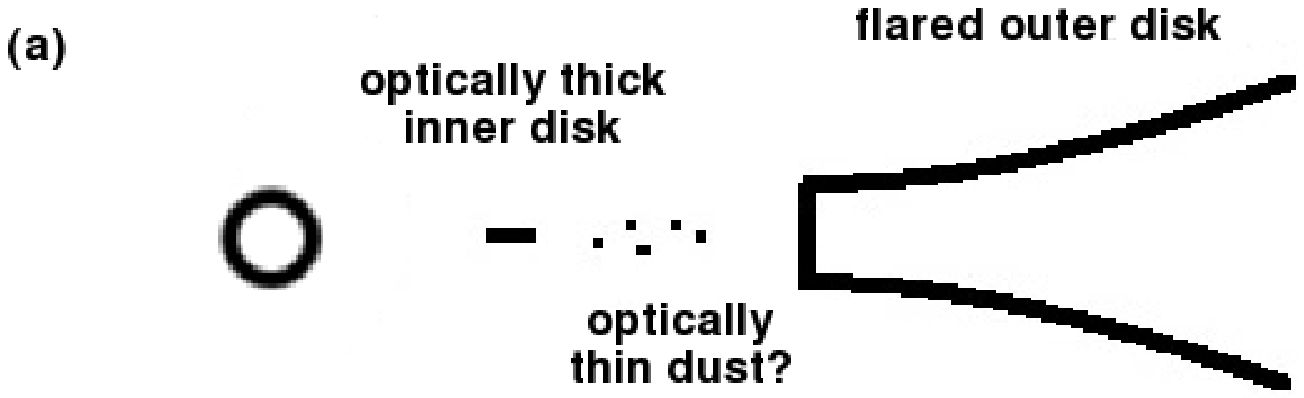}
\includegraphics[scale=.5]{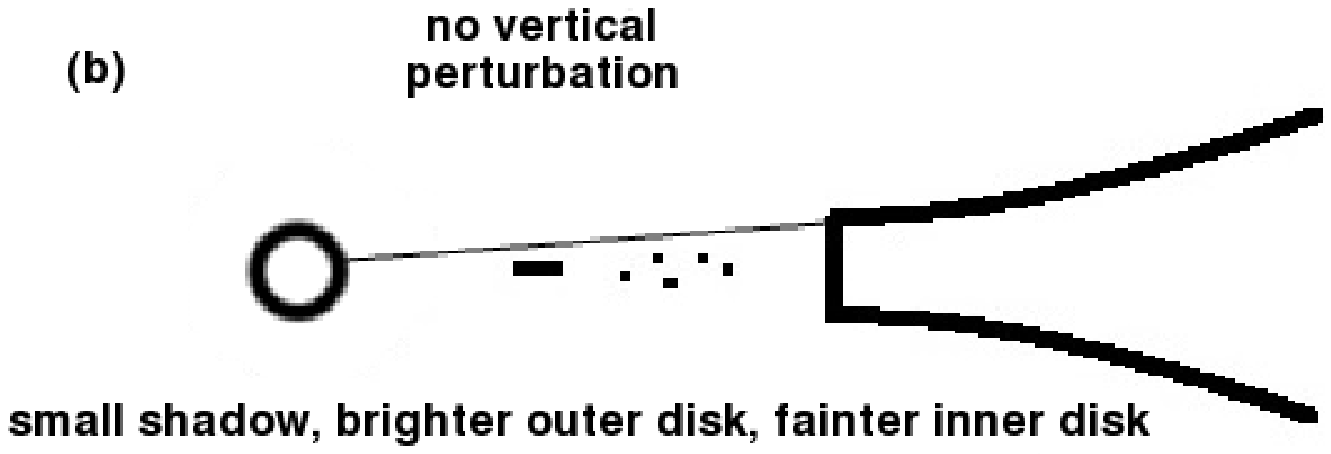}
\includegraphics[scale=.5]{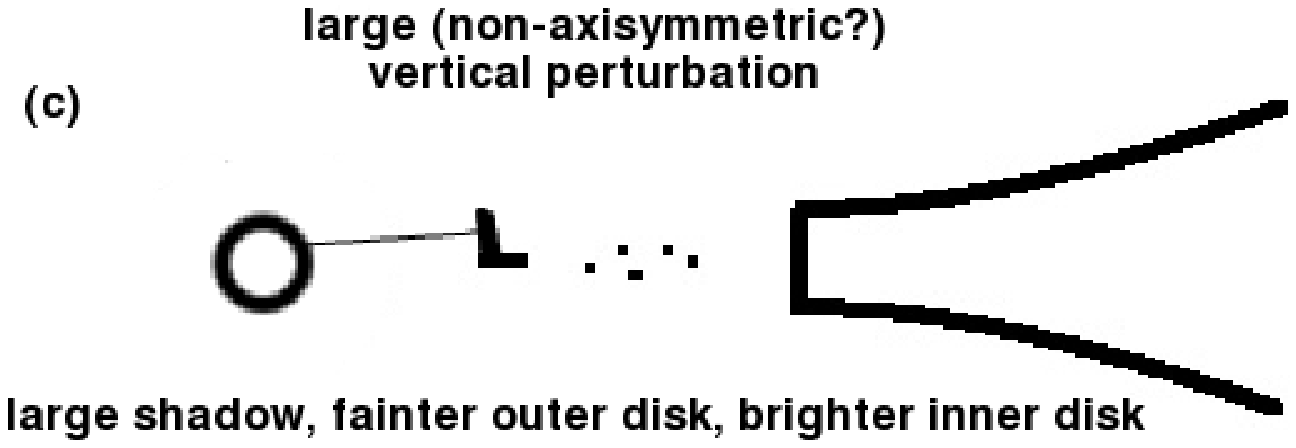}
\caption{Schematic diagrams (not to scale) describing the basic geometric model for the variability. (a) We consider an equilibrium situation that consists of an optically thick, flared, outer disk, an extended gap possibly filled with some optically thin dust, and optically thick dust near the dust destruction radius. This last bit of dust close to the star likely has a small scale height, and may not be axisymmetric, in order to match the weak average short wavelength excess seen around most of our evolved disks. (b) The variability, in its simplest form, involves an oscillation between two states. In one state, the inner disk is very flat and the outer disk is directly illuminated. (c) When a vertical perturbation exists in the inner disk, then its flux increases because its emitting area is larger, while at the same time it casts a shadow on the outer disk, reducing its temperature and flux. This perturbation may or may not be azimuthally symmetric. \label{schematics}}
\end{figure}

To describe the observations, we developed a model for disk structural perturbations to the inner disk \citep{fla10}. LRLL 31 is a pre-transition disk and the basic geometric picture is of an optically thick ring of dust at the dust sublimation radius (the inner disk) that emits predominately in the near-infrared, a large gap that may or may not be filled with optically thin dust, and an optically thick, flared outer disk that dominates the flux longward of 10\micron\ (Fig~\ref{schematics}a). \citet{esp12} recently constructed a detailed radiative transfer model that confirmed this structure. They find that the SED can be fit when including a wall of optically thick dust at the sublimation radius (0.3 AU), followed by optically thin dust extending to 14 AU beyond which is an optically thick disk. In our conceptual model of the variability, when there is no vertical perturbation to the inner disk (Fig~\ref{schematics}b) its flux is at a minimum because its emitting area is the smallest, while the long-wavelength flux is at its strongest because it is directly heated by the star. When a, possibly non-axisymmetric, perturbation increases the height of the inner disk (Fig~\ref{schematics}c) the short-wavelength emission from this region increases, because the emitting area of the inner disk is larger, while the long-wavelength emission decreases, because the regions of the outer disk responsible for this emission are being shadowed by the inner disk. This model is consistent with the wavelength dependence of the variability in the IRS spectra, the daily to weekly timescales seen during the warm-mission monitoring campaign, and the change in the 2-5\micron\ excess, which is dominated by emission from the inner disk \citep{fla11}. The presence of an optically thin gap and an optically thick, flared outer disk might not be strictly necessary to explain the variability, but we include it in the basic geometric picture because it is consistent with the general shape of the infrared SED.

\subsubsection{Gas Properties}
Our optical and near-infrared spectroscopy of the hydrogen emission lines (H$\alpha$, Pa$\beta$ and Br$\gamma$), whose flux is directly related to the accretion rate, as well as measurements from the literature, allow us to characterize the level of variability in the accretion rate and look for any connection with the infrared variability. Based on multiple epochs of the Pa$\beta$ and Br$\gamma$ emission lines, we observe a factor of five change in the accretion rate, from 0.3 to 1.6 $\times10^{-8}M_{\odot}yr^{-1}$. Accretion rates are derived from Pa$\beta$ and Br$\gamma$ EW using the JHK photometry combined with the veiling as an estimate of the continuum level \citep{fla11} and then converting from line flux to accretion rate \citep{muz98}. Accretion rate variability is also seen in multiple epochs of H$\alpha$ spectroscopy, which show large fluctuations in the EW. The accretion luminosity is small (L$_{acc}$/L$_*$$<0.1$) compared to the stellar flux. While there is evidence for a correlation between the accretion rate, as measured with the near-infrared spectra, and the infrared flux, as measured by our warm-mission monitoring, in our 2009 data we found that the change in accretion luminosity was too small to explain the structural perturbation.

\subsection{LRLL 2}
\subsubsection{Stellar Properties}
LRLL 2 is a pre-main sequence A2 star with a bolometric luminosity of 137$L_{\odot}$, a radius of 5R$_{\odot}$ \citep{luh03} and a mass of $\sim3.5M_{\odot}$ based on the \citet{sie00} 3 Myr isochrones. In our two epochs of near-infrared spectra we measure an average extinction of A$_V$=2.9 (Table~\ref{extinction}), which is consistent with previous estimates \citep{luh03} given that we use a different extinction law (R$_V$=5.5 vs. 3.1, see \citet{fla11} for details). Previous studies have found no evidence for variability in the optical, suggesting that the photospheric flux is steady \citep{coh04,lit05}.

\subsubsection{Infrared Variability}
\begin{figure*}
\center
\includegraphics[scale=.3]{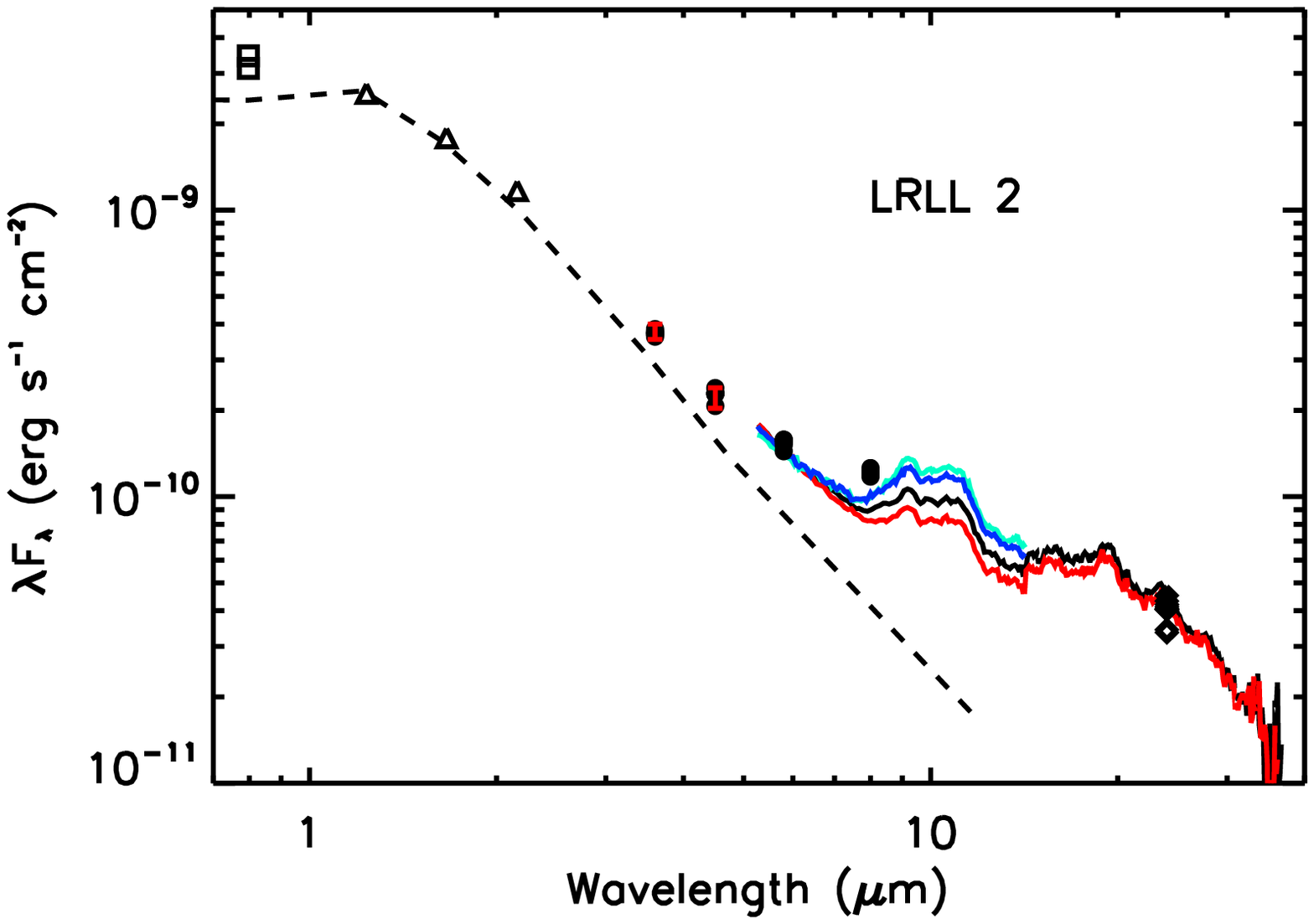}
\includegraphics[scale=.3]{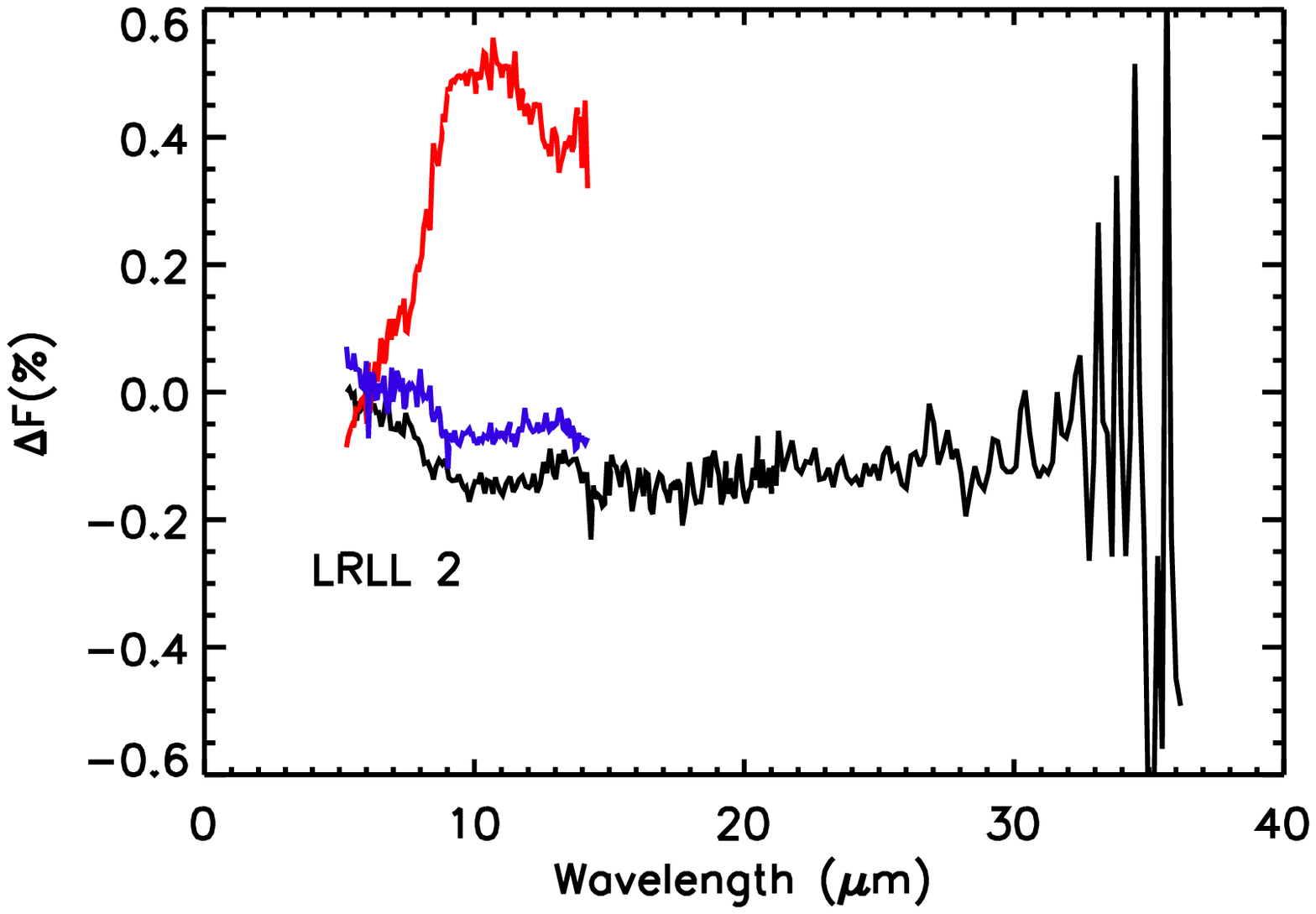}
\includegraphics[scale=.3]{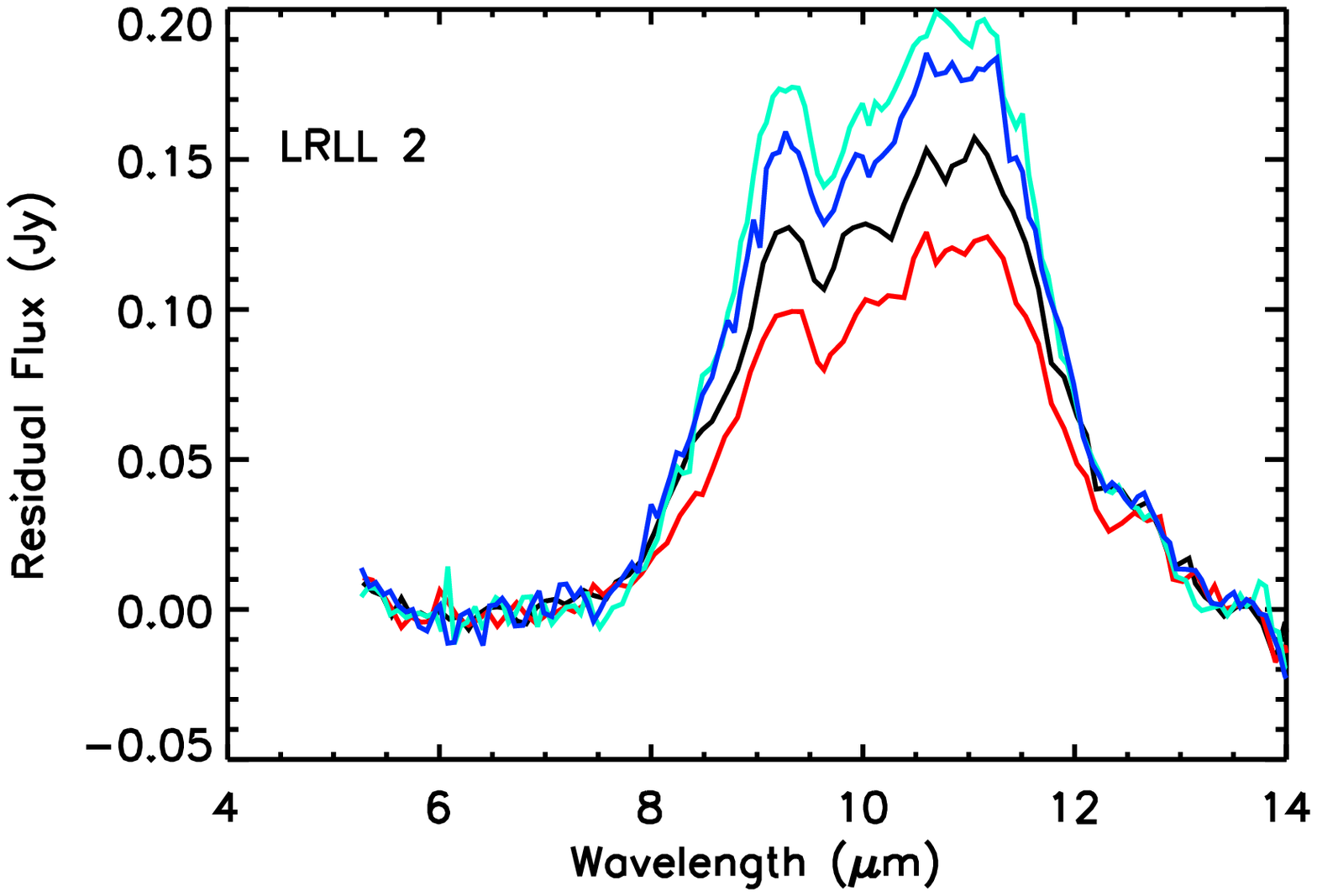}
\caption{Spitzer spectra of LRLL 2. {\it Left}: The spectral energy distribution showing the IRS data (colored lines), the IRAC cold mission data (dots), JHK photometry (triangles), optical photometry (squares) and the range of the IRAC warm-mission photometry (error bars at 3.6,4.5\micron). A reddened stellar photosphere from \citet{ken95} is shown as the dashed line for comparison. {\it Middle}: Difference spectra between different epochs of IRS spectra. Black line is the change from epoch 1 to epoch 2 (separated by one week), red line is the change from epoch 2 to epoch 3 (separated by 5 months) and the blue line is the change from epochs 3 to epoch 4 (separated by one week). {\it  Right}: Residual flux in the silicate feature after the continuum has been subtracted.  \label{lrll2_ir}}
\end{figure*}

\begin{figure*}
\center
\includegraphics[scale=.4]{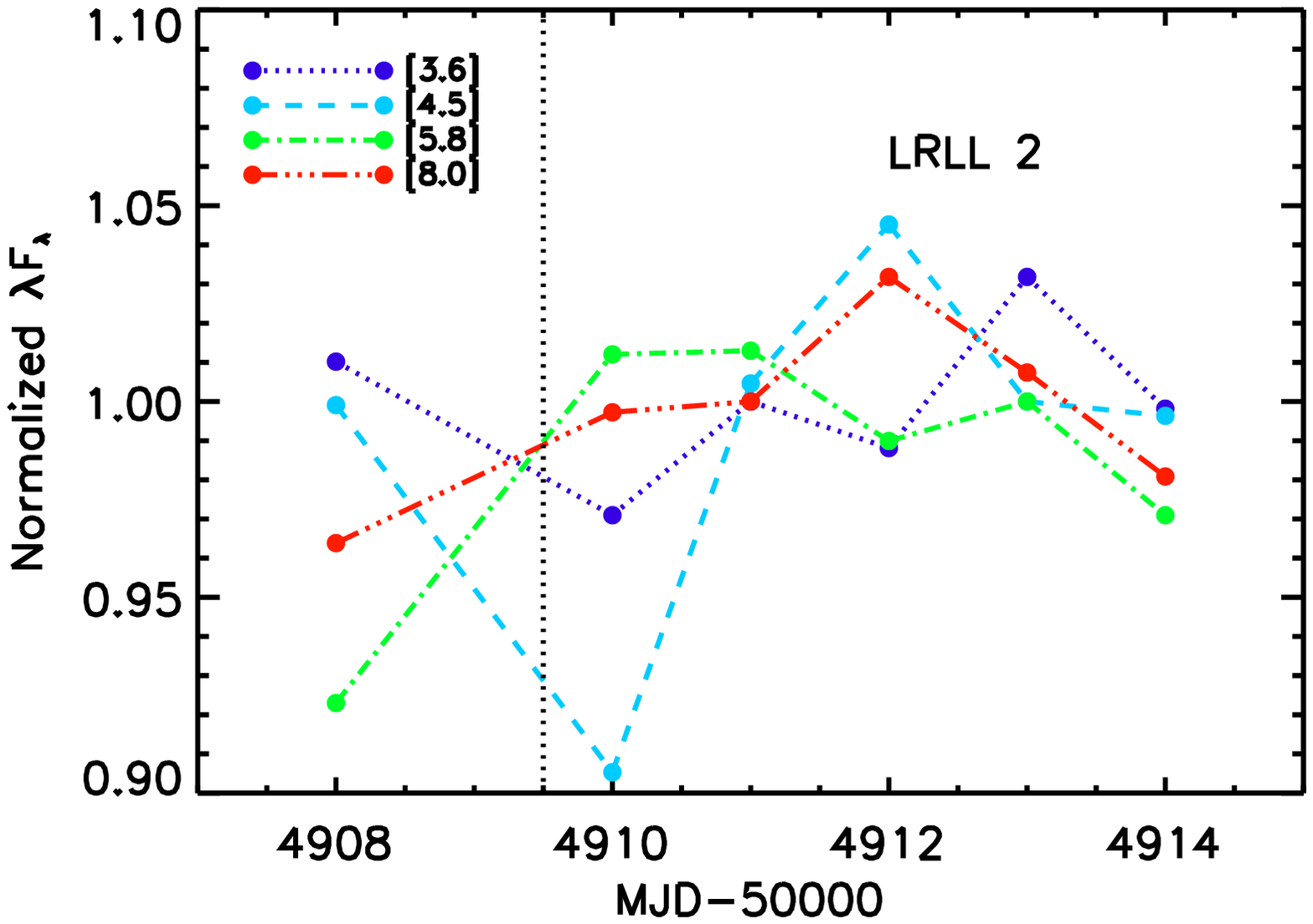}
\includegraphics[scale=.4]{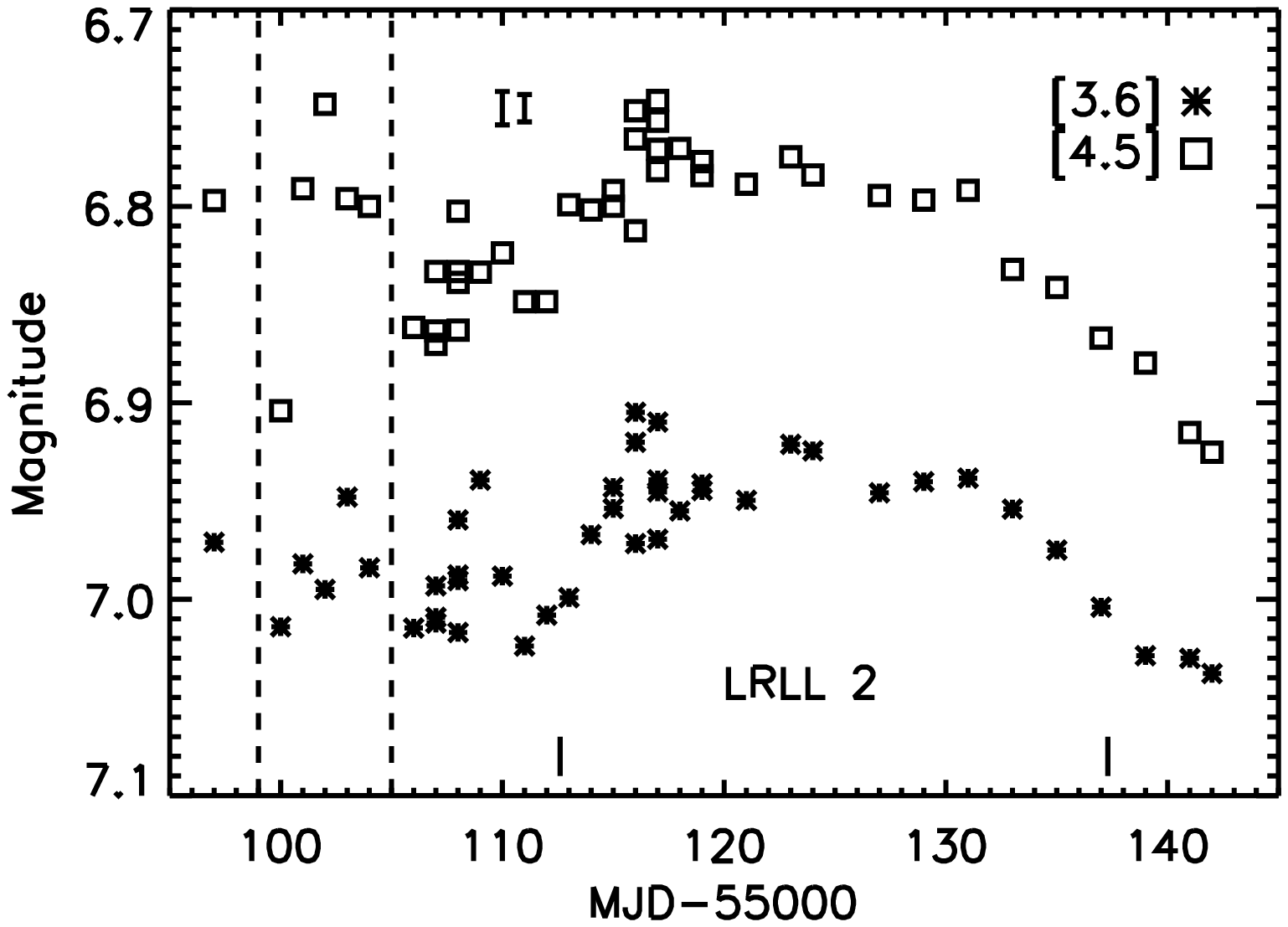}
\includegraphics[scale=.4]{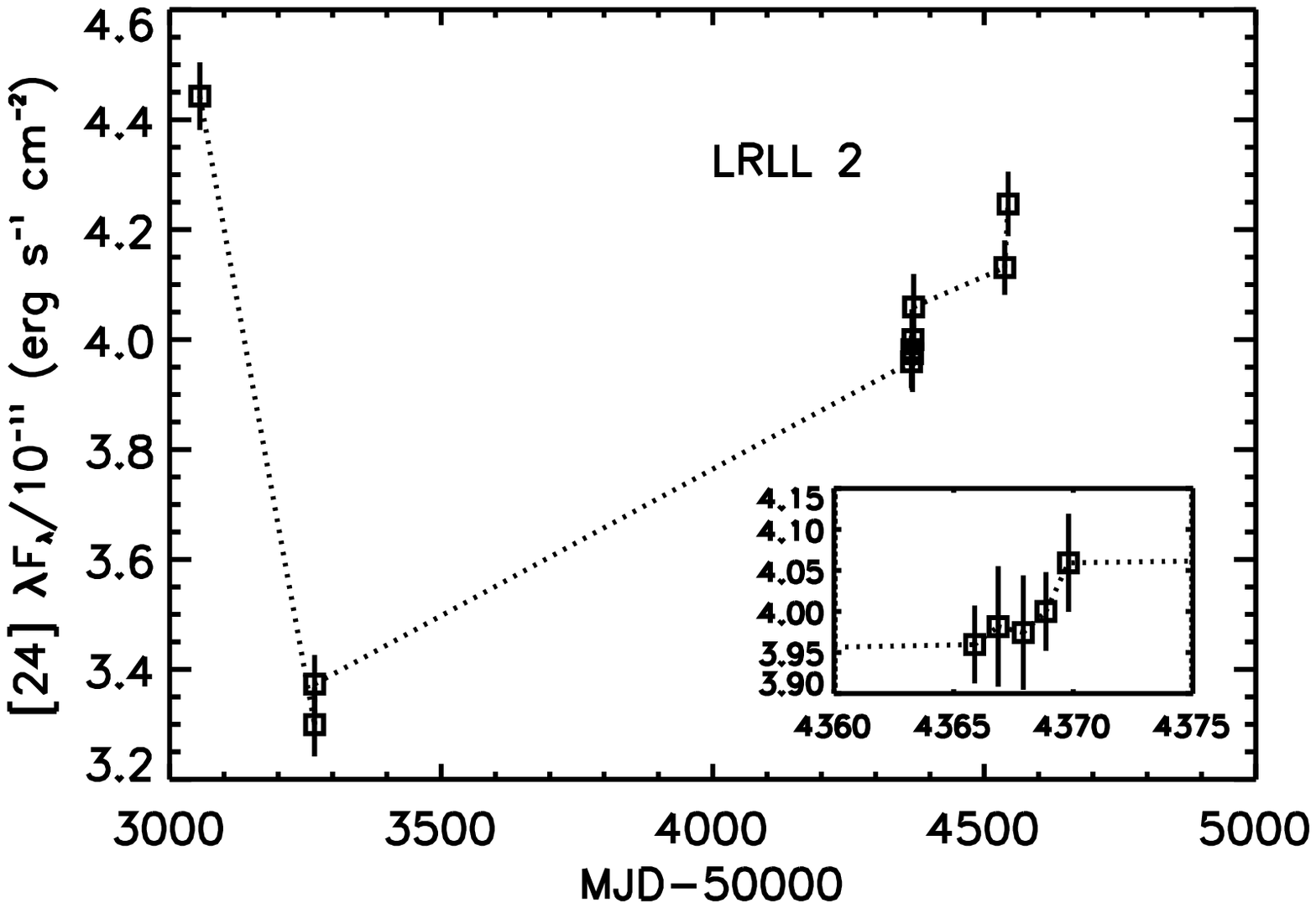}
\includegraphics[scale=.4]{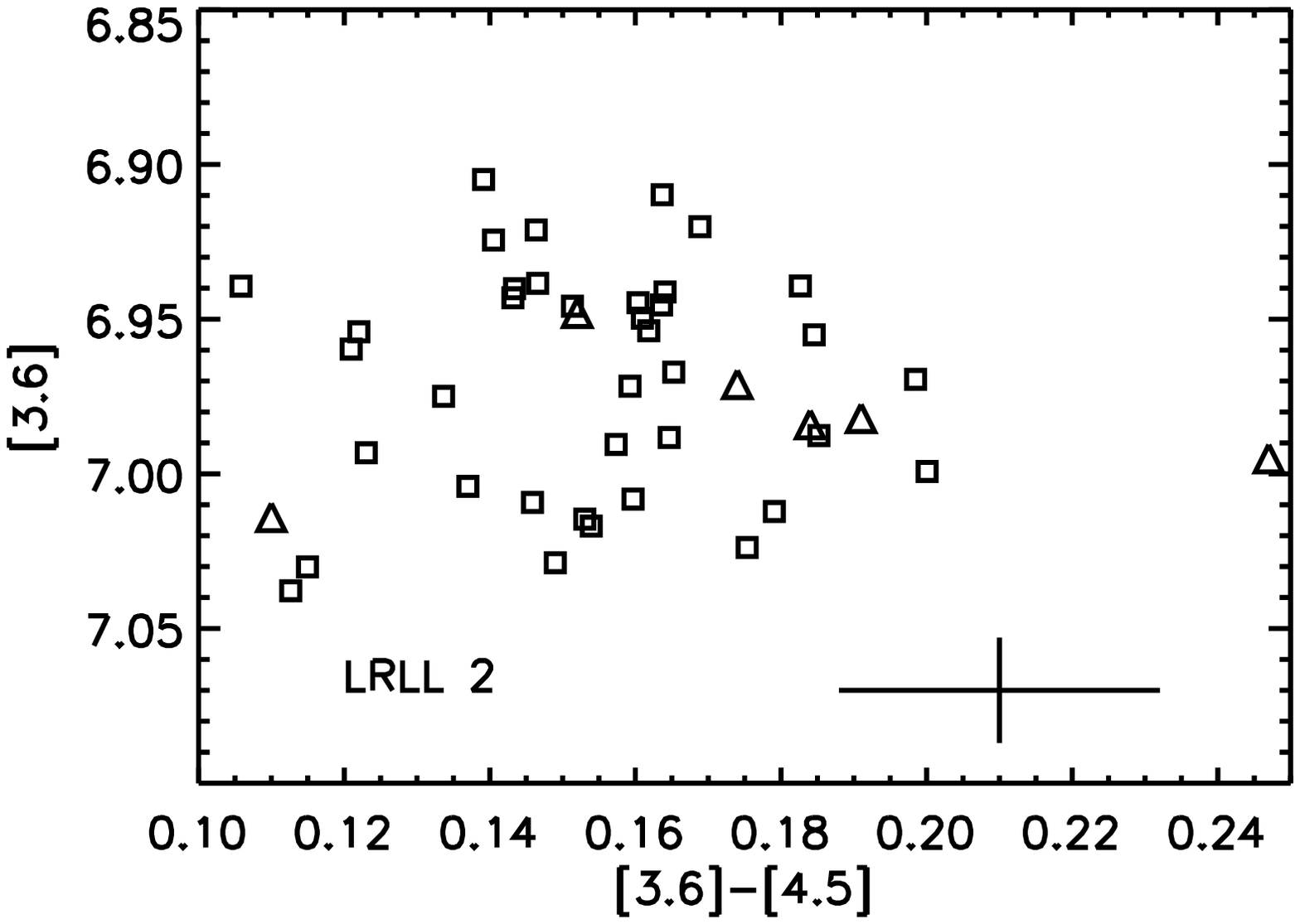}
\caption{Spitzer photometry for LRLL 2. {\it Top left} Cold mission light curves from our five consecutive days of monitoring along with earlier observations (left of dotted line). Different wavelengths are marked by different colors and lines, with the flux normalized to the median in each band. {\it Top right}: Warm-mission light curve. The error bars show the uncertainty in the 3.6 and 4.5\micron\ warm-mission photometry (left and right error bar respectively). Vertical bars show the dates when near-infrared spectra were obtained. Points to the left of the dashed line are from our five consecutive days of monitoring, as well as previous observations during the Spitzer cold-mission. {\it Bottom left}: MIPS 24\micron\ light curve. Inset zooms in on the five consecutive days of photometry. {\it Bottom right} Color-magnitude diagram with warm-mission photometry marked with squares, and cold-mission photometry marked with triangles. The error bar shows the uncertainty in the photometry. \label{lrll2_ir2}}
\end{figure*}

The IRS spectra of LRLL 2 show large ($\sim20\%$) variations in as little as one week and the change in the SED shows a wavelength dependence where the SED appears to pivot at $\lambda=6\micron$ (Fig~\ref{lrll2_ir} and \ref{lrll2_ir2} summarize all of the Spitzer data). There is a systematic offset between the LL and SL spectra ($\sim10\%$) and we reduce the LL spectra by this amount when comparing spectra from different epochs. To determine if the silicate flux changes in addition to the continuum we subtract a second order polynomial fit to the continuum (defined as $5<\lambda<7\micron$ and $13<\lambda<14\micron$) from the silicate feature to produce the result in Figure~\ref{lrll2_ir}. The flux of the silicate emission feature changes between these spectra, and is proportional to the long-wavelength continuum, similar to LRLL 31. The 24$\micron$ photometry shows flux variations up to 25\%, similar to that seen in the IRS spectra. Our five consecutive days of 24\micron\ monitoring show little change, suggesting that the largest variations are on longer timescales. This is consistent with the five consecutive days of 3-8$\micron$ photometry, which show almost no variation. Our 3.6,4.5$\micron$ monitoring offers a more complete picture of the timescale and shows slow variations ($\sim0.1$ mag) on weekly timescales with no detectable periodicity. 

\begin{figure*}
\center
\includegraphics[scale=.45]{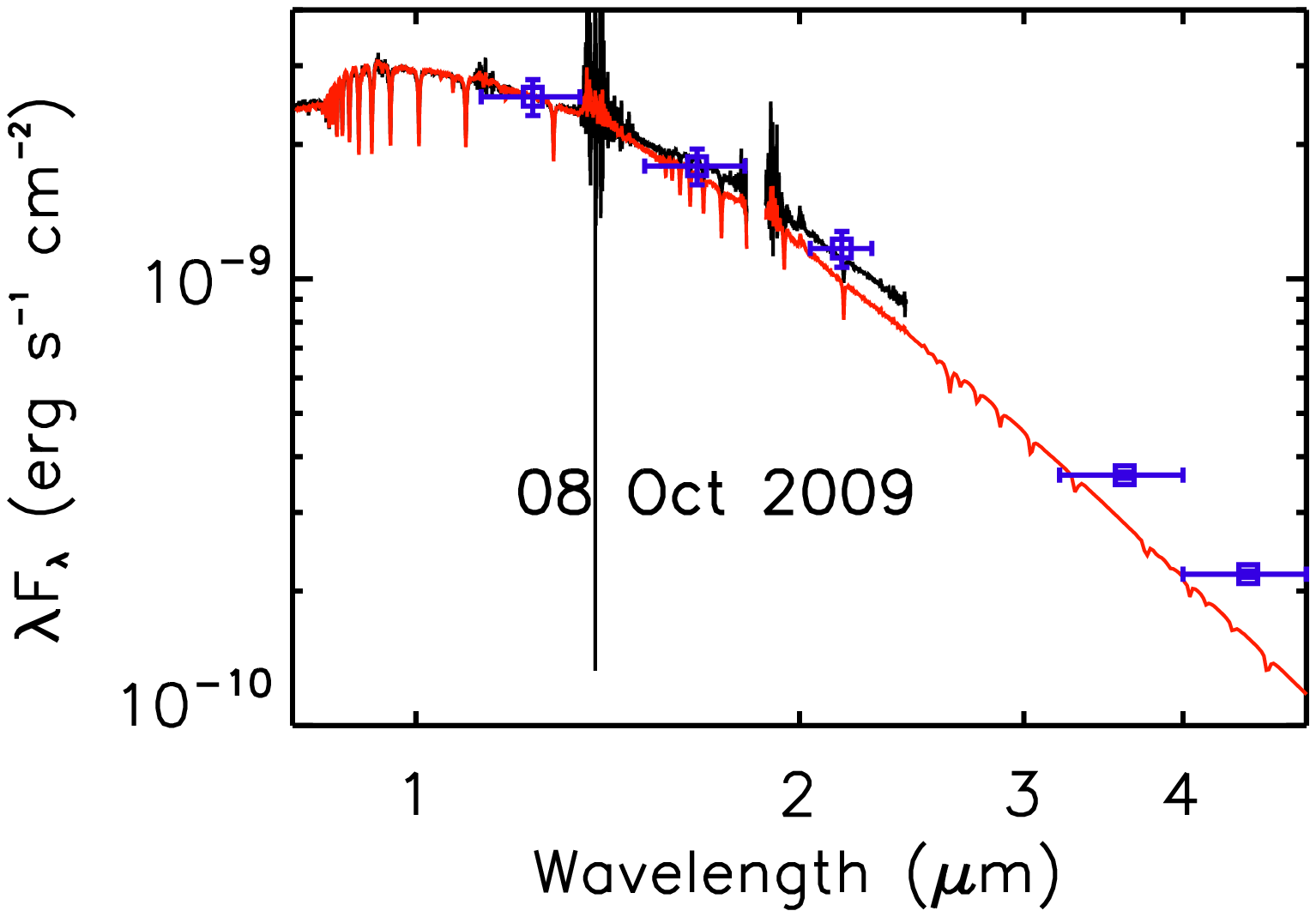}
\includegraphics[scale=.45]{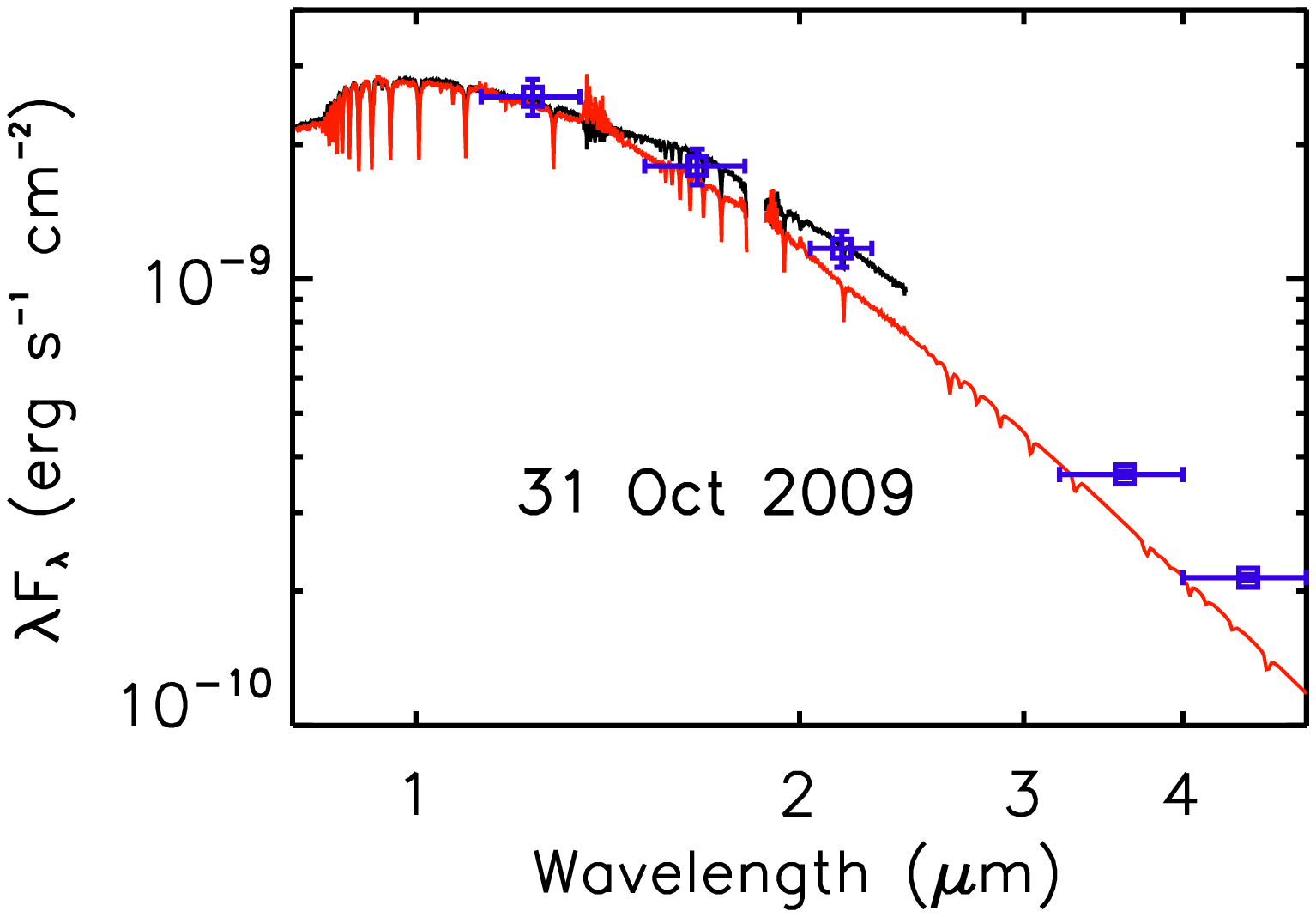}
\caption{Spex spectra for LRLL 2. Spectra have been scaled to 2MASS photometry. IRAC warm-mission photometry has been included where available. A stellar photosphere (HD 57928 + a Kurucz model) shown in red has been reddened by A$_V$=2.9 and scaled to the spectra at J band for comparison. Blue points are the photometry, and the error bars in the X-axis show the size of the passbands. \label{l2_spex+phot}}
\end{figure*}

These long-wavelength data trace material many AU from the star while our ground-based spectroscopy is sensitive to much warmer dust close to the star, assuming thermal equilibrium temperatures. Based on the shape of the 0.8-2.5$\micron$ spectra, there appears to be excess emission in K band (Fig~\ref{l2_spex+phot}). We estimate the veiling to be $\sim0.2$ in the K-band based on the shape of the spectra\footnote{Veiling cannot be determined from line EWs because of the absence of non-hydrogen photospheric absorption lines in the A2 stellar spectrum}. This is weaker than is typically seen around stars with dust in a puffed inner rim of a disk. We do not have enough spectral information to determine if the excess is due to hot dust at the dust destruction radius, or optically thin material further out, but we can say that the excess, including the 3.6 and 4.5\micron\ photometry, is consistent with optically thick dust at T$>$1000K. \citet{esp12} find that the infrared SED can be fit with optically thick material that extending inward to the dust destruction radius (1.68 AU). There model does not require an optically thin gap and instead the disk extends continuously out from the dust destruction radius and the diminished infrared flux relative to a normal disk is due to significant dust settling.

\subsubsection{Gas Properties}
\begin{figure*}
\center
\hbox{\includegraphics[scale=.5]{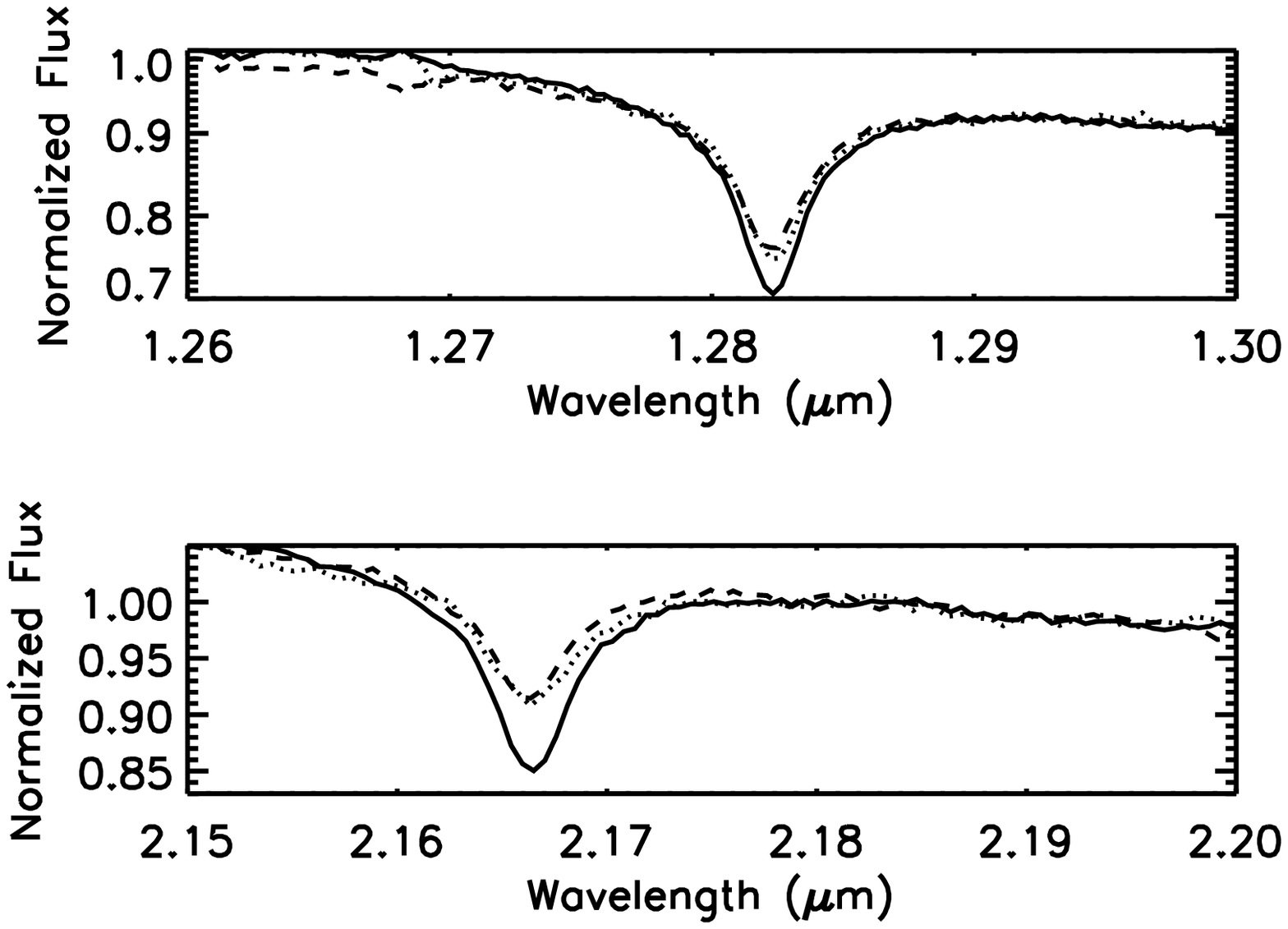}
\includegraphics[scale=.5]{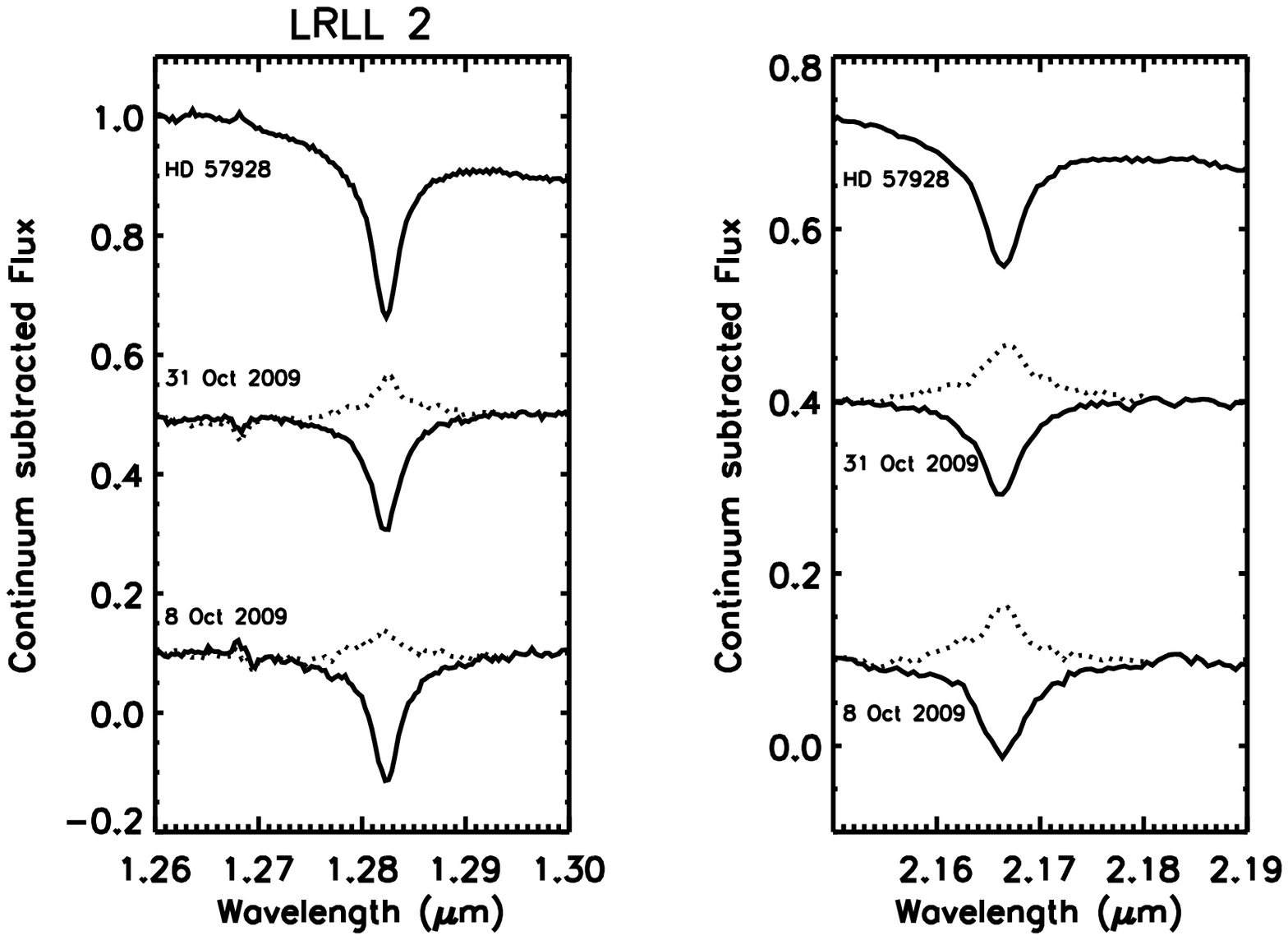}}
\caption{Pa$\beta$ and Br$\gamma$ lines for LRLL 2. On the left the two epochs (dashed and dotted lines) are directly compared, along with a WTTS standard (solid line). There is little evidence for variability between these two epochs. On the right the spectra (solid lines) are shown after the photospheric absorption has been subtracted (dotted lines). Once the strong intrinsic absorption has been removed, there is evidence for emission indicating ongoing accretion. \label{l2_lines}}
\end{figure*}

During two epochs of observations separated by 3 weeks in 2009 we see little change in the strength of the Pa$\beta$ and Br$\gamma$ lines (Fig~\ref{l2_lines}). The line strengths are weaker than simple photospheric absorption, suggesting additional emission within the system, most likely due to hot gas in the accretion flow. In Figure~\ref{l2_lines} we display the lines after the photosphere has been subtracted, showing the emission more clearly, while Table~\ref{line_strength_table} lists the measured equivalent widths (EW). The difference in EW between the two epochs is most likely due a difference in our determination of the continuum rather than real fluctuations in the accretion rate. A slight mismatch in spectral type can have a significant effect on our subtraction of the line, especially in the line wings, but does not strongly affect the conclusion that there is some emission in the line due to an accretion flow. Accretion rates derived from these infrared lines are listed in Table~\ref{accretion}. Recently \citet{don11} have shown that the relationship between Br$\gamma$ line flux and accretion rate derived from low-mass stars, which is used here, is also valid for higher mass Herbig Ae/Be stars. The average accretion rate is 3$\times10^{-7}M_{\odot}$yr$^{-1}$ and the accretion luminosity is small relative to the stellar luminosity (L$_{acc}$/L$_*$=0.06). The uncertainty in the accretion rate is a factor of 2 due to possible spectral type mismatch, uncertainties in the location of the continuum and the uncertainty in the conversion from line flux to accretion rate.

\subsection{LRLL 21}
\subsubsection{Stellar Properties}
LRLL 21 is a K0 star originally classified by \citet{luh03}, whose stellar flux varies substantially. We were able to obtain seven epochs of near-infrared photometry concurrent with our ground-based infrared spectra, which includes three epochs in 2008 separated by one week and 4 epochs in 2009 spread out over one month. Details on the reduction of these data, along with how the luminosity is derived from the J band photometry, can be found in \citet{fla11}. We also utilize 2MASS photometry taken in 1998. We find large year-to-year variations in the J band flux of 0.4 mag (Table~\ref{nir_phot}) but no daily or weekly fluctuations larger than 0.1 mag. In 2009, when we have contemporaneous near-infrared and 3.6,4.5\micron\ photometry, we find no significant change in the J band flux from Oct 8 to Nov 8 despite a large increase in the mid-infrared flux over this same period. Optical photometry from \citet{luh03} (I=13.21) and \citet{cie06} (R=14.69) is consistent with the low stellar flux state seen in 2008. \citet{cie06} find small short-term optical fluctuations likely due to cool spots rotating across the surface of the star and deduce a rotation period of 2.5 days. This is consistent with the lack of large daily and weekly changes in the stellar flux in our near-infrared photometry. Our estimates of the luminosity from the J band fluxes range over 2.52-3.66L$_{\odot}$ (Table~\ref{lstar}), which corresponds to M$\sim$1.8$M_{\odot}$ based on the \citet{sie00} isochrones. We measure a constant extinction (A$_V$=4.2, Table~\ref{extinction}) during each epoch of near infrared photometry, suggesting that the obscuration is not variable, unless the obscuring source has large enough grains that it does not redden the central star, although the presence of very large grains in the disk is inconsistent with the strong silicate emission feature. 

There is also evidence that this system is a binary. Based on our high-resolution optical spectra, we find two peaks in the cross-correlation function, one at -15.7 km/sec and one at 66.8 km/sec (Table~\ref{velocity}). \citet{dah08} measure a radial velocity of +17.01 km/sec, but do not mention any sign of binarity. It is possible that \citet{dah08} observed the system during a phase when the two components were aligned along the line of sight and there was no significant difference in their radial velocities. 

\begin{figure*}
\center
\includegraphics[scale=.3]{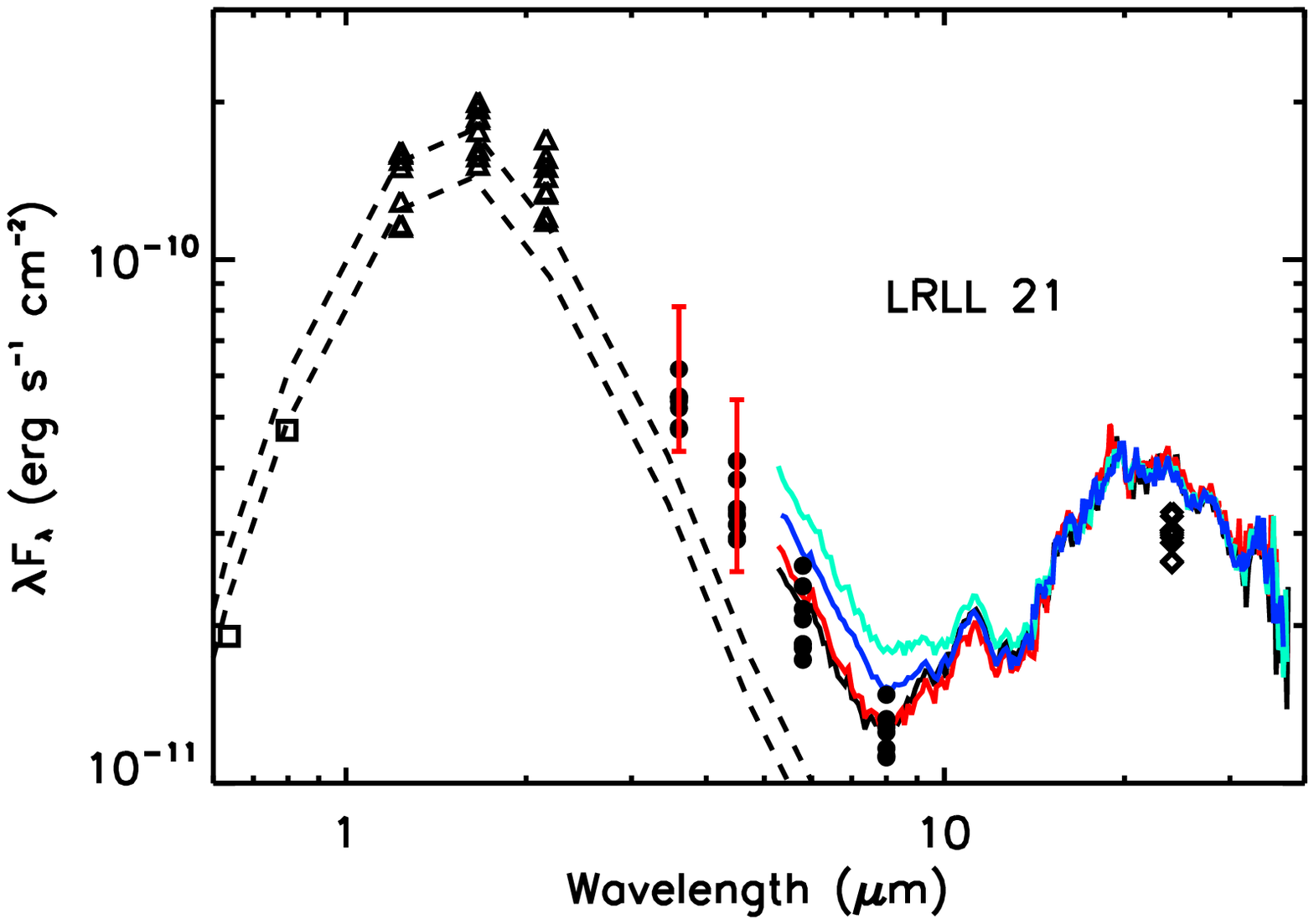}
\includegraphics[scale=.3]{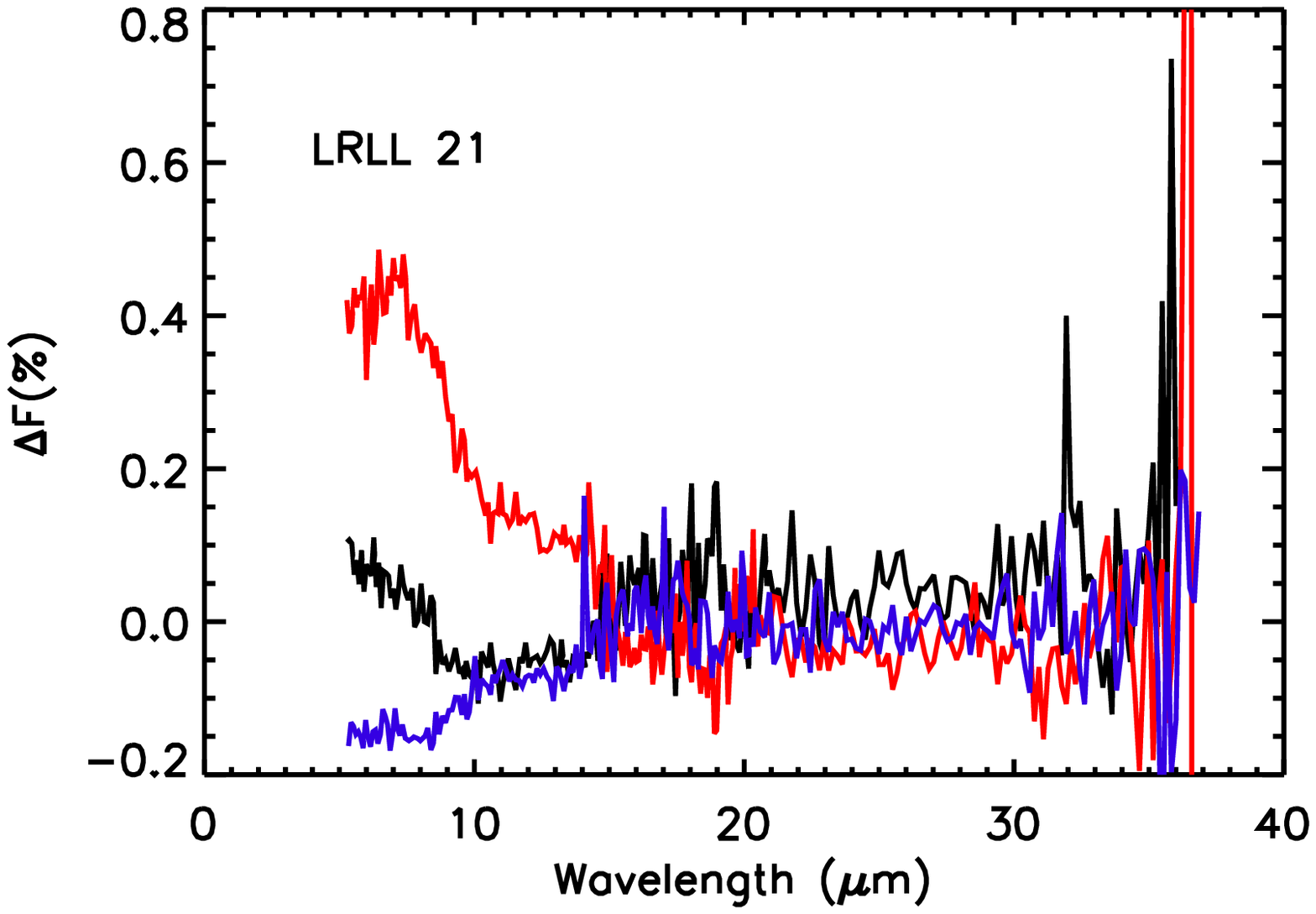}
\includegraphics[scale=.3]{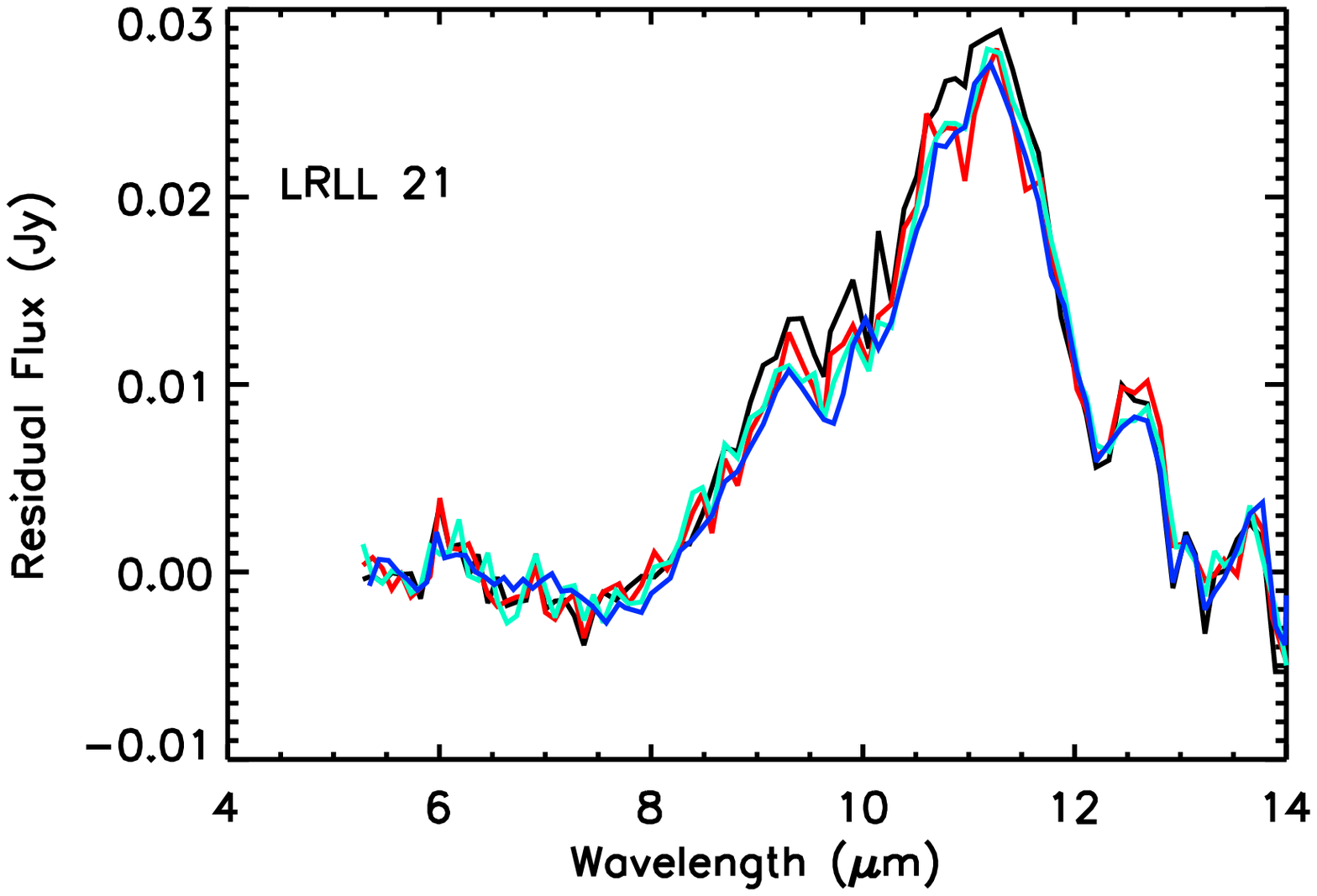}
\caption{Same as Figure~\ref{lrll2_ir} but for LRLL 21. Two photospheres are shown in the SED plot, corresponding to the high and low state of the stellar flux.\label{lrll21_ir}}
\end{figure*}

\begin{figure*}
\center
\includegraphics[scale=.4]{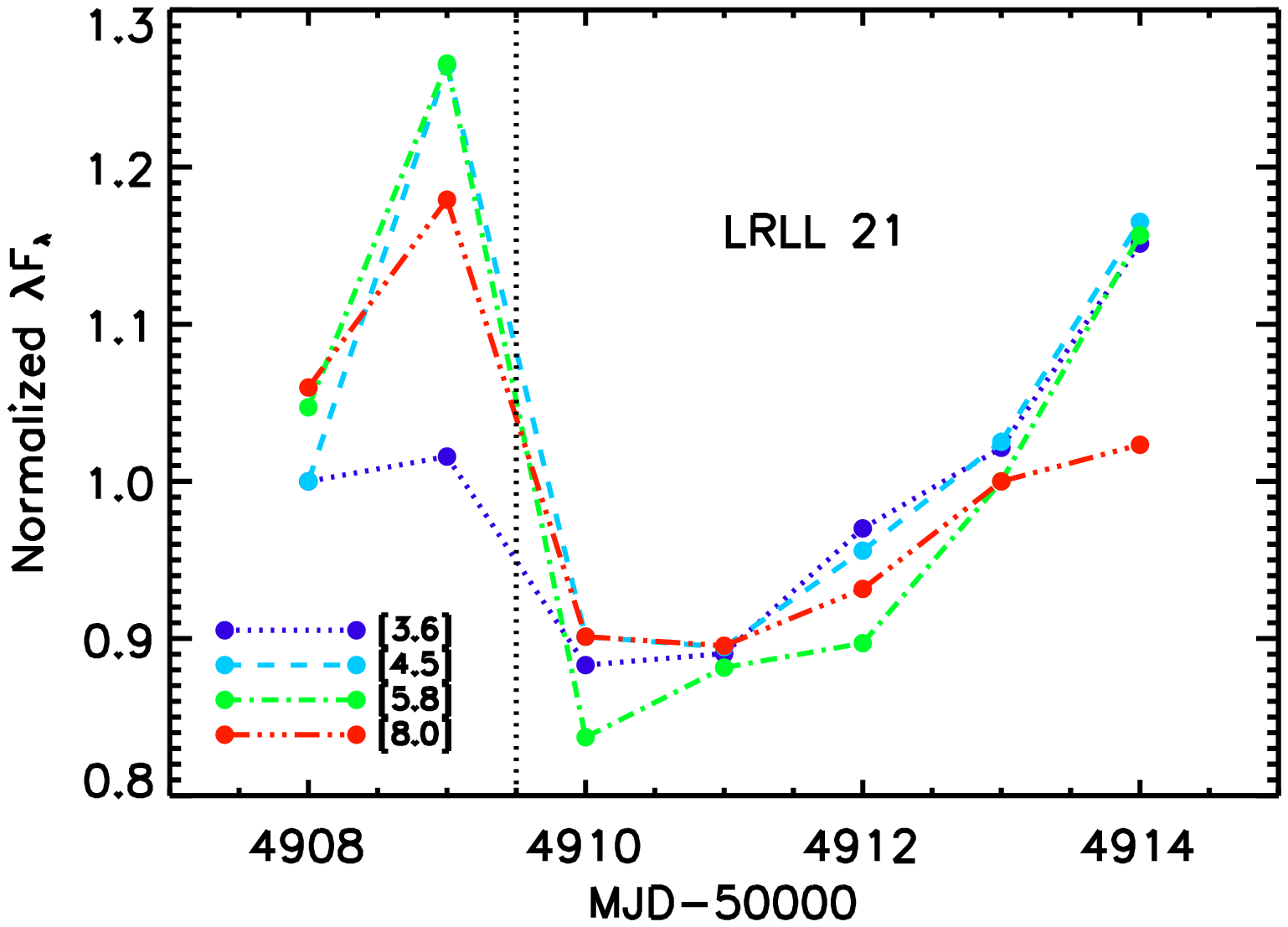}
\includegraphics[scale=.4]{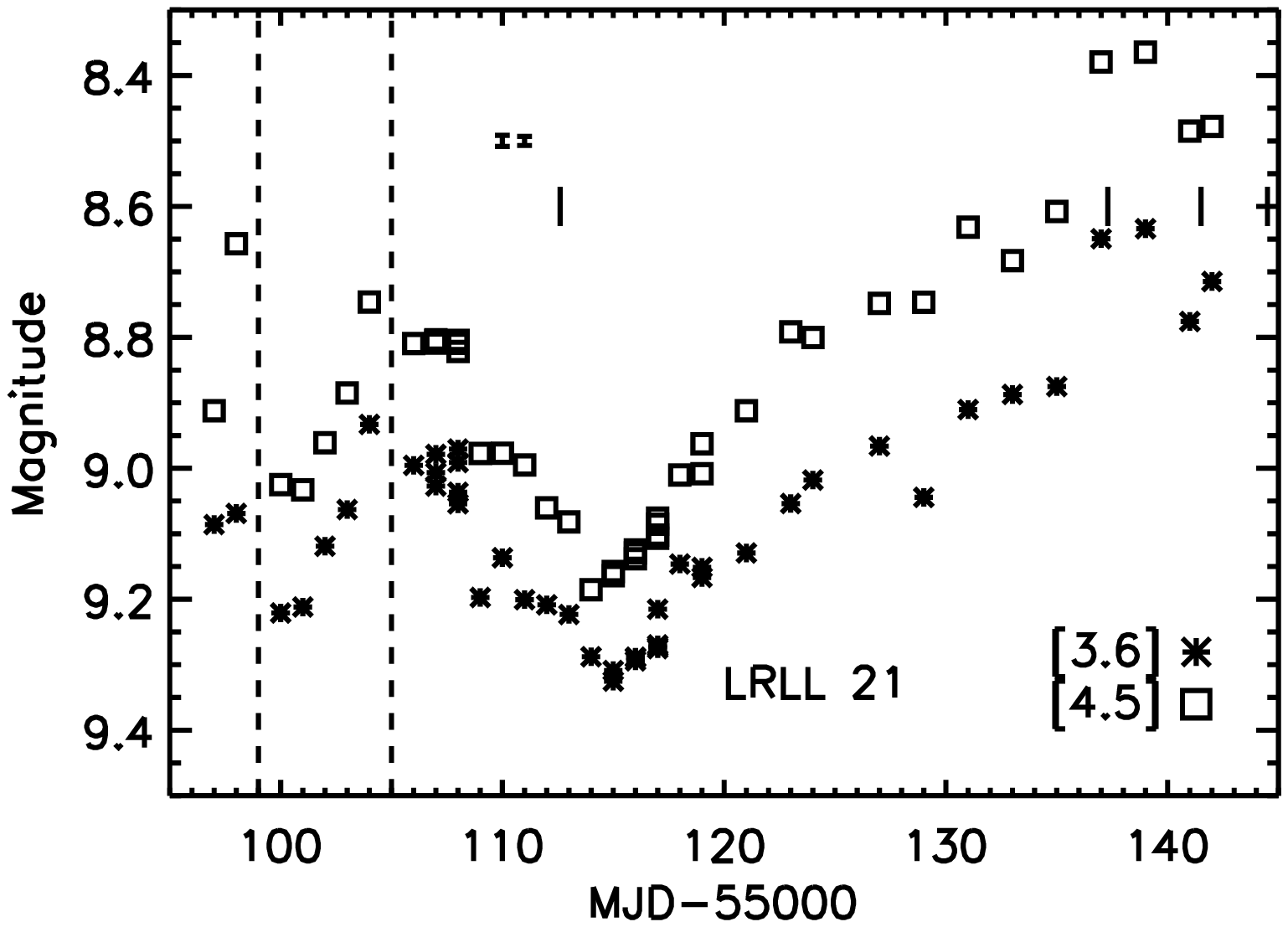}
\includegraphics[scale=.4]{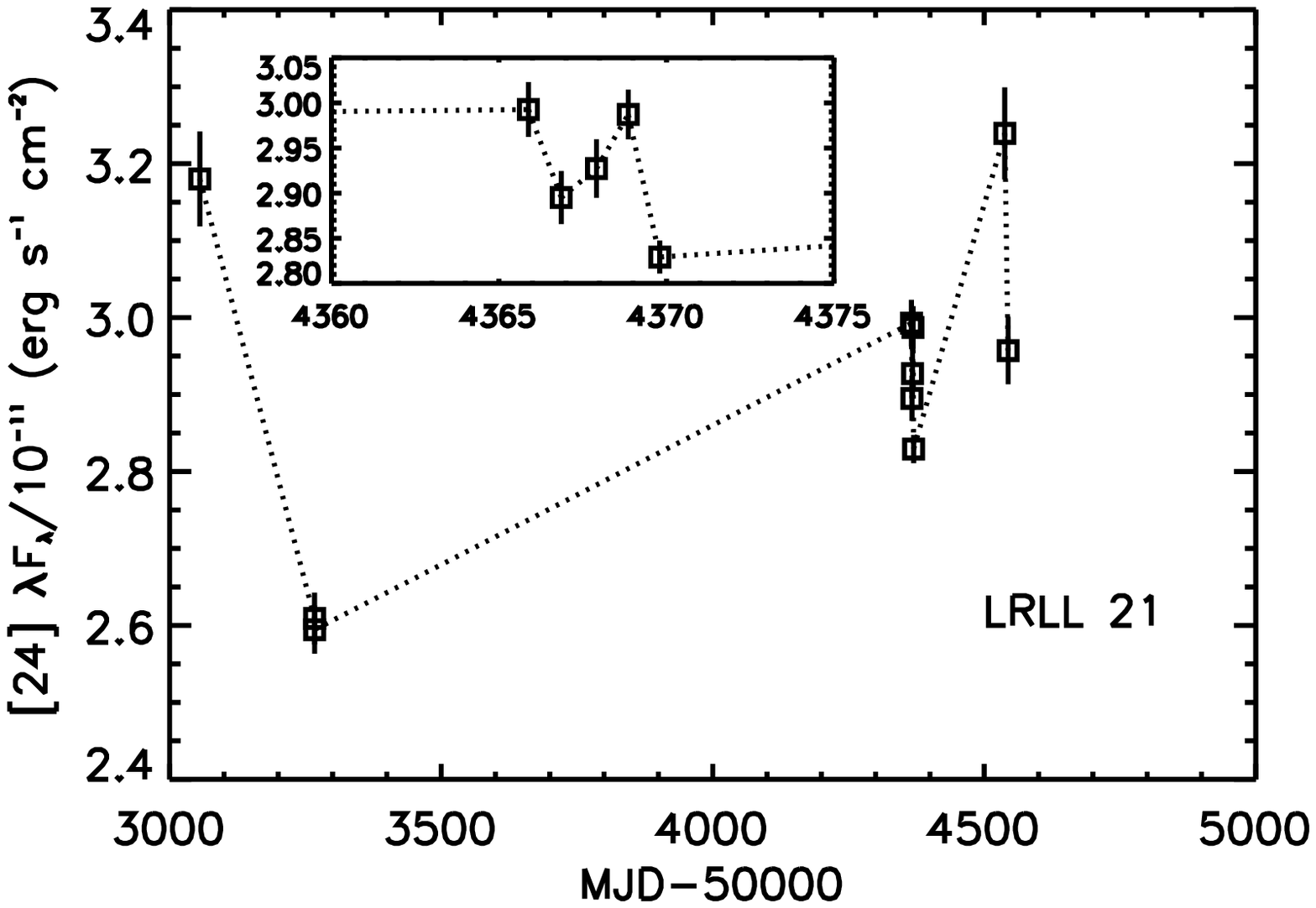}
\includegraphics[scale=.4]{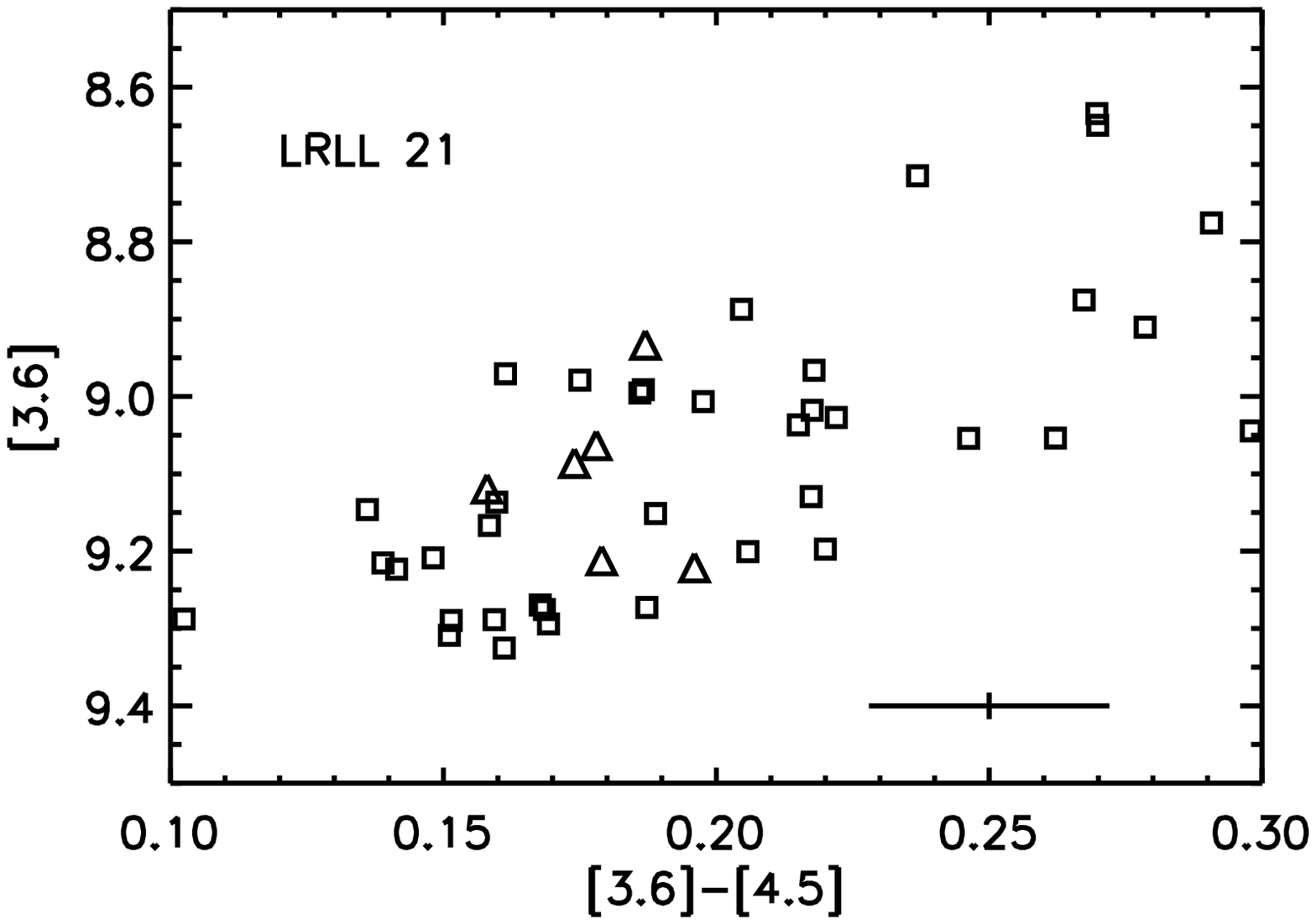}
\caption{Same as Figure~\ref{lrll2_ir2} but for LRLL 21. \label{lrll21_ir2}}
\end{figure*}

\subsubsection{Infrared Variability}
In the IRS spectra, LRLL 21 displays large variations ($\sim40\%$) at $\lambda<15\micron$ in observations separated by one week with no fluctuations ($<8\%$) at longer wavelengths (Fig~\ref{lrll21_ir}). However, the 24$\micron$ photometry does show fluctuations up to $20\%$ (Fig~\ref{lrll21_ir2}), and this flux is systematically lower than the IRS spectra, suggesting that the long wavelength flux does vary. The flux of the silicate feature does not change with time, despite the change in the underlying continuum. Our five consecutive days of 3-8$\micron$ photometry, as well as our 3.6,4.5$\micron$ monitoring show rapid variations even from one day to the next, although the predominate timescale appears to be weeks and months with no sign of periodicity. Of all the stars in our sample, this one shows the largest fluctuations.

\begin{figure}
\includegraphics[scale=.5]{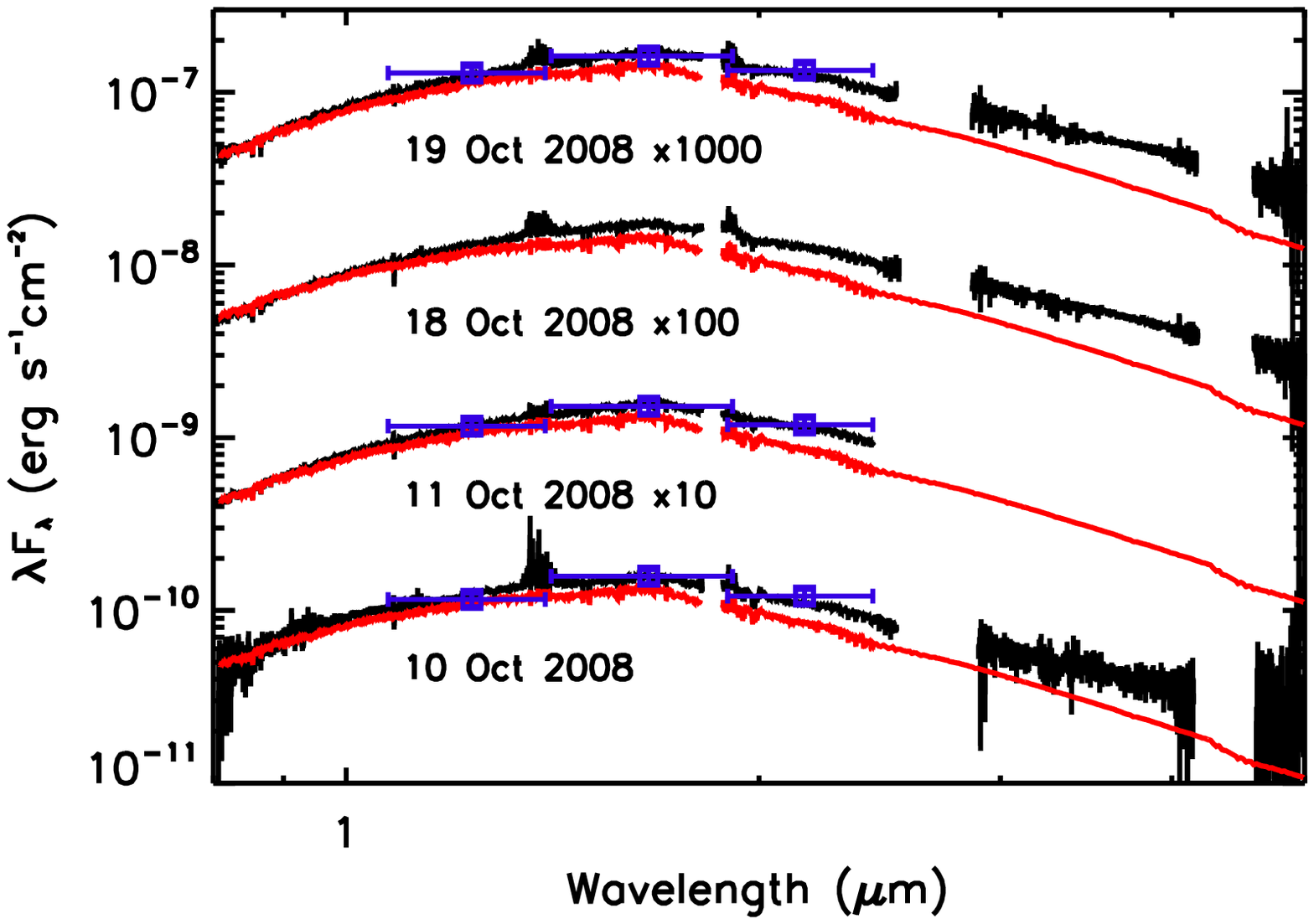}
\includegraphics[scale=.5]{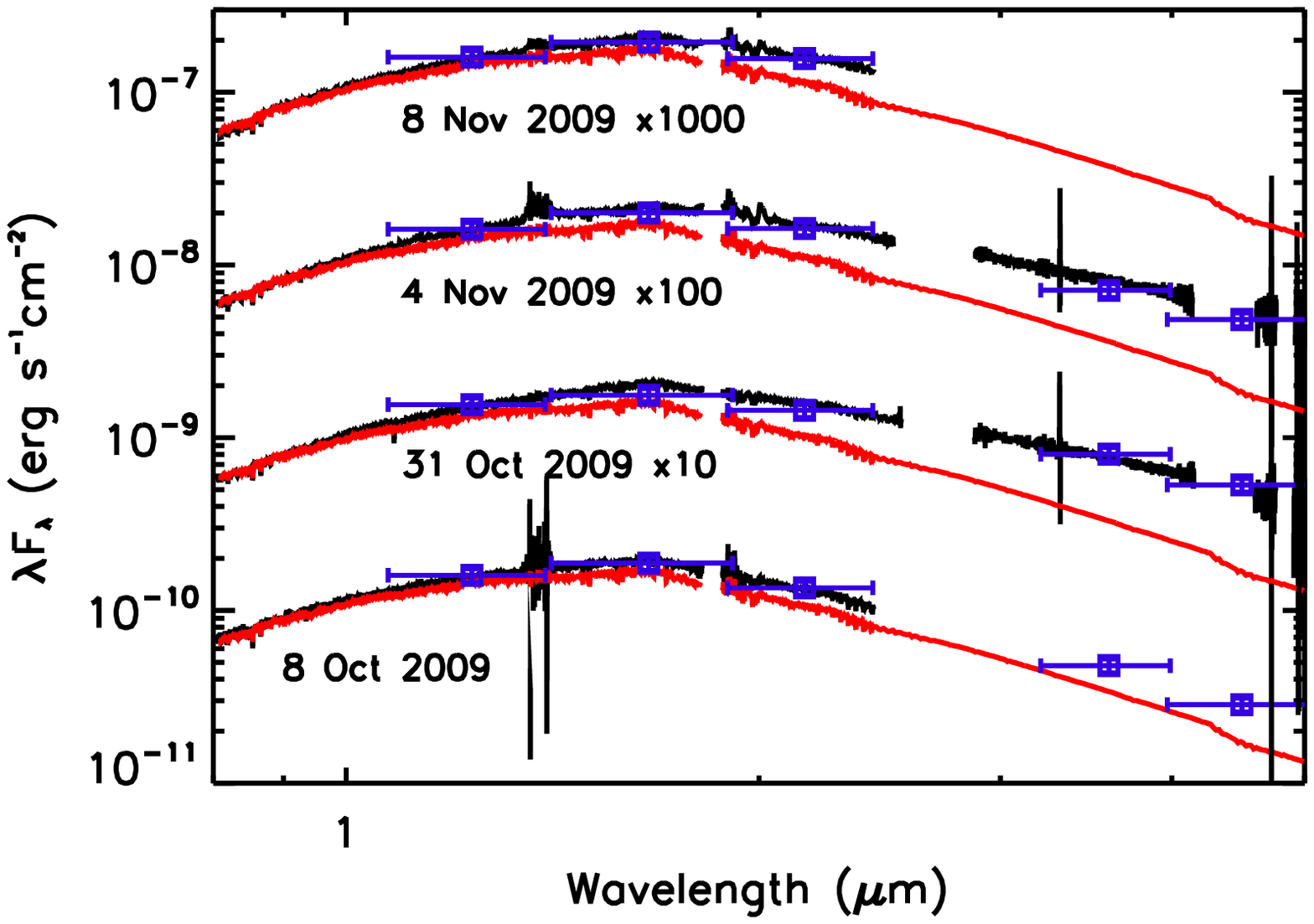}
\caption{Spex spectra (black line) and photometry (blue points) for LRLL 21. Spectra have been scaled to match photometry. A stellar photosphere (LkCa 19 + Kurucz model) shown in red has been reddened by the A$_V$ measured on each night to LRLL 21 and scaled to the short wavelength flux of the spectra. Spectra have been scaled by a constant factor for clarity.\label{l21_spex+phot}}
\end{figure}

\begin{figure}
\includegraphics[scale=.5]{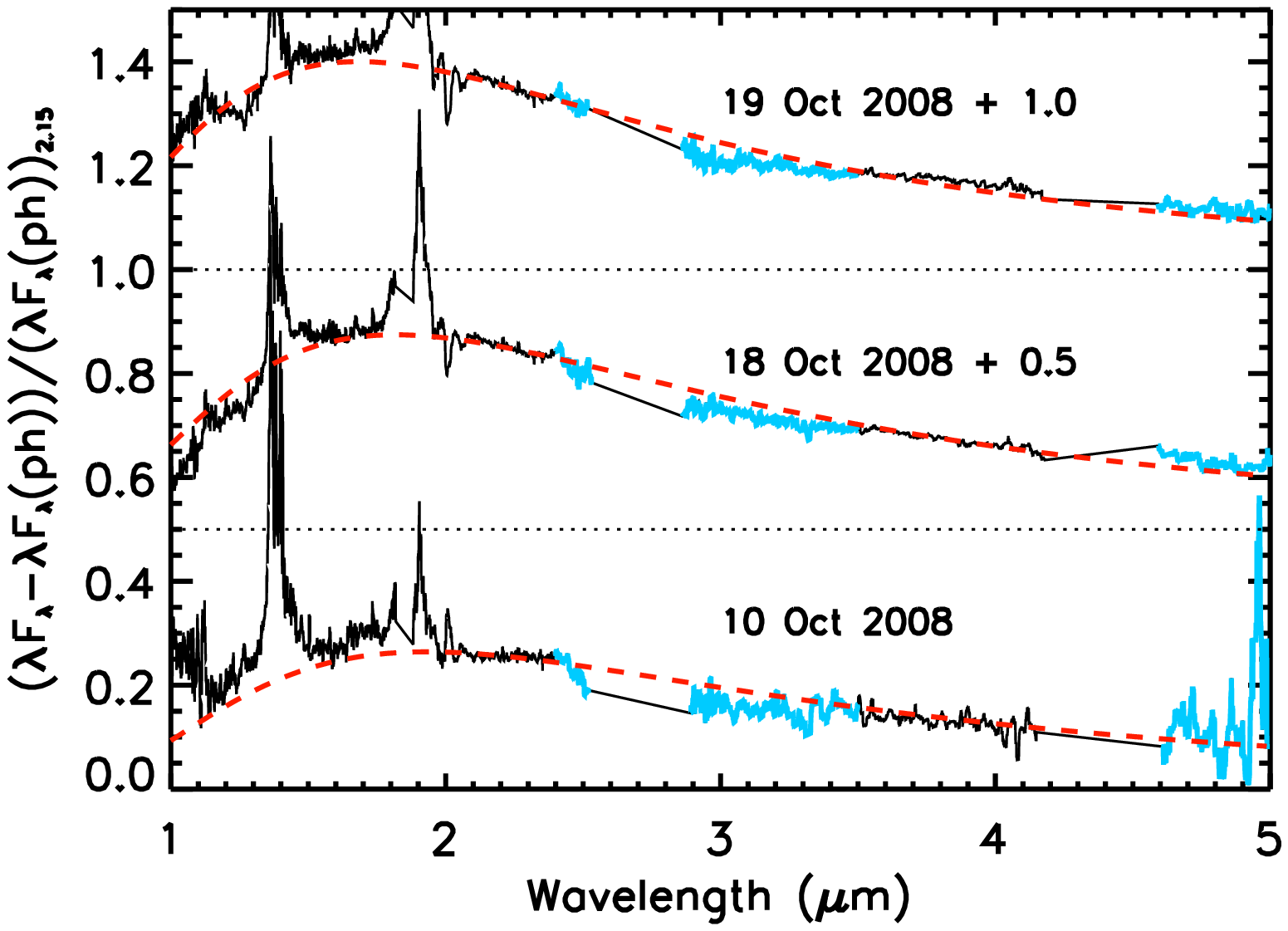}
\includegraphics[scale=.5]{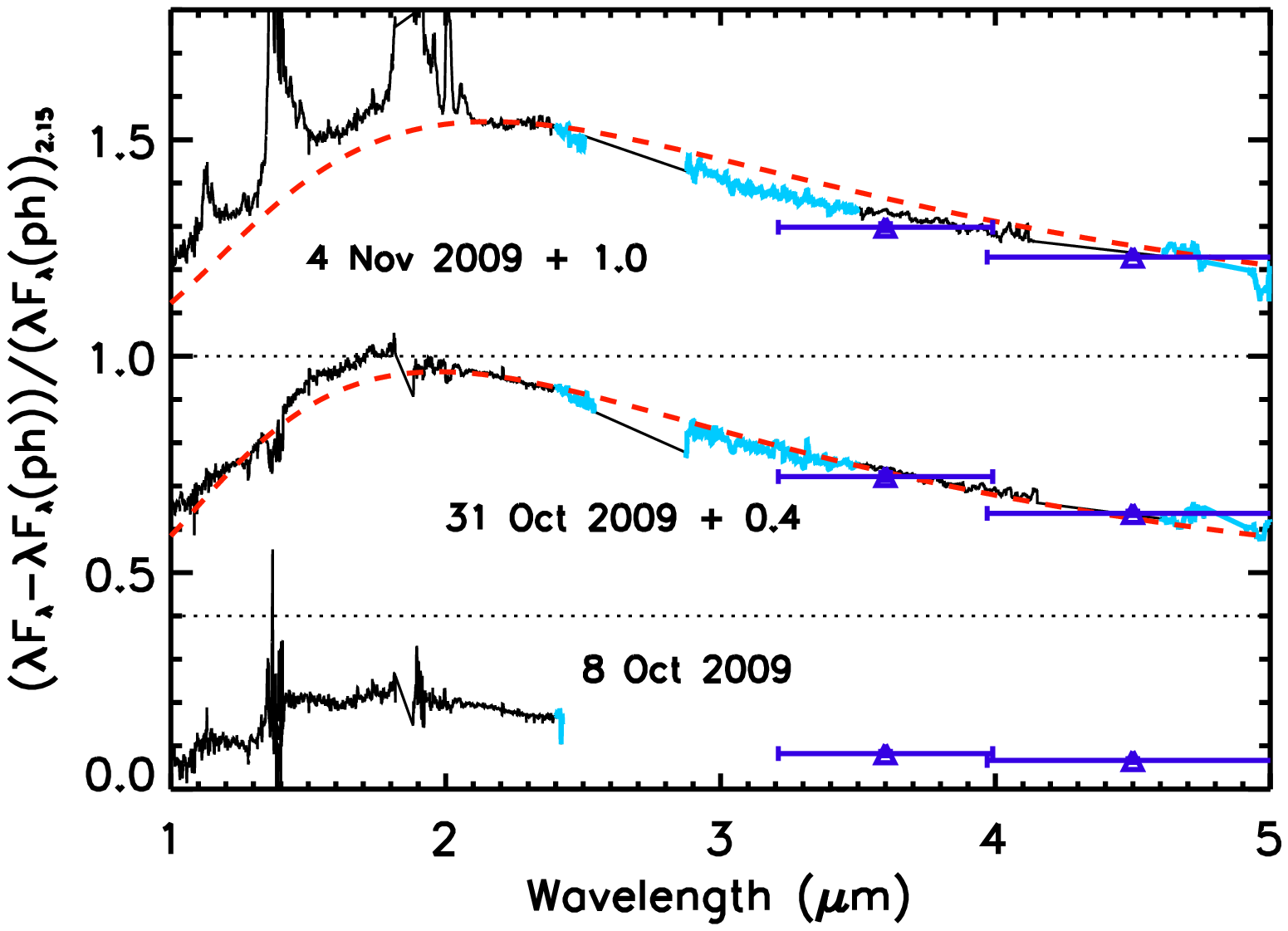}
\caption{The infrared excess of LRLL 21. Each plot shows the difference between LRLL 21 and the standard, normalized to the photospheric flux at 2.15\micron. Some have been moved upward for clarity and the horizontal dotted line shows the zero point for each excess spectra The best fit blackbody is shown with a red dashed line and the 3.6,4.5\micron\ photometry is included when available. Each spectrum has been smoothed by a median filter 0.01\micron\ wide to reduce the noise in the continuum. The parts of the spectra marked in blue are strongly affected by the telluric correction. Note that the scale changes for each plot. The strength of the excess shows large variations from night to night, although the temperature stays roughly constant.\label{l21_excess}}
\end{figure}

\begin{figure}
\includegraphics[scale=.5]{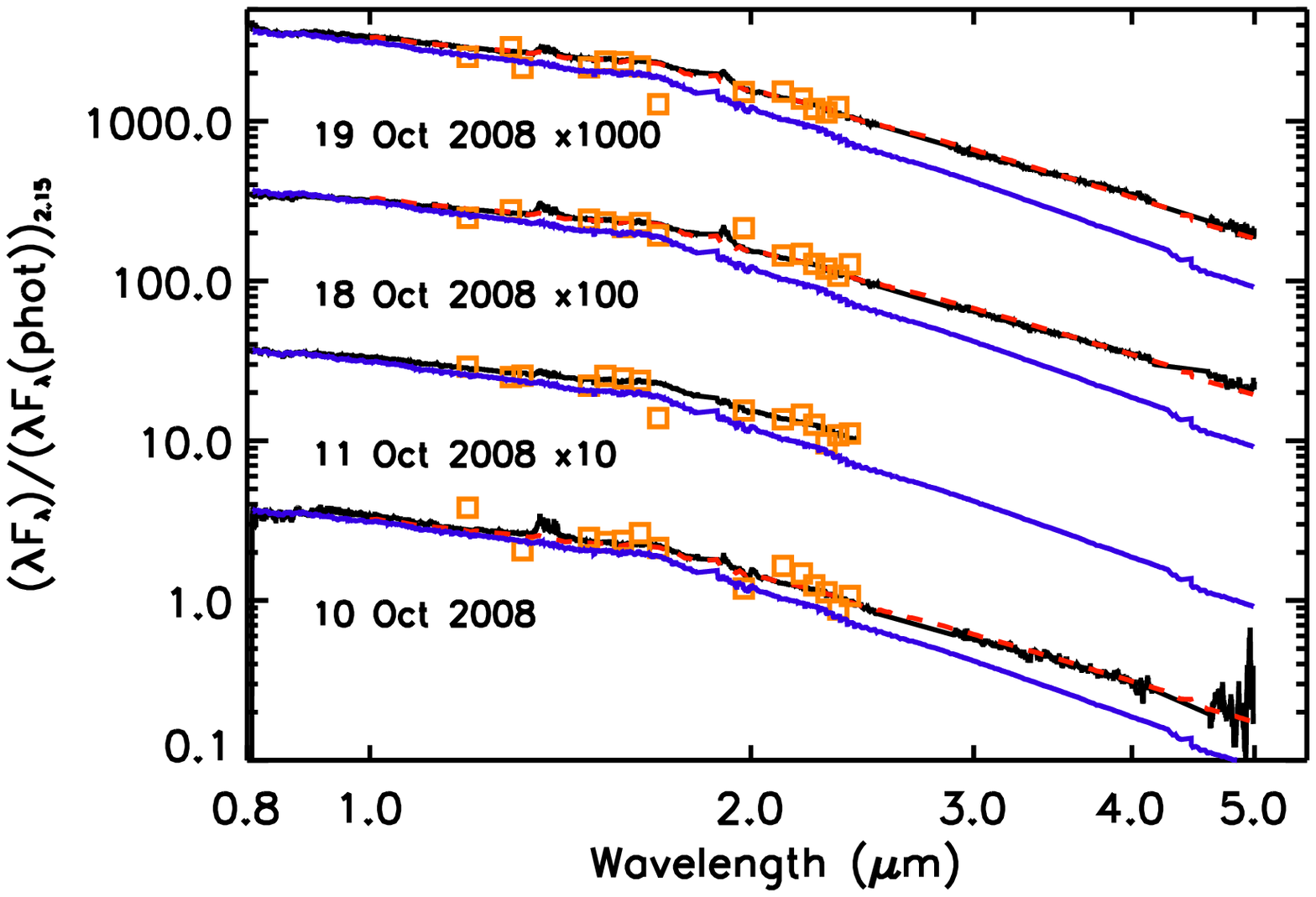}
\includegraphics[scale=.5]{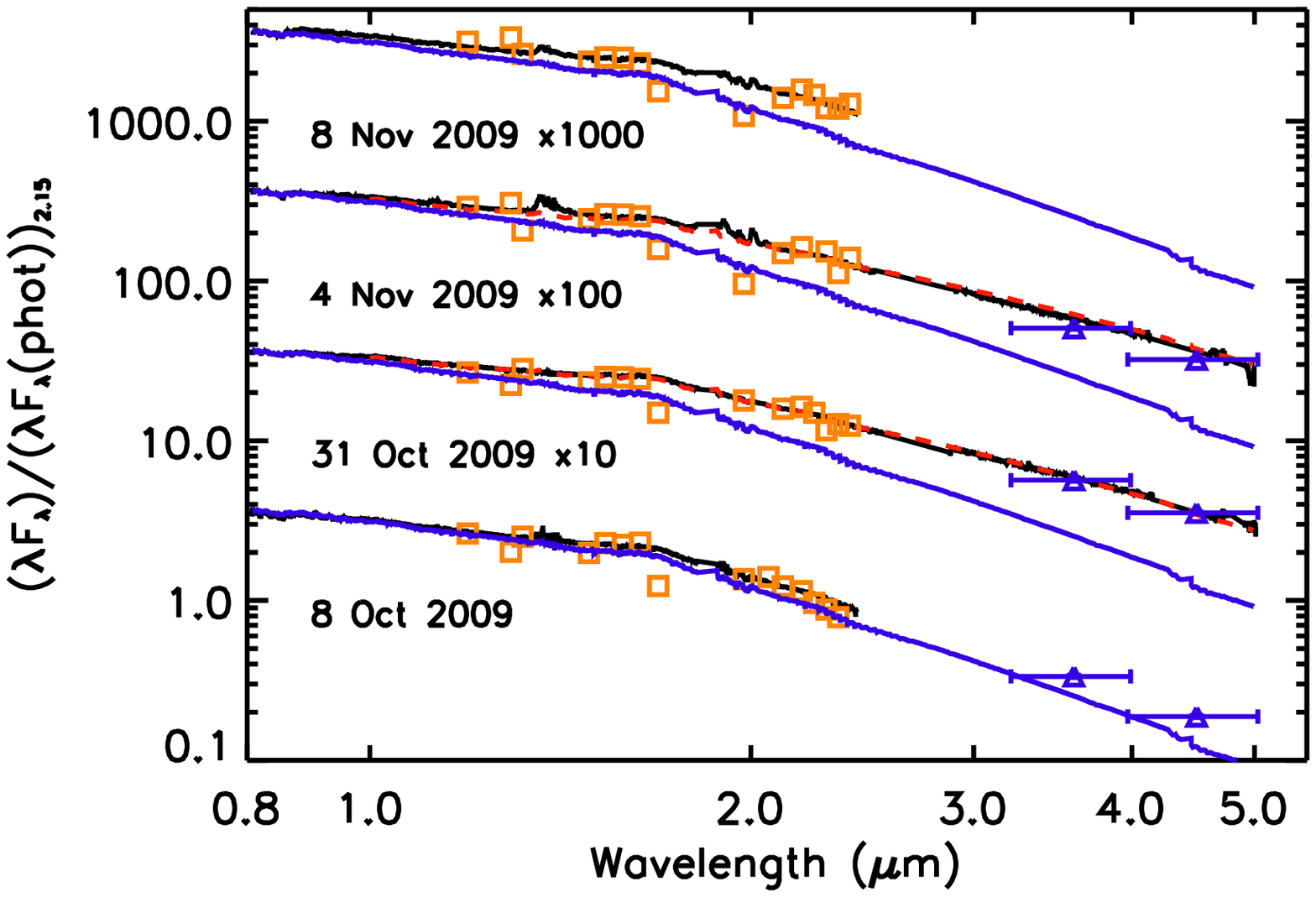}
\caption{Dereddened spectra of LRLL 21, along with the veiling measurements (orange squares), 3.6 and 4.5\micron\ photometry where available (blue triangles) and a stellar photosphere for comparison (blue line). Red dashed line is the best fit photosphere+blackbody emission for those days where we have spectra beyond 2.5\micron. LRLL 21 shows a large range in the strength of the excess throughout our observations.\label{l21_excess2}}
\end{figure}

For LRLL 21 we have ground-based spectra (Fig~\ref{l21_spex+phot}) allowing us to measure the shape of the excess emission from 0.8-5\micron, which in most T Tauri stars is dominated by emission from the optically thick inner wall at the dust destruction radius \citep{muz03}. Here we briefly summarize our procedure for deriving the excess spectrum with more details in \citet{fla11}. First we measure the veiling ($r=F_{excess}/F_{photosphere}$) by comparing the LRLL 21 spectra to that of a WTTS of the same spectral type (LkCa 19) in fifteen small bins from 0.8 to 2.5\micron. We fit a line to these veiling measurements as a function of wavelength to derive the K-band (2.15\micron) veiling. We use this value to normalize the LRLL 21 spectra to the photosphere level, whose shape is estimated using our K0 WTTS standard along with a Kurucz model extension (T$_{eff}$=5250, log g =2.5) beyond 2.5\micron. The LRLL 21 spectra and the photosphere are then subtracted to produce a spectrum of the infrared excess. The 0.8-5$\micron$ excess emission, shown in Figures~\ref{l21_excess},\ref{l21_excess2}, is consistent with an optically thick disk with a temperature of 1900 K at every epoch (Table~\ref{ir_excess_table}), while the strength of the emission rapidly varies with K-band veiling measurements ranging from 0.19 to 0.56 (Table~\ref{ir_excess_table}). The typical uncertainties in the derived temperature and veiling are 200K and 0.1 respectively. \citet{esp12} are able to fit the infrared SED with a disk model that includes an optically thick inner disk ring at the sublimation radius (0.13 AU), consistent with our near-infrared spectra, followed by a gap of optically thin dust that extends to 9 AU. In 2009 we see a large increase in veiling from Oct 8 to Oct 31, which is consistent with the large increase in infrared flux seen in the 3.6 and 4.5\micron\ photometry during the same time period. The dust temperature we derive is slightly higher than expected for the sublimation of silicate dust, which occurs closer to 1500 K, similar to the high temperature we observed in LRLL 31 \citep{fla11}. Not taking account of a possible massive companion (discussed above) could lead us to overestimate the dust temperature by $\sim$100 K and underestimate the K-band veiling by $\sim$0.1.  It is also possible that the inner disk is not entirely optically thick. If the inner disk is azimuthally asymmetric, with part of the disk being optically thick while part is optically thin, then trying to fit this emission as an optically thick blackbody will lead to an overestimation of the temperature by a few hundred kelvin. Since the temperature stays constant, these systematic effects are also likely constant and should not effect our interpretation of the variability. 

The dominant change in the system is in the strength of the disk emission, not the temperature of the dust. The [3.6]-[4.5] color of the system gets redder as the flux increases (Fig~\ref{lrll21_ir2}). This is consistent with blackbody emission from dust, on top of the stellar continuum emission, increasing in strength while not changing in temperature.  Despite being higher than expected, the fact that the dust temperature is close to the sublimation temperature suggests that the inner edge of the disk is set by dust sublimation and not the dynamical carving of a companion. The radius at which this occurs depends on the grain properties and the flux striking the surface, which in the case of LRLL 21 is dominated by the stellar flux (see discussion below). Since the stellar flux is relatively constant on the timescales over which the infrared flux varies, the radius of the inner disk should be constant. The changing flux implies that the emitting area of the disk is varying, and if the radius of the inner disk is constant then the scale height must be varying. 

Since the 2-5\micron\ excess is reprocessed stellar flux, the ratio of the excess to the stellar flux represents the covering fraction of the inner disk, which we can use to quantify the strength of the excess on different epochs. We calculate the total dust emission by integrating a blackbody at the measured strength and temperature. For those days in 2009 when we do not have 2-5\micron\ spectra we instead use the 3.6 and 4.5\micron\ photometry to estimate the strength of the excess and assume a temperature of T$_{dust}$=1900K. The results are reported in Table~\ref{ir_excess_table}; for LRLL 21 the covering fraction varies from $\sim4\%$ up to 13\%, while typical T Tauri stars have inner disks with covering fractions of 12\% \citep{dul01}. On average the covering fraction of the inner rim of LRLL 21 is lower than a typical T Tauri star, which may be due to advanced grain growth and settling at the very inner edge of the disk, although the strong silicate emission feature is inconsistent with significant grain growth where this flux arises. Another possibility is that the inner rim does not azimuthally extend all the way around the star, as was suggested earlier to explain the high measured dust temperature. If instead the inner disk has patches of optically thin material interspersed between optically thick dust, then the total covering fraction could be reduced. This may explain the periods when the emission is especially low, which would otherwise require an extremely flat disk seen close to edge on. The lowest measured covering fraction would correspond to only one-quarter of the inner disk being occupied by optically thick material, while the rest was either optically thin or devoid of dust. 

\begin{figure}
\center
\includegraphics[scale=.5]{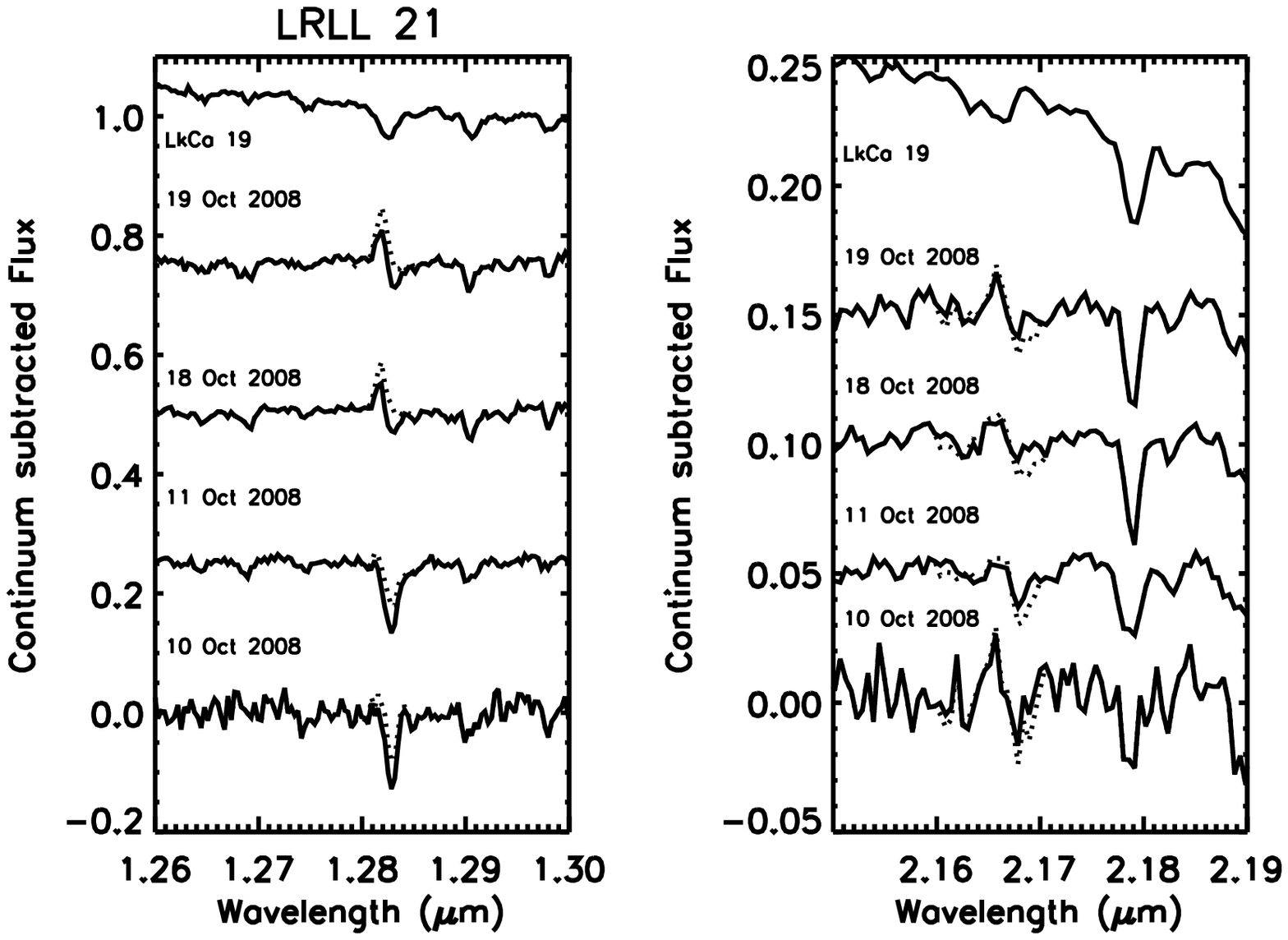}
\includegraphics[scale=.5]{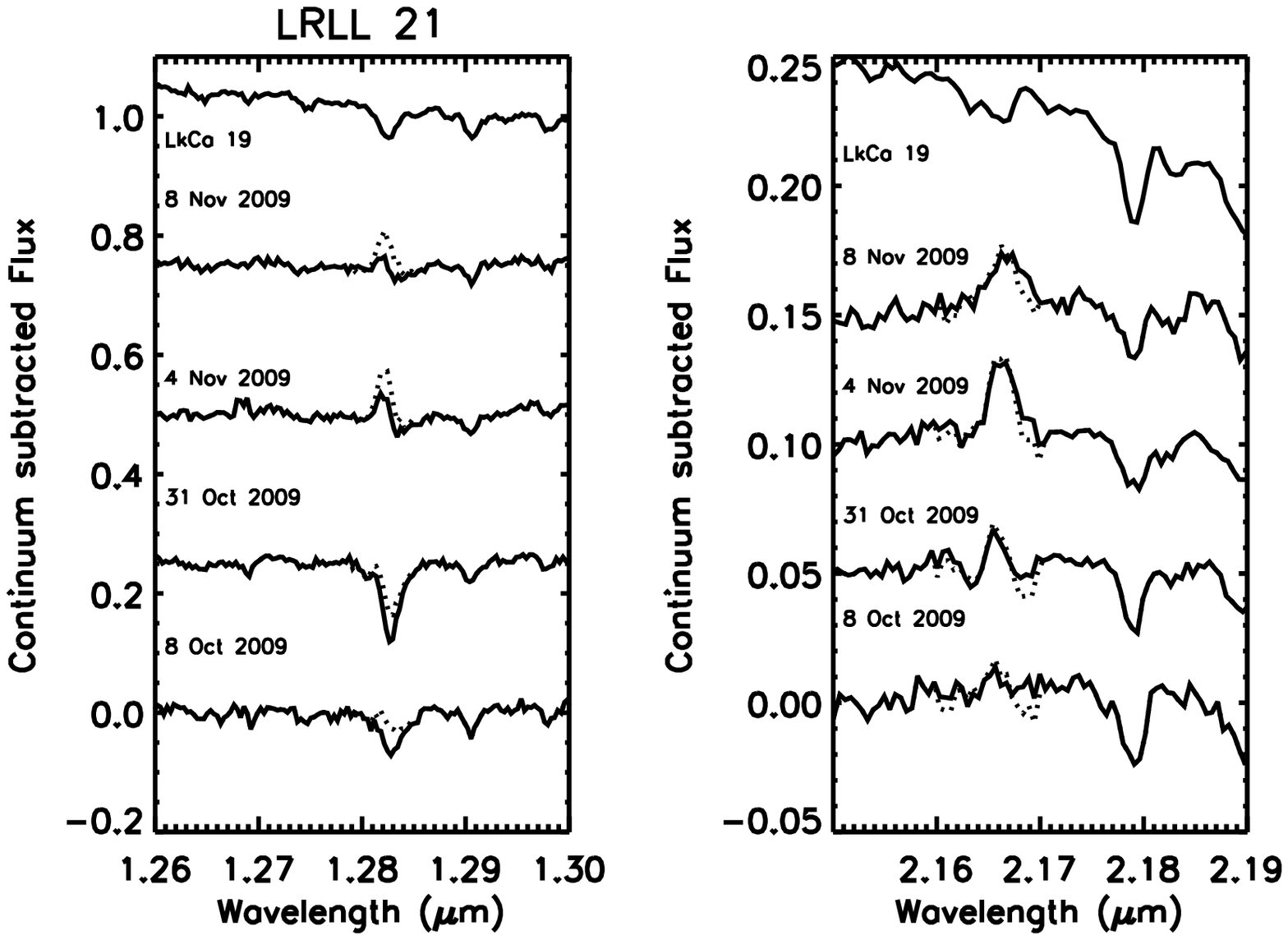}
\caption{Line measurements from 2008 (top two panels) and 2009 (bottom two panels). Both the Pa$\beta$ line ($\lambda=1.282\micron$) and the Br$\gamma$ line ($\lambda=2.166\micron$) are shown for each day. The solid line is the observed spectrum while the dashed line has  had the photospheric absorption subtracted from the line. At the top of each panel is the K0 WTTS standard. There are large variations in the emission lines, mainly on timescales of weeks.\label{l21_lines}}
\end{figure}

\subsubsection{Gas Properties}
\begin{figure}
\hbox{\includegraphics[scale=.5]{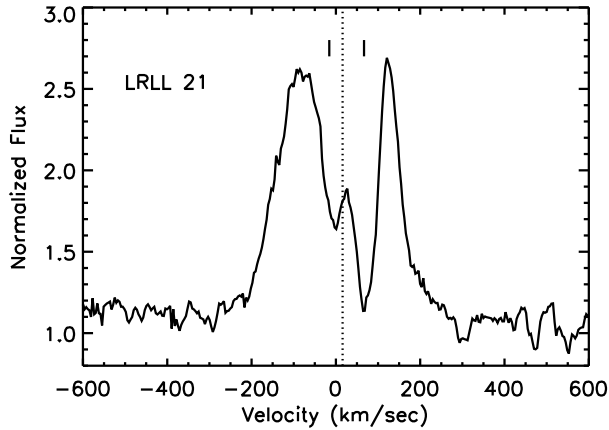}}
\caption{Optical spectra of the H$\alpha$ line for LRLL 21. Dashed line is the median velocity of the cluster, while the two tick marks are placed at the velocities of the two binary components. \label{l21_halpha}}
\end{figure}

Based on eight total epochs of near infrared spectra, including four in 2008 over the course of 8 days and four in 2009 over the course of one month, we find large fluctuations in the Pa$\beta$ and Br$\gamma$ lines, often changing from absorption to emission (Fig~\ref{l21_lines}). The emission line strengths of Pa$\beta$ and Br$\gamma$ reach a peak of -1.25 and -1.11\AA\ respectively, with accretion rates varying from undetectable to $10^{-8}M_{\odot}$yr$^{-1}$ (Table~\ref{accretion}). We place upper limits of a few times $10^{-9}M_{\odot}$yr$^{-1}$ on the accretion rate based on the uncertainty in the spectra on those days when no emission was detected. There are no day-to-day changes seen in the 2008 spectra but there are large changes from one week to the next, as is also seen in the 2009 data. \citet{dah08} and \citet{luh03} measure the EW of the H$\alpha$ line to be -4.48 and -4.7\AA\ respectively, while our high-resolution MMT spectrum  (Fig~\ref{l21_halpha}) shows an EW of -7.1\AA, all above the spectral type dependent H$\alpha$ EW threshold for accretors \citep{whi03}, and our low-resolution spectrum measures -1.59\AA, consistent with large fluctuations in the accretion rate. In our high-resolution H$\alpha$ spectrum the full velocity width at 10\%\ maximum height is 422 km/sec, above the 270 km/sec threshold for active accretors \citep{whi03}, although this may be influenced by the possible presence of a companion. The high-resolution profiles of the H$\alpha$ line from \citet{dah08} and our MMT spectra show strong central absorption, possibly due to two separate H$\alpha$ emission lines from the two components of the binary system as well as possible wind absorption. The possible binary nature of this system prevents us from using the resolved line width to derive the accretion rate, as has been done in other studies \citep{nat04}. Our low-resolution optical spectrum, taken in 2009 when the Pa$\beta$ line was in absorption, shows especially weak H$\alpha$ emission compared to what is typically seen around actively accreting stars and is likely entirely chromospheric rather than accretion emission  \citep{whi03}, although it is possible that unresolved absorption leads to a very small EW even with ongoing accretion. This very weak accretion measure is similar to the behavior of the infrared lines, which show no emission in some epochs, possibly also due to unresolved absorption.  Even during the most active accretion, the ratio of accretion to stellar luminosity (L$_{acc}$/L$_*$=0.06) is small.

\subsection{LRLL 58}
\subsubsection{Stellar Properties}
LRLL 58 is a M1.25 star with a luminosity of 0.72L$_{\odot}$ and a radius of 2.1R$_{\odot}$ \citep{luh03}. Its mass is $\sim0.7$M$_{\odot}$ based on its position on the HR diagram relative to the \citet{sie00} 3 Myr isochrone. \citet{coh04} and \citet{cie06} detect periodic variability in the optical with a period of 7.4 days, consistent with the rotation of star spots across the stellar surface. The I band flux is roughly constant over years (I=14.10,14.32,14.18 from \citet{coh04}, \citet{cie06}, \citet{lit05} respectively) suggesting that there are no large fluctuations in stellar flux. From our hectochelle data we measure a radial velocity of 14.2 km/sec and an upper limit on the rotational velocity of 15.0 km/sec (Table~\ref{velocity}). Based on two epochs of near-infrared spectra, its extinction is constant at A$_V$=3.4 (Table~\ref{extinction}). Unlike LRLL 21, there is no substantial variation in the stellar flux of LRLL 58.

\begin{figure*}
\center
\includegraphics[scale=.3]{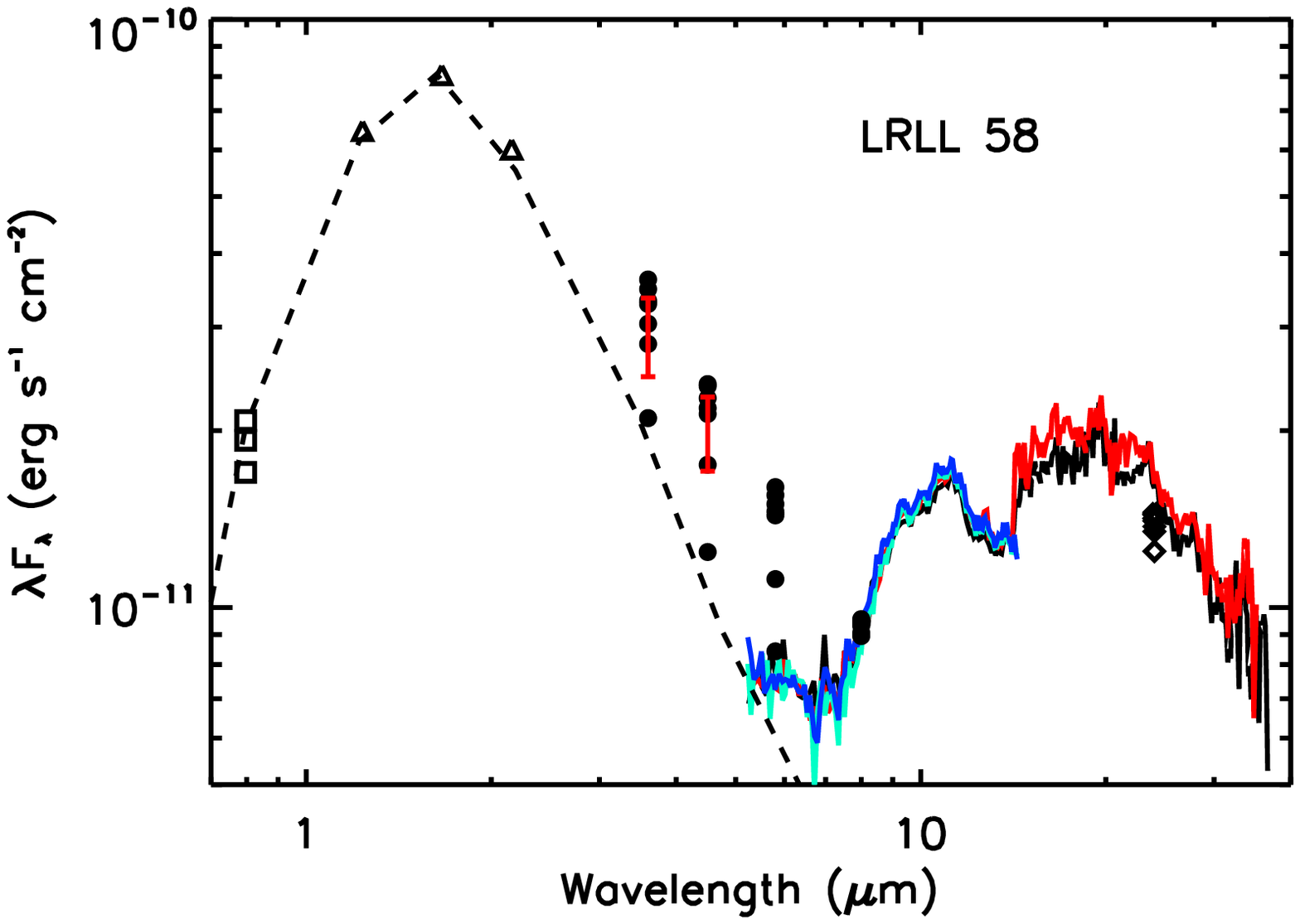}
\includegraphics[scale=.3]{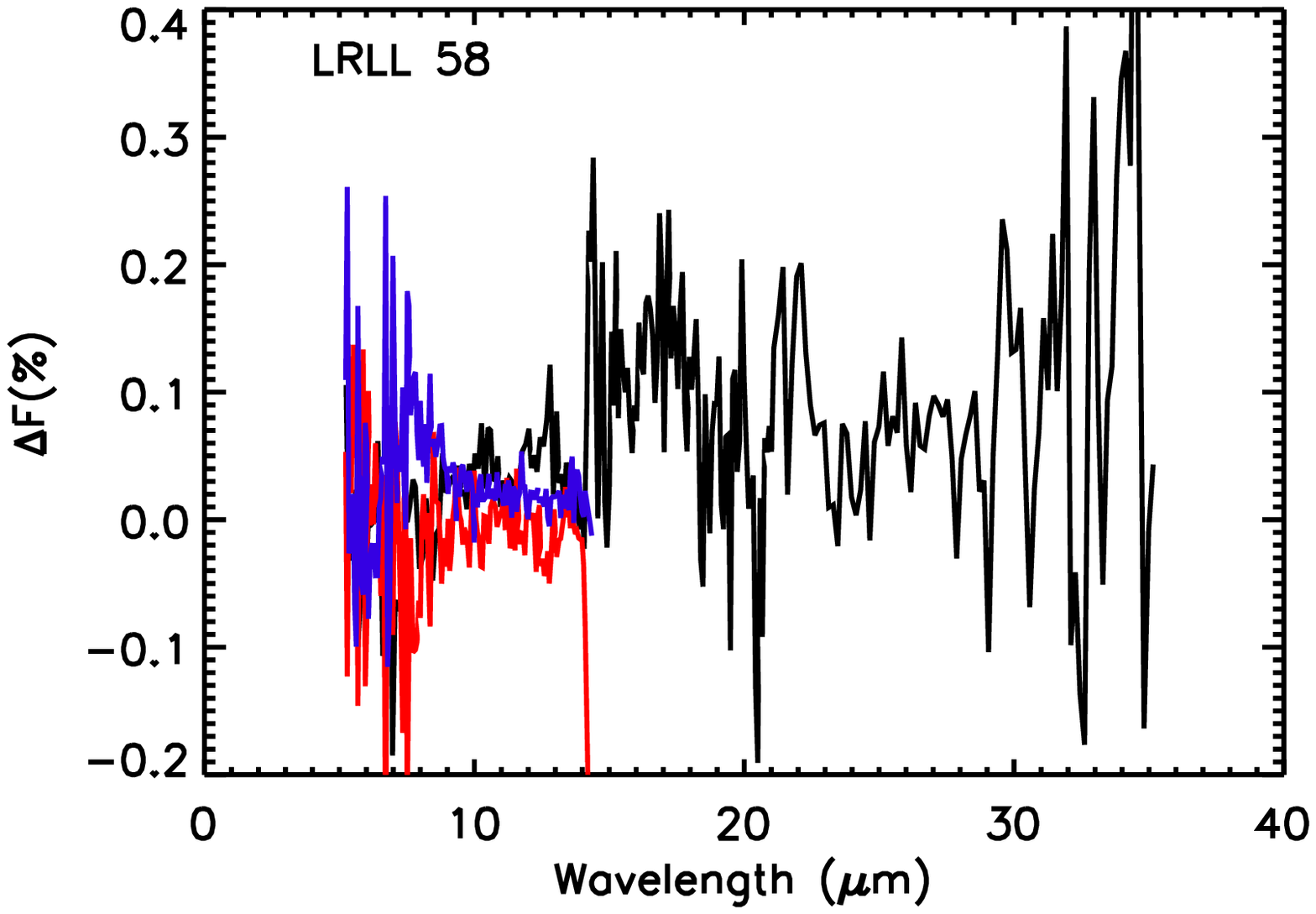}
\includegraphics[scale=.3]{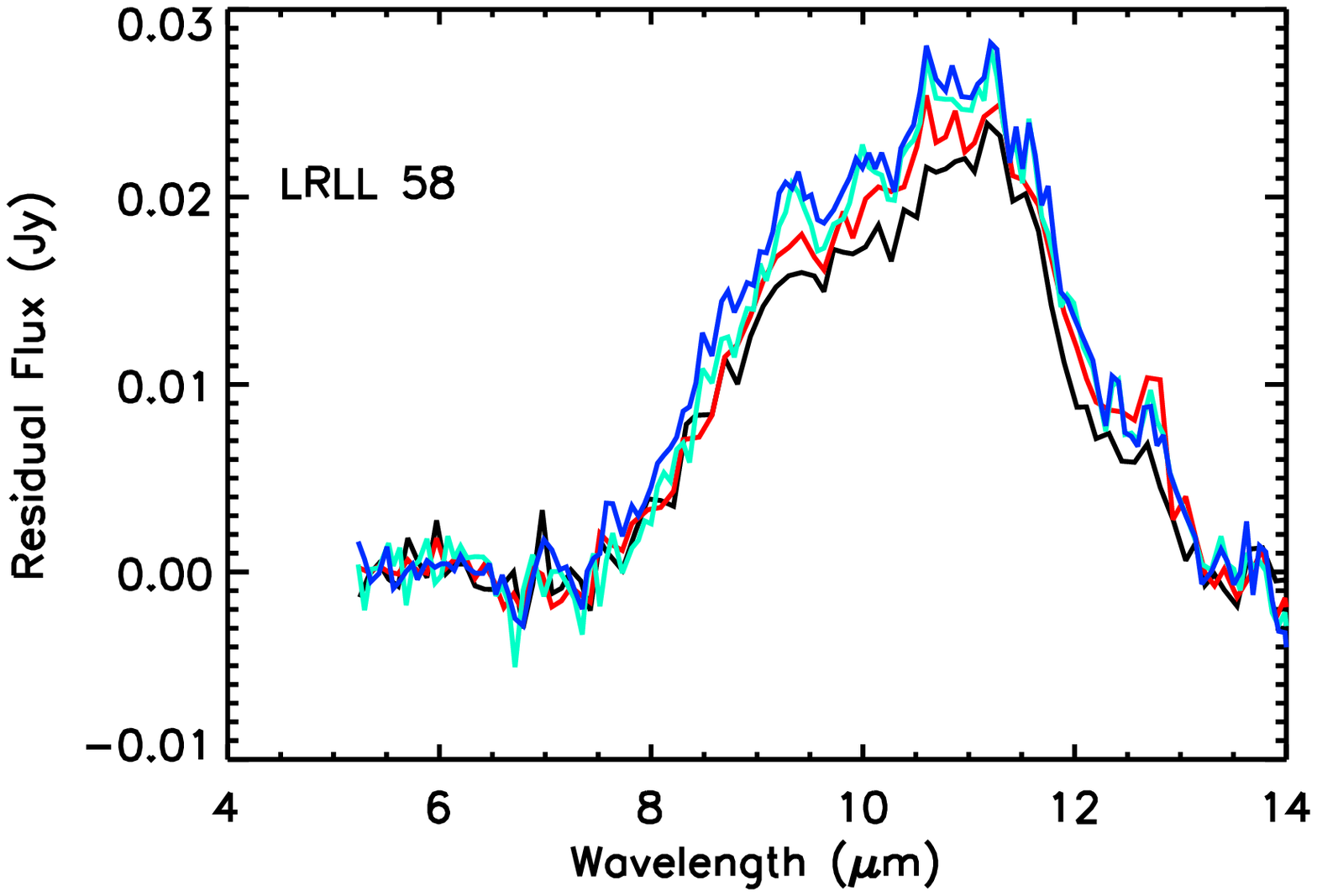}
\caption{Same as Figure~\ref{lrll2_ir} but for LRLL 58.\label{lrll58_ir}}
\end{figure*}

\begin{figure*}
\center
\includegraphics[scale=.4]{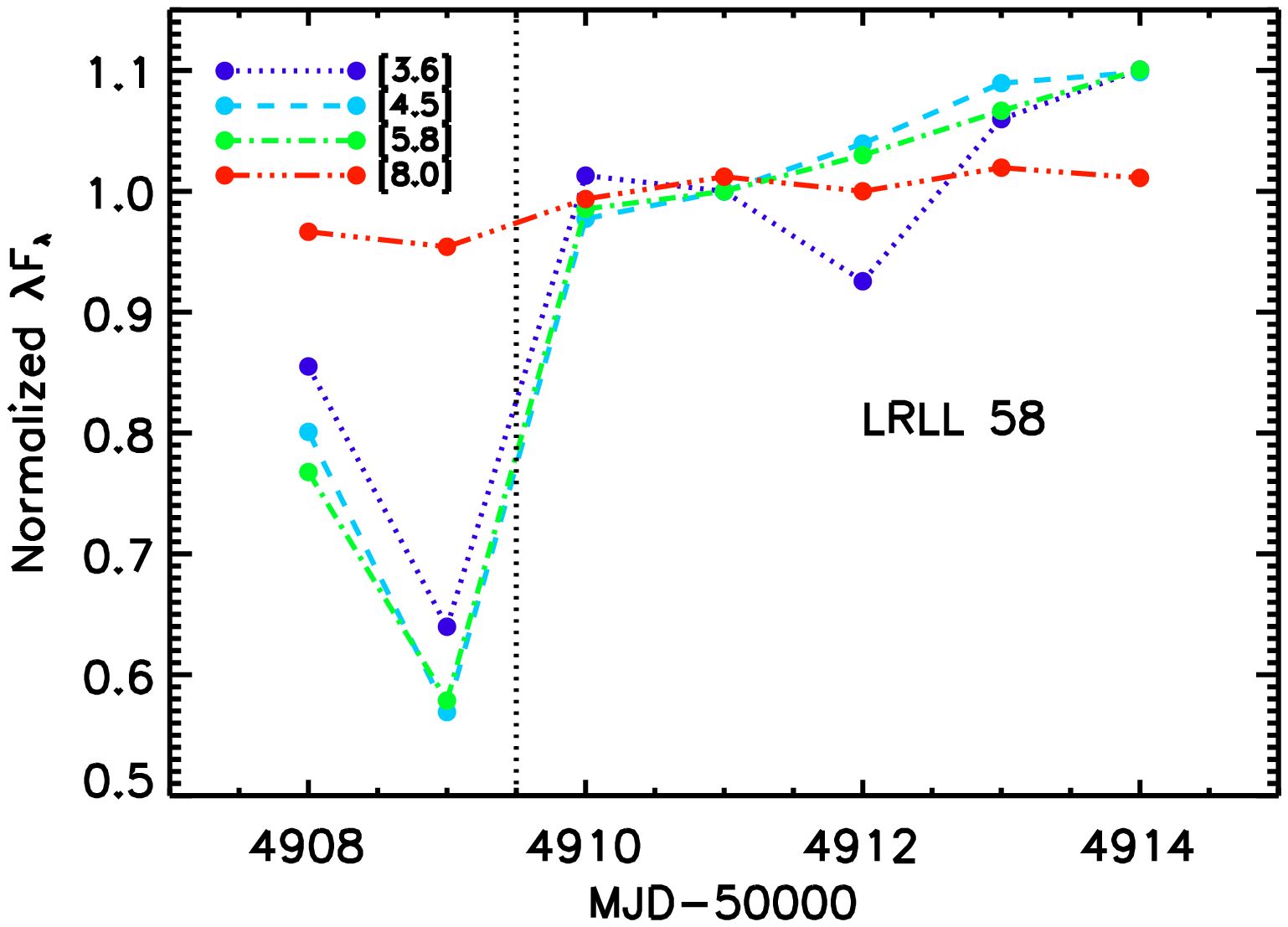}
\includegraphics[scale=.4]{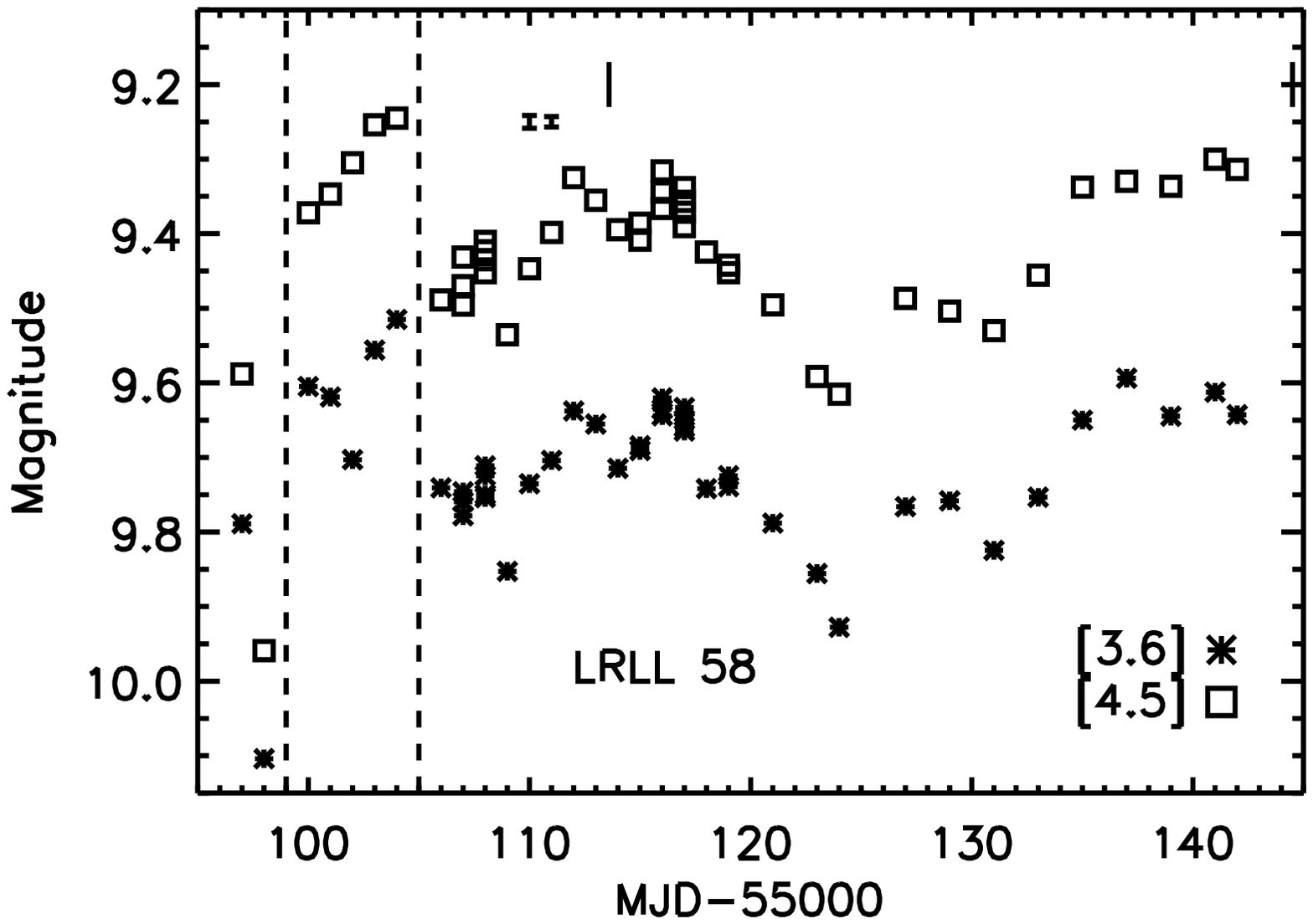}
\includegraphics[scale=.4]{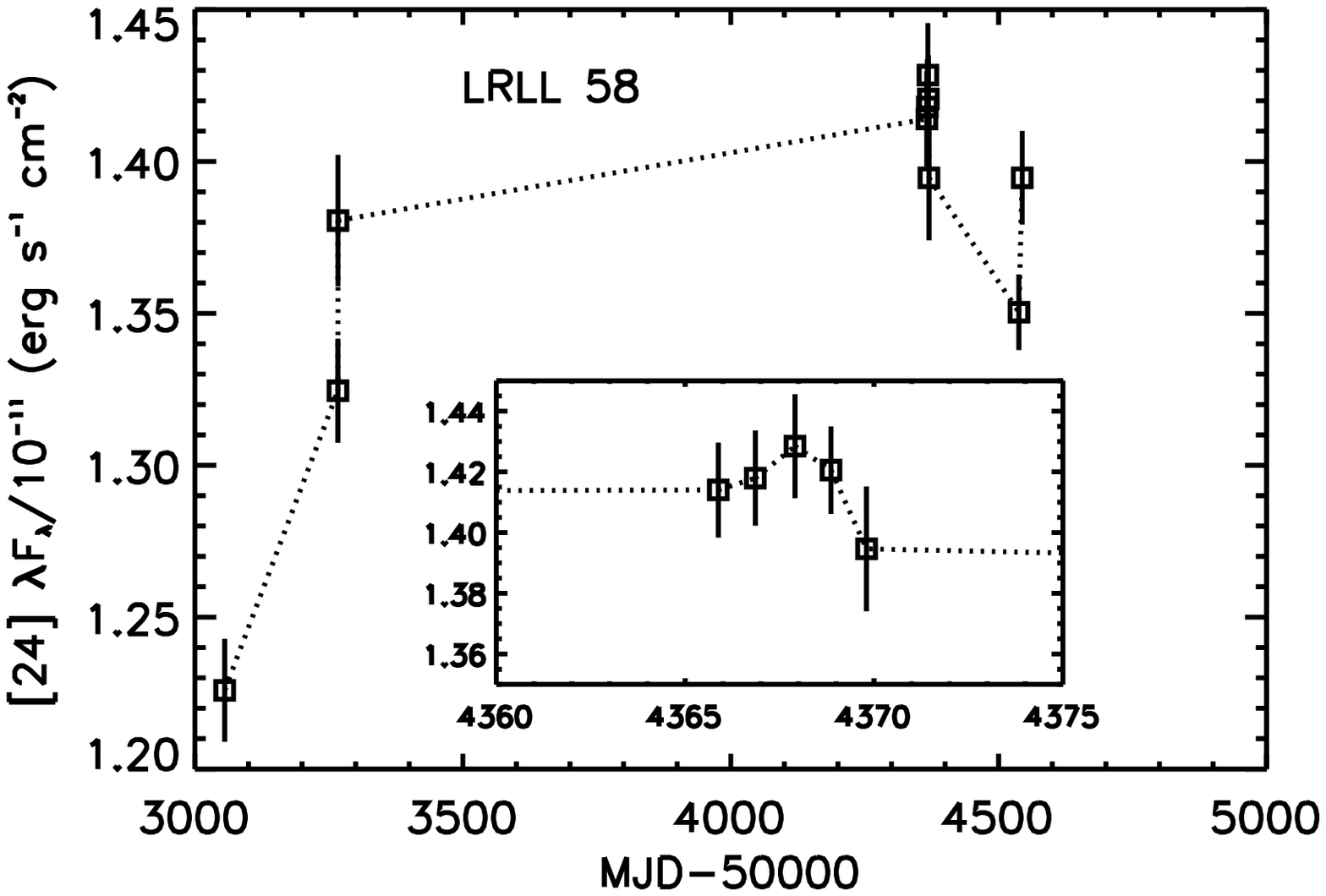}
\includegraphics[scale=.4]{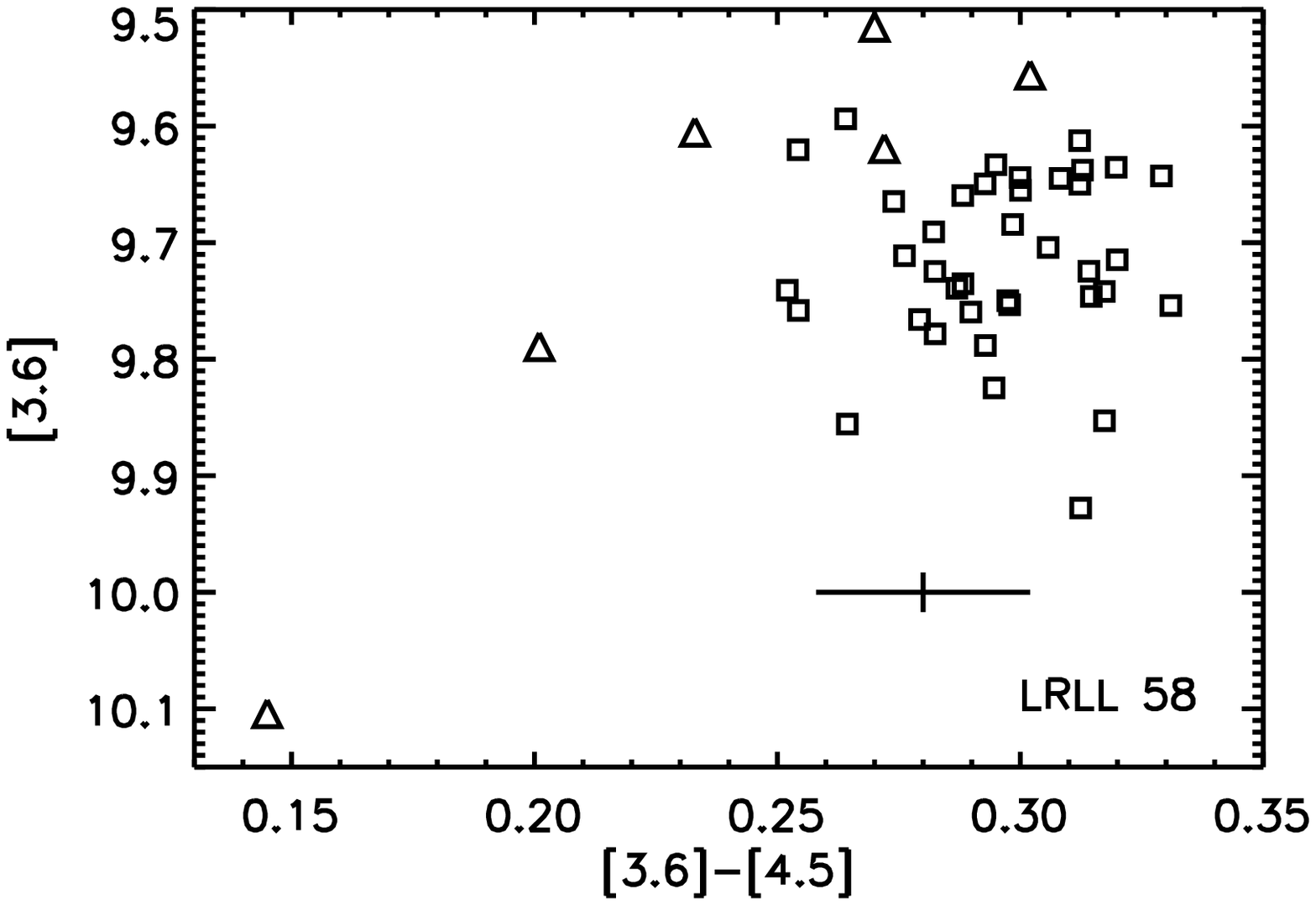}
\caption{Same as Figure~\ref{lrll2_ir2} but for LRLL 58. \label{lrll58_ir2}}
\end{figure*}

\subsubsection{Infrared Variability}
\begin{figure}
\includegraphics[scale=.45]{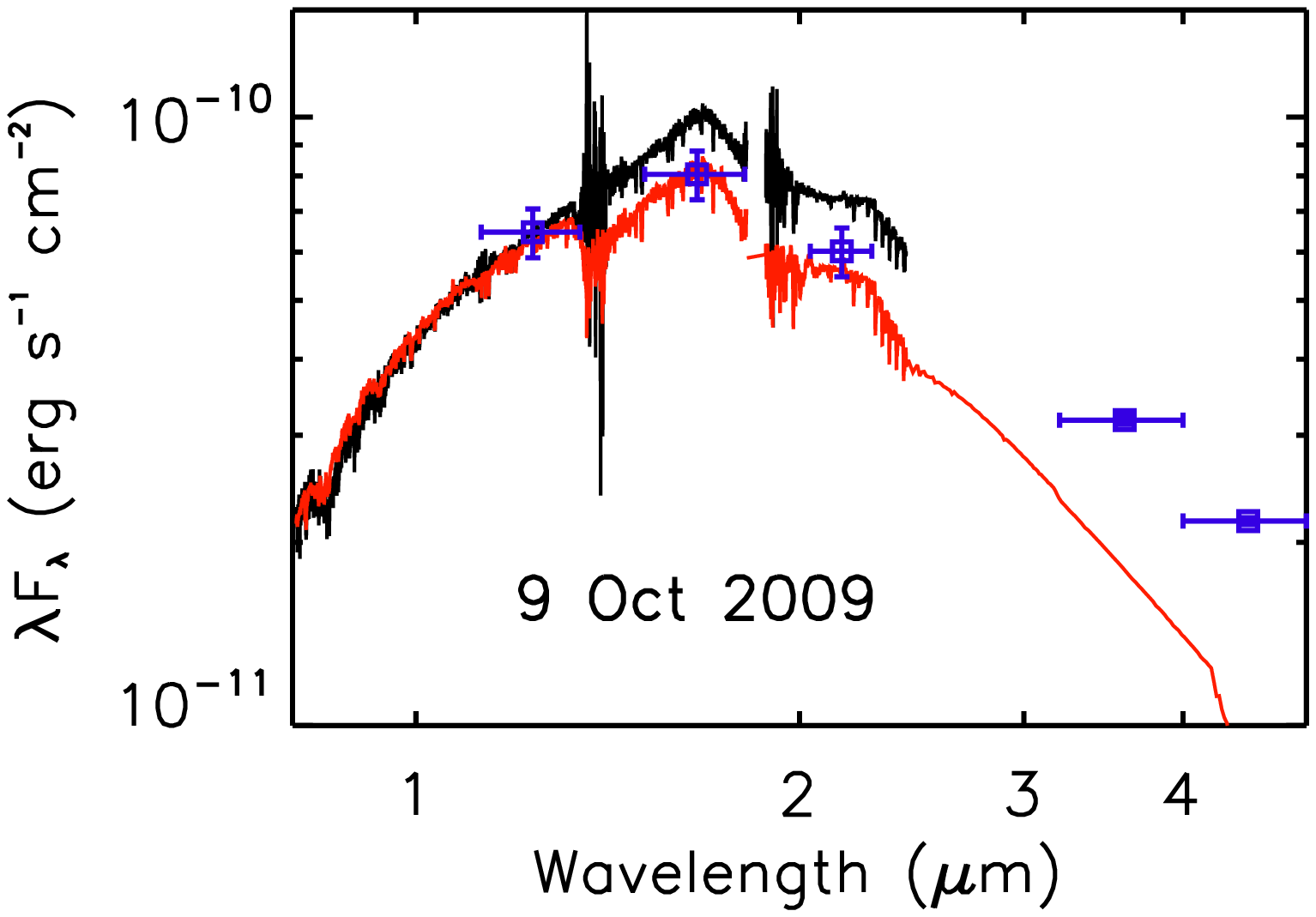}
\includegraphics[scale=.45]{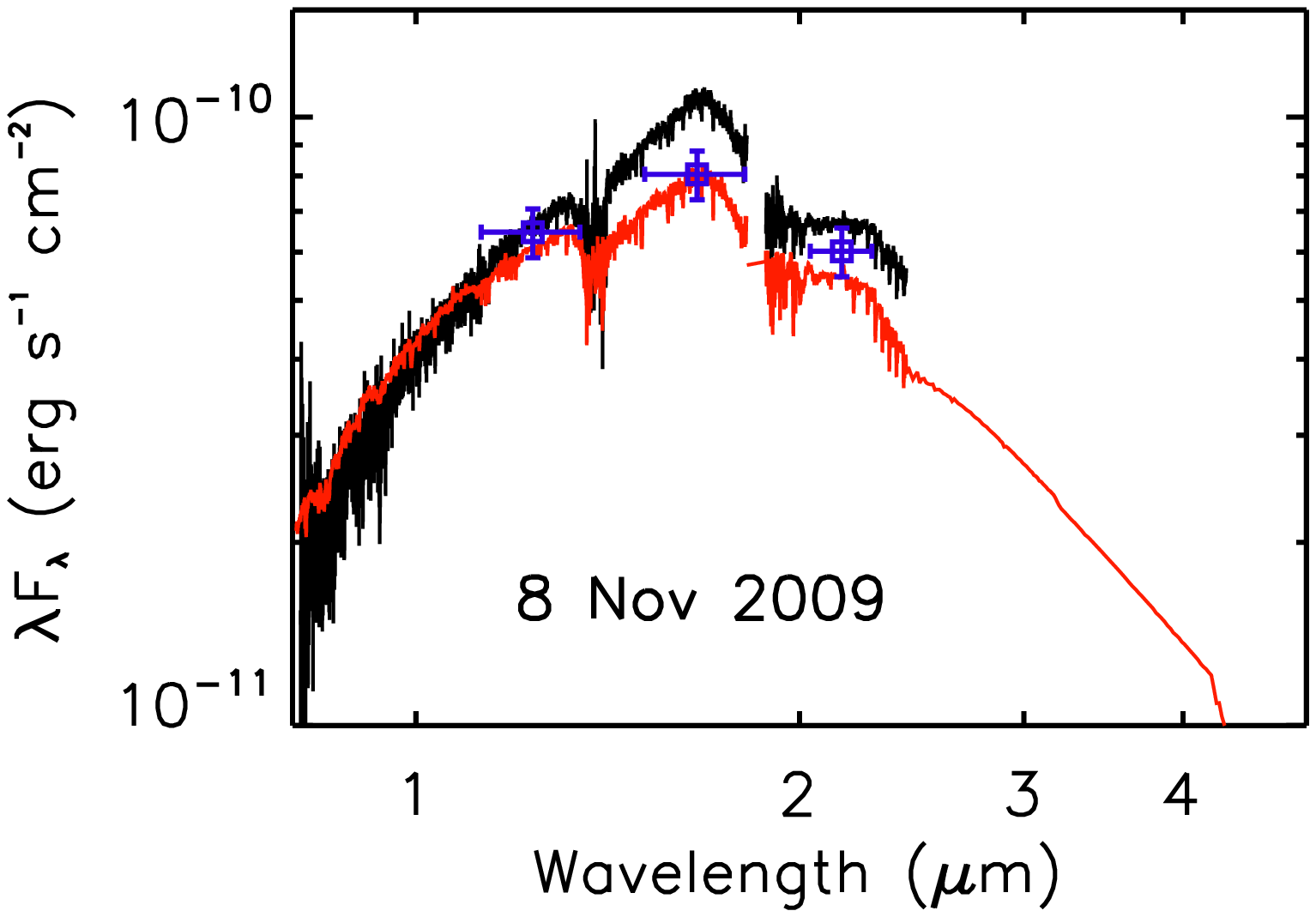}
\caption{Spex spectra (black line) and photometry (blue points) for LRLL 58. Spectra have been scaled to match 2MASS photometry. A stellar photosphere (JH 108 + Kurucz model) shown in red has been reddened by the A$_V$ measured on each night to LRLL 58 and scaled to the short wavelength flux of the spectra. The mismatch between the spectra and the 2MASS photometry is likely a sign of variability between these epochs.\label{l58_spex+phot}}
\end{figure}

\begin{figure}
\includegraphics[scale=.45]{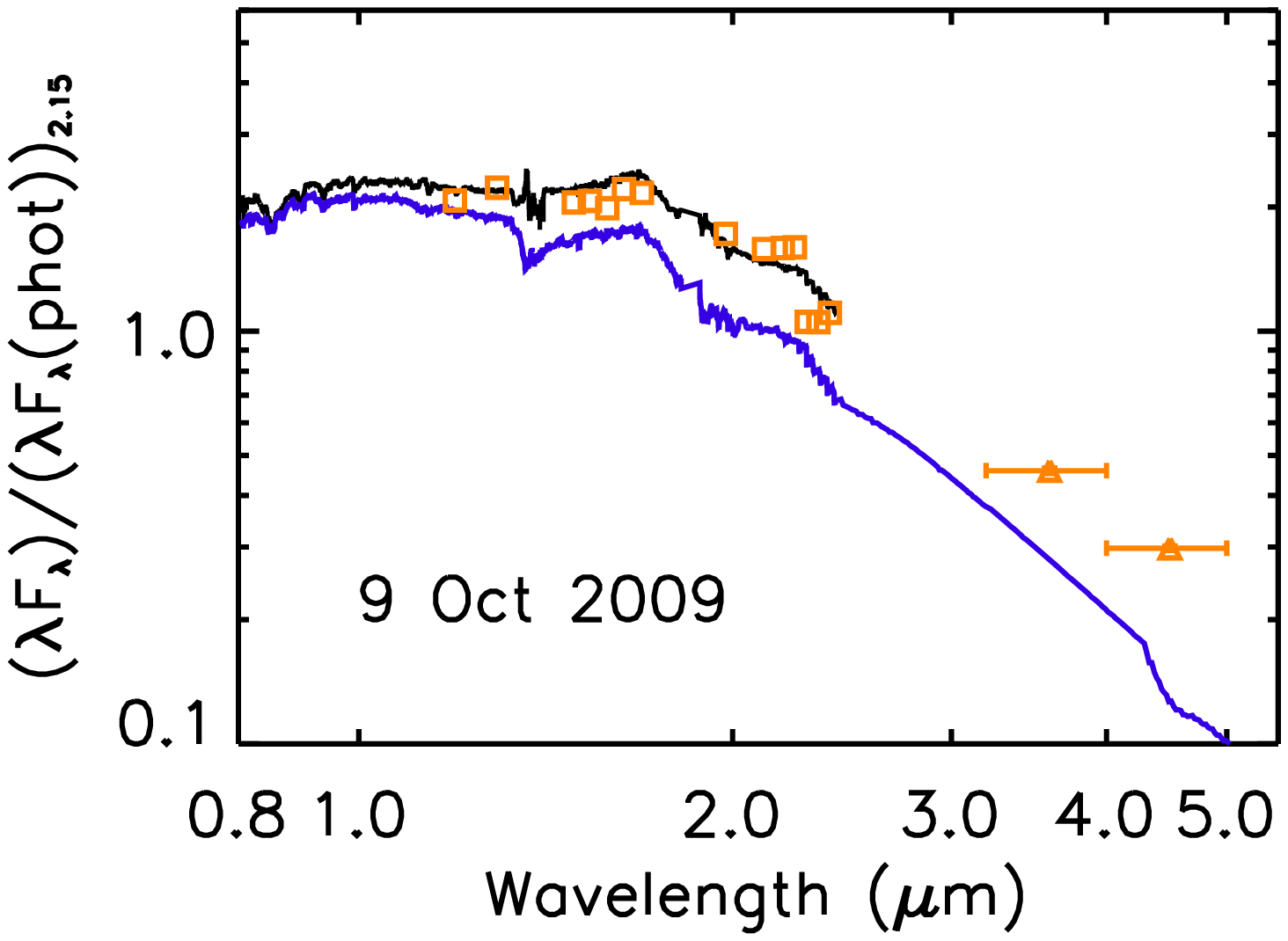}
\includegraphics[scale=.45]{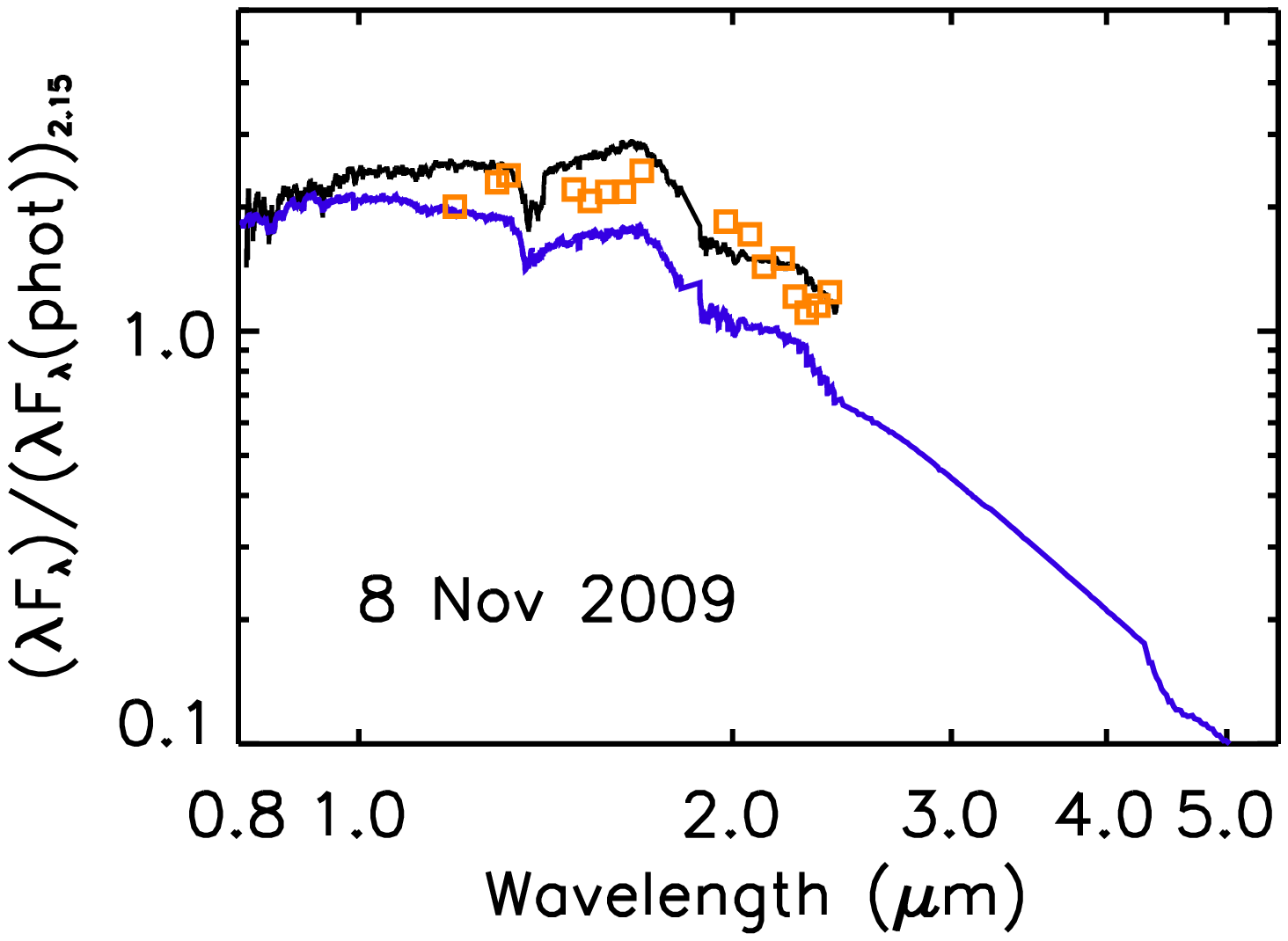}
\caption{Dereddened spectra of LRLL 58, along with the veiling measurements (orange squares), 3.6 and 4.5\micron\ photometry where available (orange triangles) and a stellar photosphere for comparison (blue line). LRLL 58 shows a strong excess throughout our spectra. \label{l58_excess}}
\end{figure}

In the IRS spectra, LRLL 58 displays no variations ($<5\%$ for $\lambda<14\micron$ and $<20\%$ for $\lambda\ge14\micron$), although the long-wavelength end has large systematic uncertainties due to uncertain background subtraction (Figure~\ref{lrll58_ir}). These systematic uncertainties are the likely source of the mismatch between the SL and LL spectra at 14\micron, which we correct for by decreasing the LL spectra by a constant multiplicative factor when comparing different epochs. The 24$\micron$ photometry shows significant fluctuations (14\%), which fall below our detection limit in the IRS spectra, on predominately weekly and longer timescales (Fig~\ref{lrll58_ir2}). The five consecutive days of 3-8$\micron$ photometry show a steady increase in flux, while the 3.6 and 4.5$\micron$ monitoring confirms that large ($0.3$ mag), daily fluctuations occur, with no evidence for periodicity. When comparing the IRS spectra with the 2-8$\micron$ photometry there appear to be very large fluctuations that have not been captured during any single monitoring campaign. The 3-8\micron\ photometry shows a distinct excess in all but one epoch, while the IRS spectra is consistent with almost no excess at these wavelengths. Only at 8\micron\ are the photometry and the spectra consistent with the same flux level. In the 3-8$\micron$ monitoring there is a distinct wavelength dependence: as the wavelength increases the size of the fluctuation decreases. This is consistent with the SED variations between the IRS spectra and the photometry suggesting that the SED pivots at $\lambda=8\micron$, similar to LRLL 21 and 31. Significant veiling is present in the 0.8-2.5$\micron$ spectra (r$_K$=0.46,0.48, Table~\ref{ir_excess_table},Fig~\ref{l58_spex+phot},\ref{l58_excess}). As with LRLL 2 we cannot constrain the exact source of this excess, but it is consistent with hot dust emission ($>1200$K) from close to the star. As with LRLL 21 we can estimate the covering fraction of the hot dust using our near-simultaneous near-infrared spectra and Spitzer photometry on 9 Oct 2009. We find a covering fraction of $\sim$13\% (Table~\ref{ir_excess_table}), which is comparable to what is seen around typical T Tauri stars. This represents the inner disk at its maximum flux, while there are some epochs when the 2-8\micron\ flux drops almost to the photosphere, suggesting a very small covering fraction.

\subsubsection{Gas Properties}
\begin{figure*}
\center
\includegraphics[scale=.5]{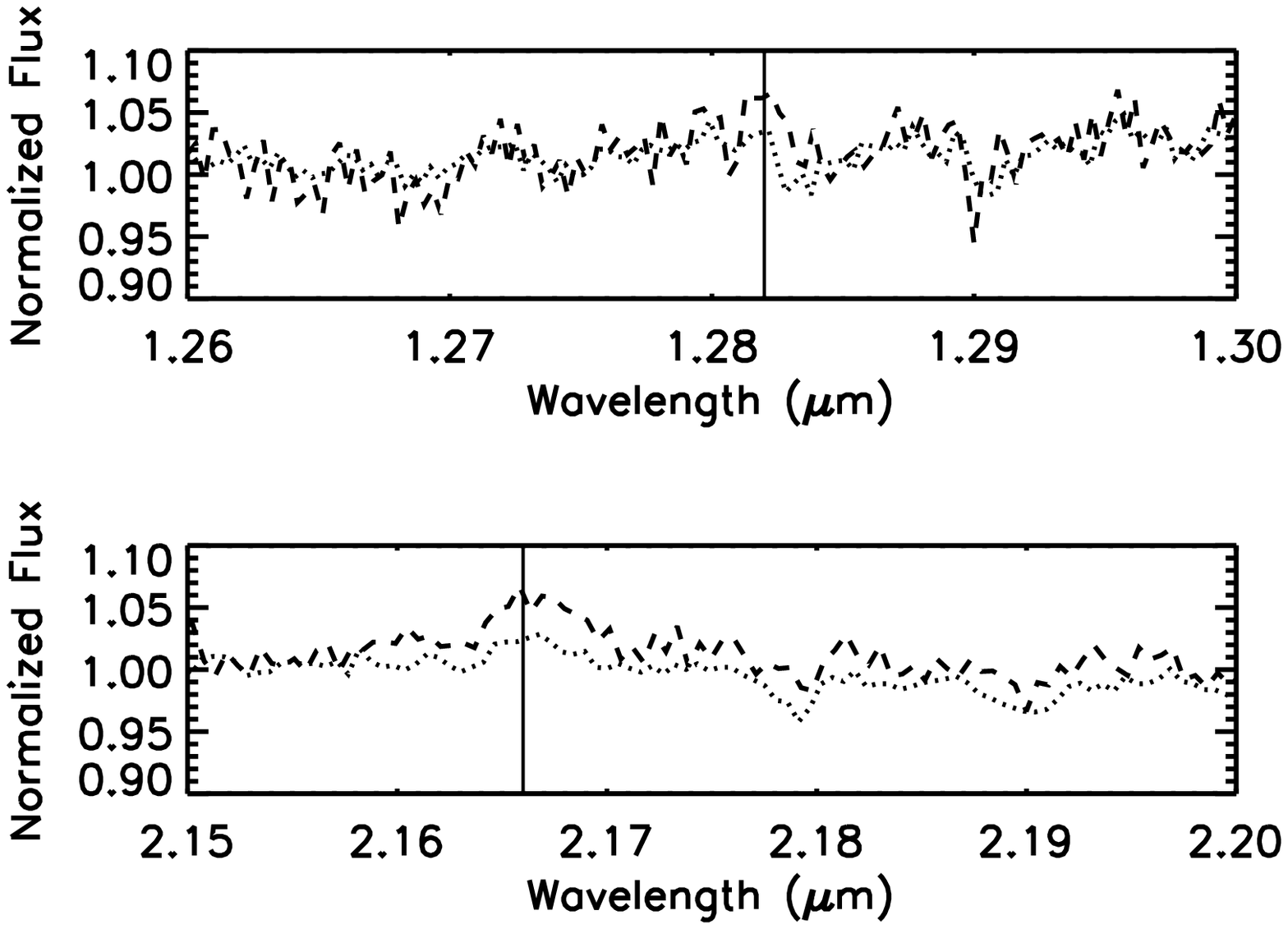}
\includegraphics[scale=.5]{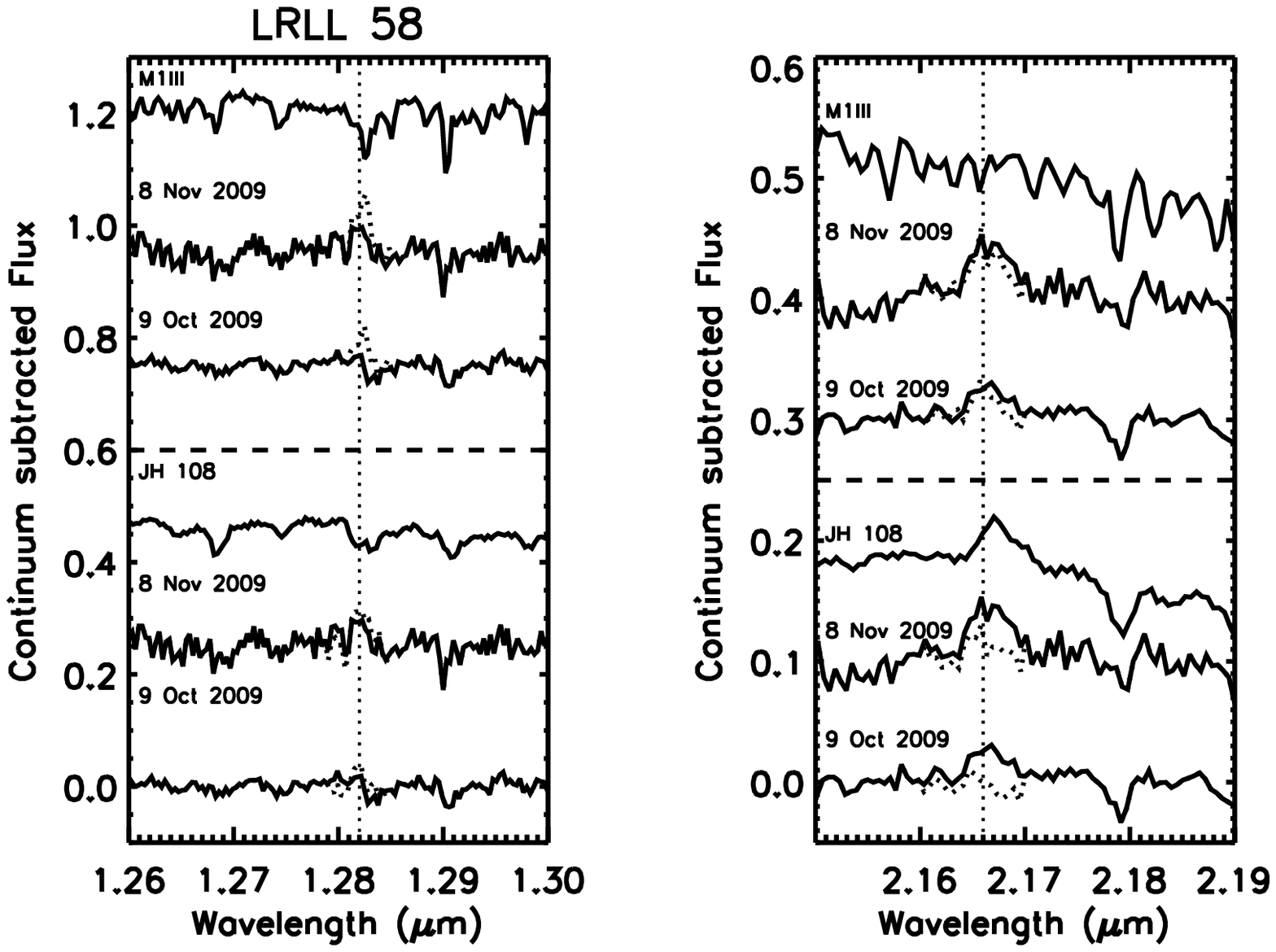}
\caption{Same as Figure~\ref{l2_lines}, but for LRLL 58. There is some evidence for a change in the line strengths between the two epochs, and ongoing accretion in both epochs, although the uncertainties in the derived accretion rates are large. When subtracting by a photospheric standard we consider both a WTTS and a giant from the IRTF spectral library, since the WTTS seems to exhibit Br$\gamma$ emission. \label{l58_lines}}
\end{figure*}

\begin{figure}
\hbox{\includegraphics[scale=.5]{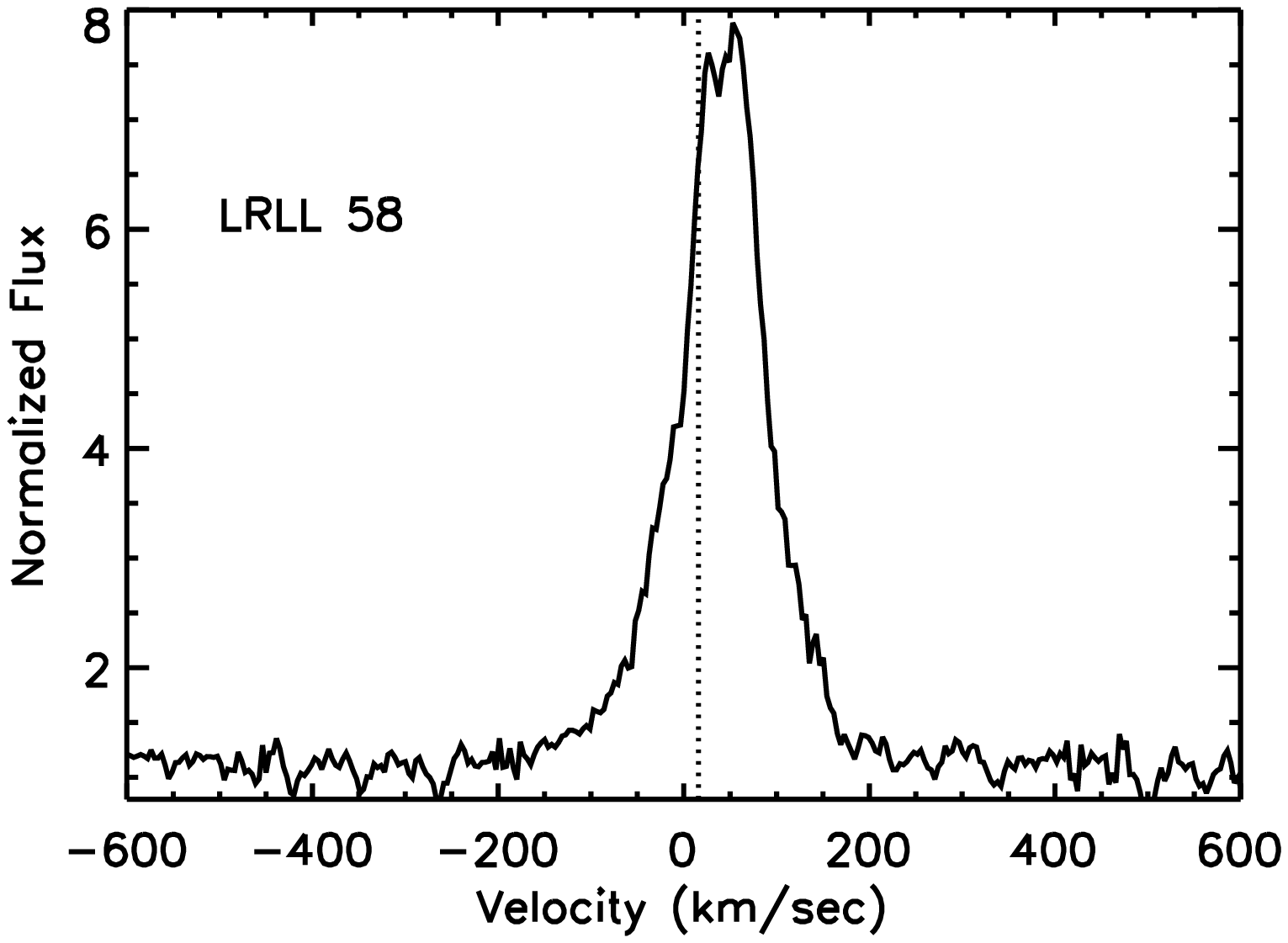}}
\caption{Optical spectra of the H$\alpha$ line for LRLL 58. Dashed line is the mean velocity of the cluster. \label{l58_halpha}}
\end{figure}

During our two near-infrared observations in 2009 separated by roughly one month the Pa$\beta$ and Br$\gamma$ emission appears to change. Fig~\ref{l58_lines} shows that in the second epoch the lines are slightly stronger, although the noisy continuum makes it difficult to calculate accurate EWs.  An additional complication is that our M1 WTTS standard JH 108, which is used to estimate the photospheric absorption, appears to show emission near the Br$\gamma$ line (Fig~\ref{l58_lines}), making it difficult to estimate the emission for these lines. The equivalent width of the H$\alpha$ line varies from -9\AA\ \citep{her98} to -20.8\AA\ from our low-resolution spectra and is -14.9\AA\ (full-width at 10\%\ of 240km/sec) in our high-resolution spectrum (Fig~\ref{l58_halpha}), indicating that there have been large fluctuations in the accretion flux. The small H$\alpha$ line strength measured by \citet{her98} is close to the boundary between accreting and non-accreting sources \citep{whi03}, suggesting that the accretion rate can drop to very low levels. Our high-resolution spectrum displays a broad line width indicative of ongoing accretion. Based on our measured accretion rate from the infrared lines ($\dot{M}\sim5\times10^{-9}M_{\odot}yr^{-1}$), the accretion luminosity is a small proportion (L$_{acc}$/L$_*$$\lesssim0.1$) of the total luminosity that is responsible for heating the disk. The accretion rate derived from the width of the H$\alpha$ line \citep{nat04} is $\dot{M}\sim3\times10^{-11}M_{\odot}yr^{-1}$, which confirms that the accretion luminosity is very small. Given the variable accretion flow and the systematic uncertainties in these accretion measures, it is difficult to evaluate the consistency in the accretion rate measured by the infrared lines and the resolved width of the H$\alpha$ line.

\subsection{LRLL 67}
\subsubsection{Stellar Properties}
LRLL 67 is a M0.75 star with a luminosity of 0.48L$_{\odot}$, a radius of 1.6R$_{\odot}$ \citep{luh03}, and a mass of $\sim0.6$M$_{\odot}$ \citep{sie00}. \citet{dah08} measured the radial velocity to be 15.09 km/sec, with v$\sin$i=13.59 km/sec, \citet{nor06} find v$_r$=16.9 km/sec with v$\sin$i$<$11.0 km/sec and our hectochelle data measure v$_r$=15.0km/sec and v$\sin$i$<$15.0 km/sec (Table~\ref{velocity}). The lack of significant variation in the radial velocity, suggests that there is no massive companion in this system. \citet{cie06} did not find evidence for periodic optical fluctuations. We do not have any near-infrared spectra of this star with which to estimate the extinction and we adopt the value from \citet{luh03} (A$_V$=2.4). As with LRLL 2 and 58, there is no evidence for large fluctuations in the stellar flux.

\begin{figure*}
\center
\includegraphics[scale=.3]{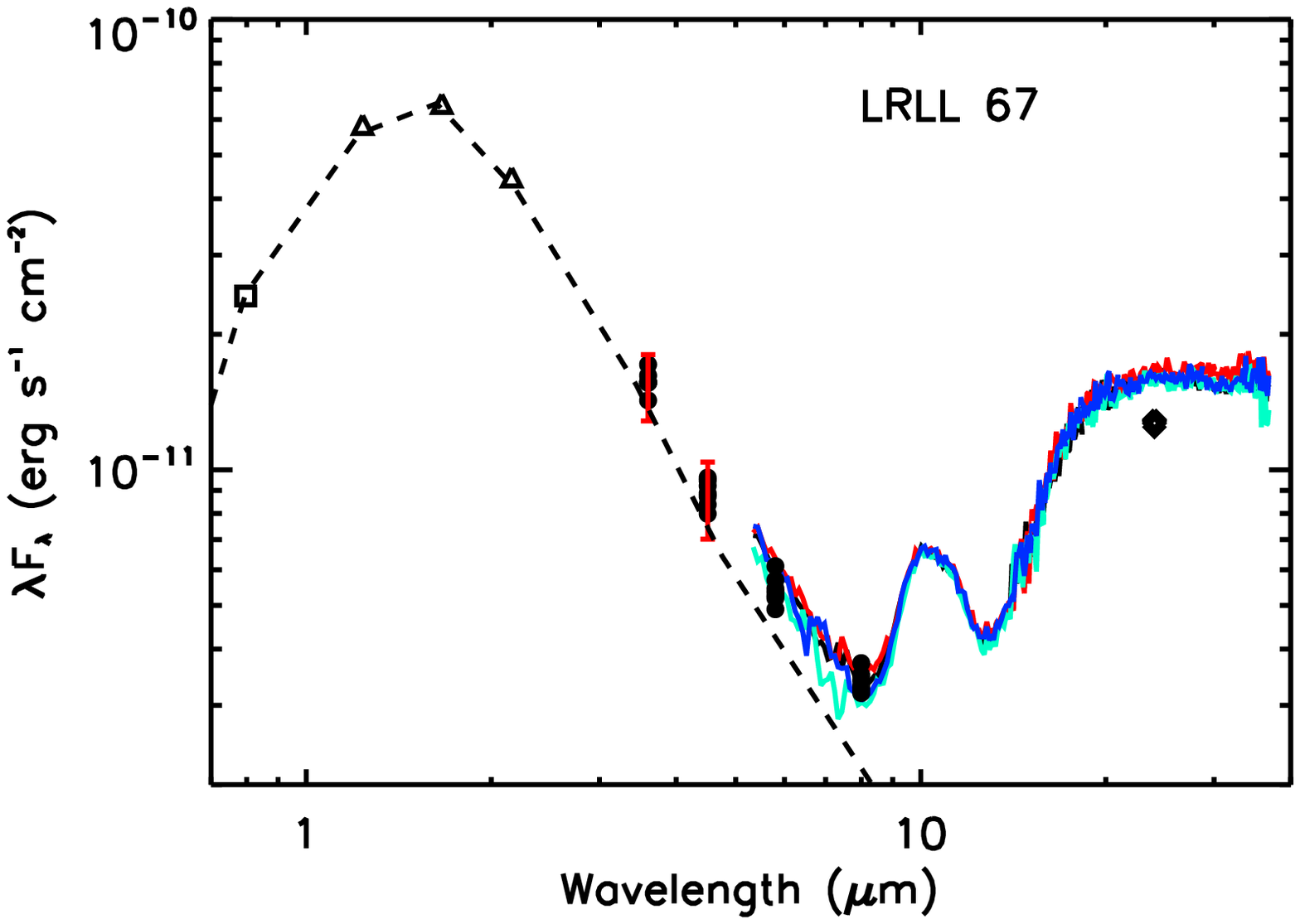}
\includegraphics[scale=.3]{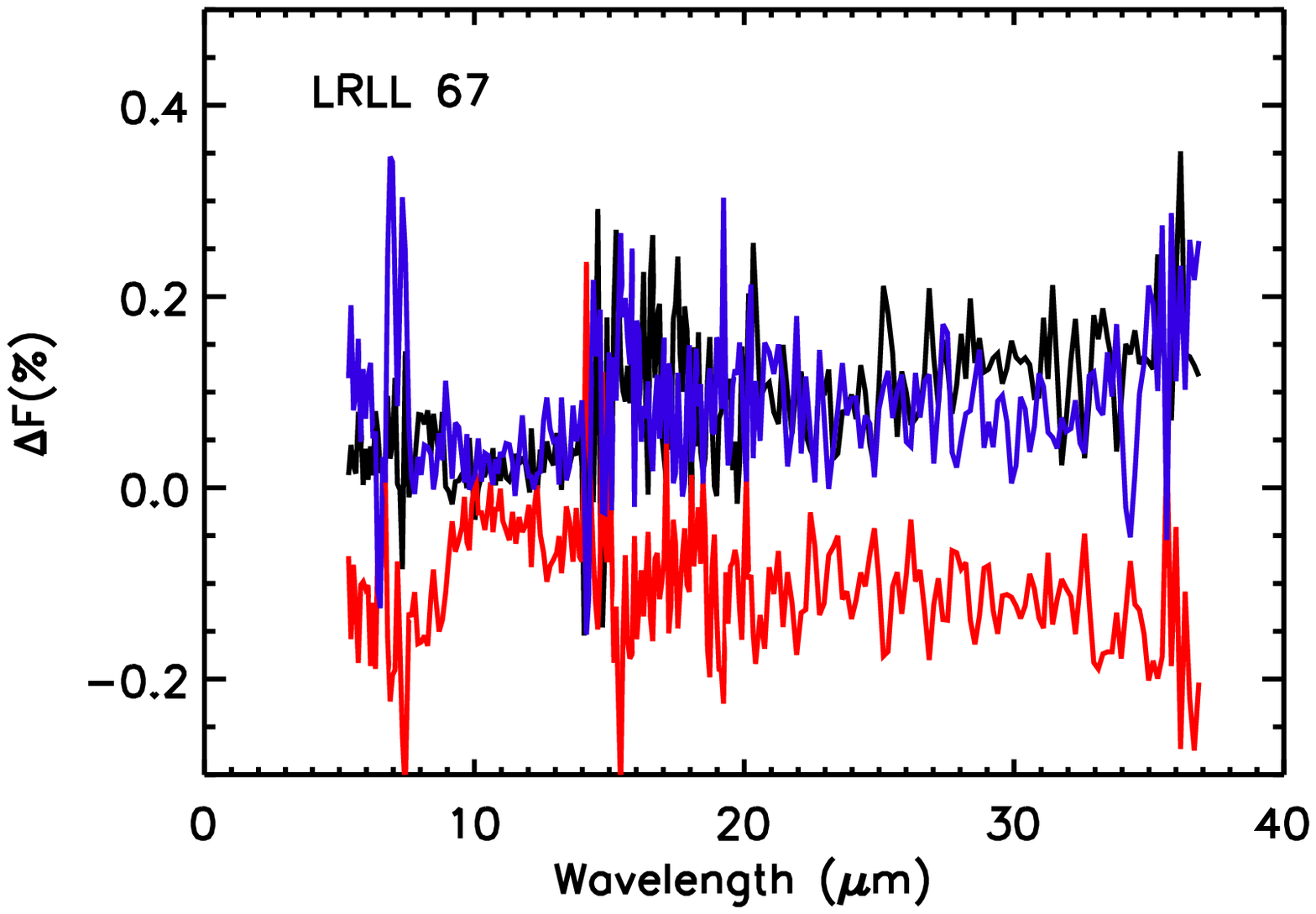}
\includegraphics[scale=.3]{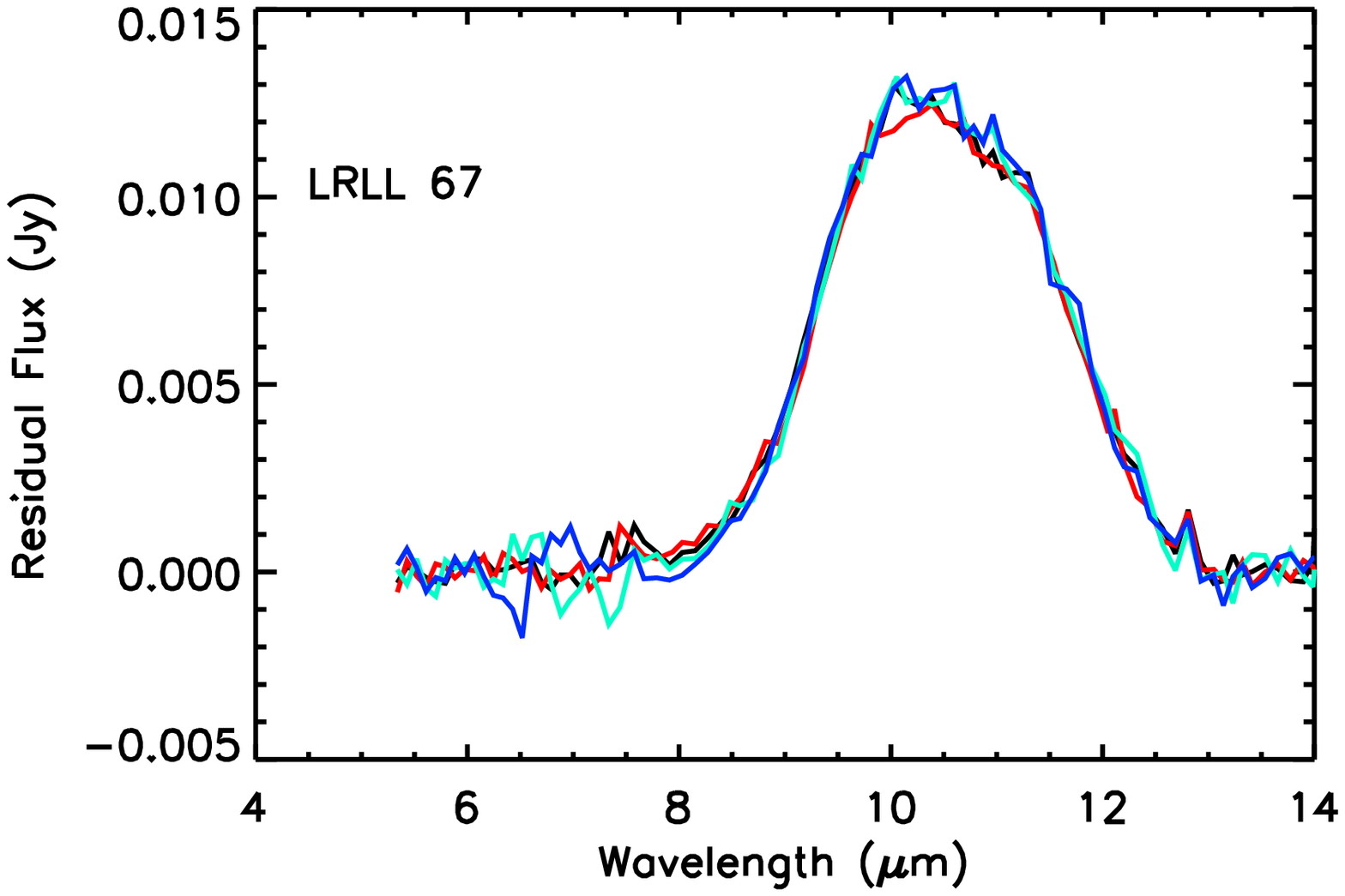}
\caption{Same as Figure~\ref{lrll2_ir} but for LRLL 67.\label{lrll67_ir}}
\end{figure*}

\begin{figure*}
\center
\includegraphics[scale=.4]{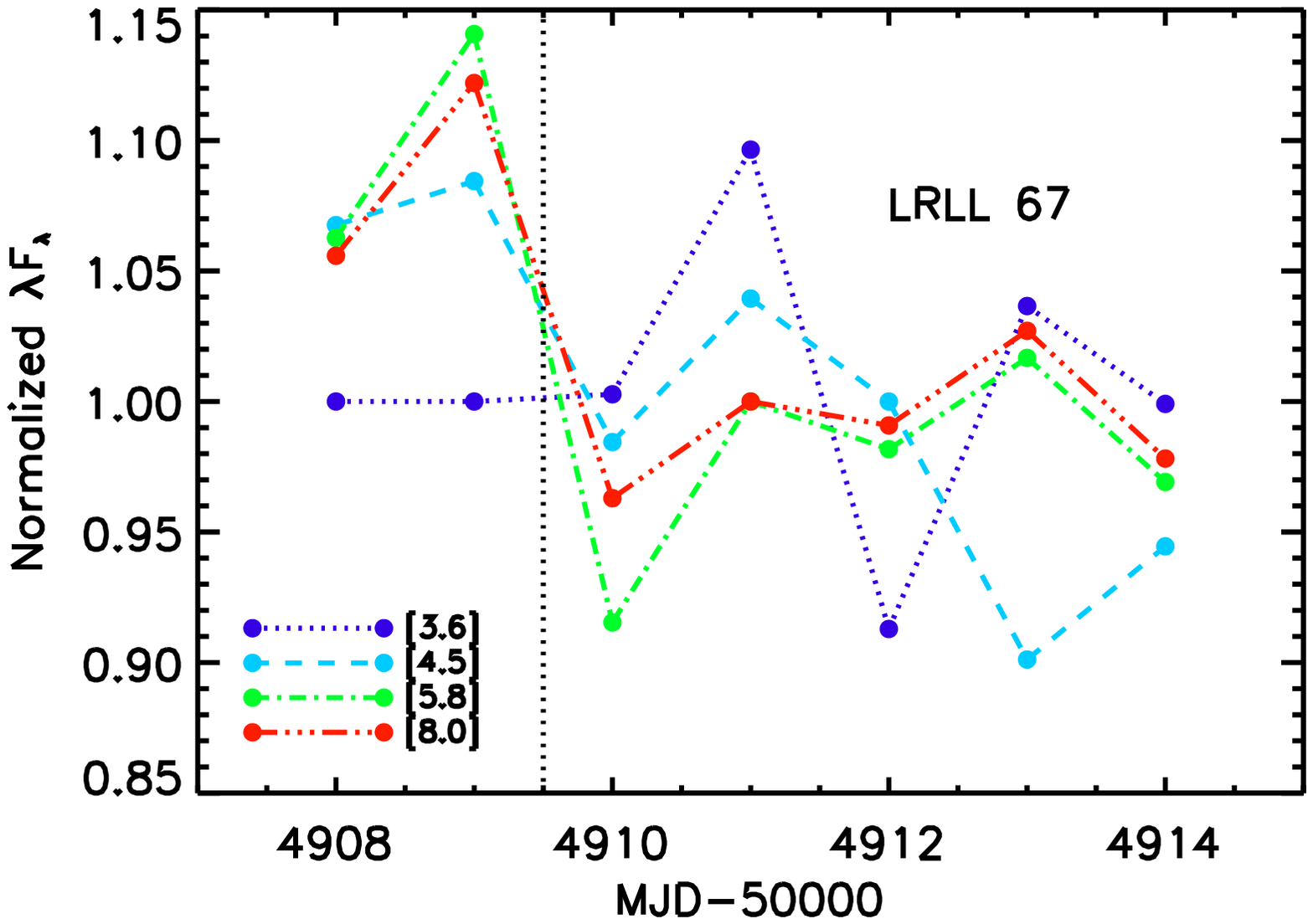}
\includegraphics[scale=.4]{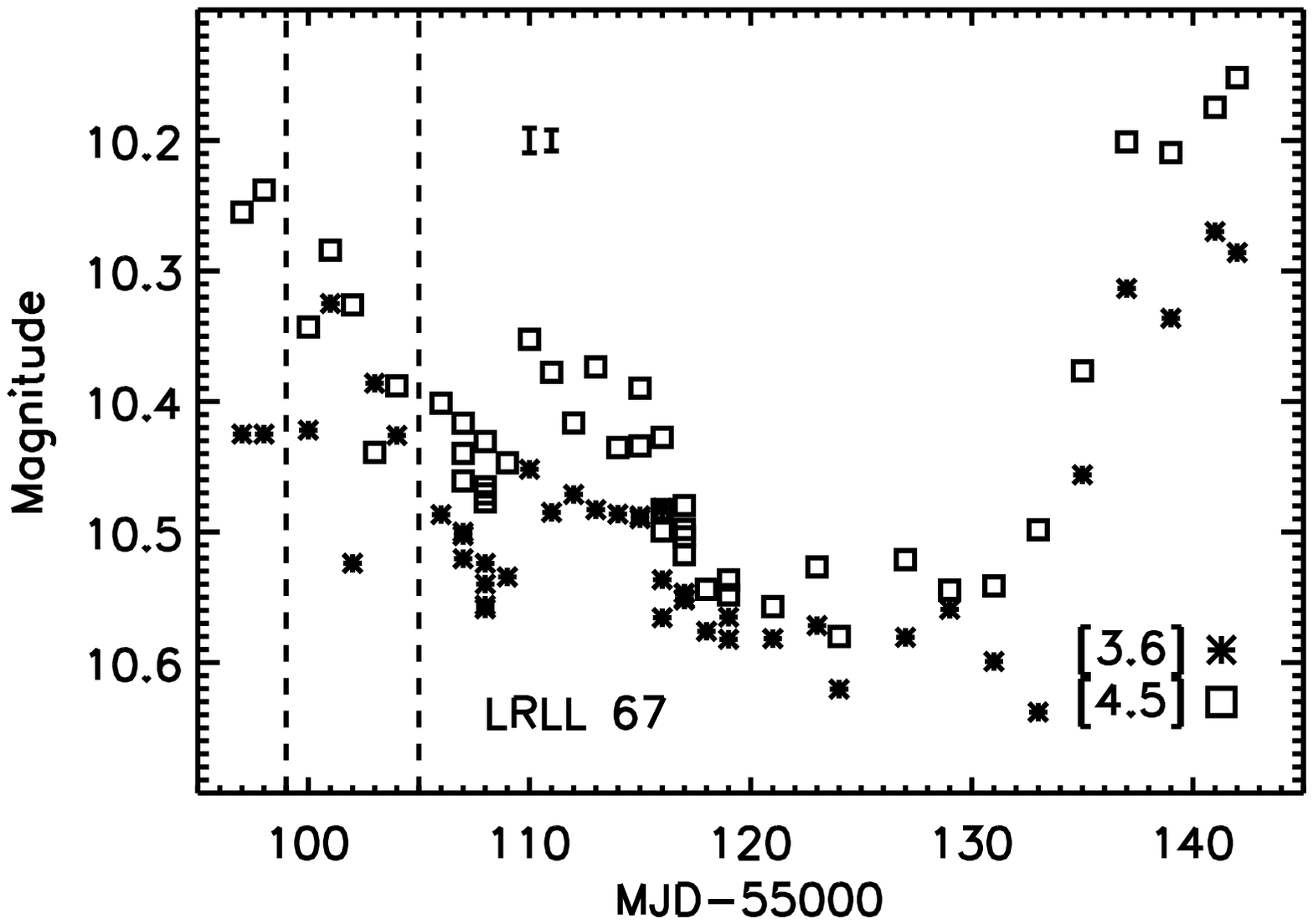}
\includegraphics[scale=.4]{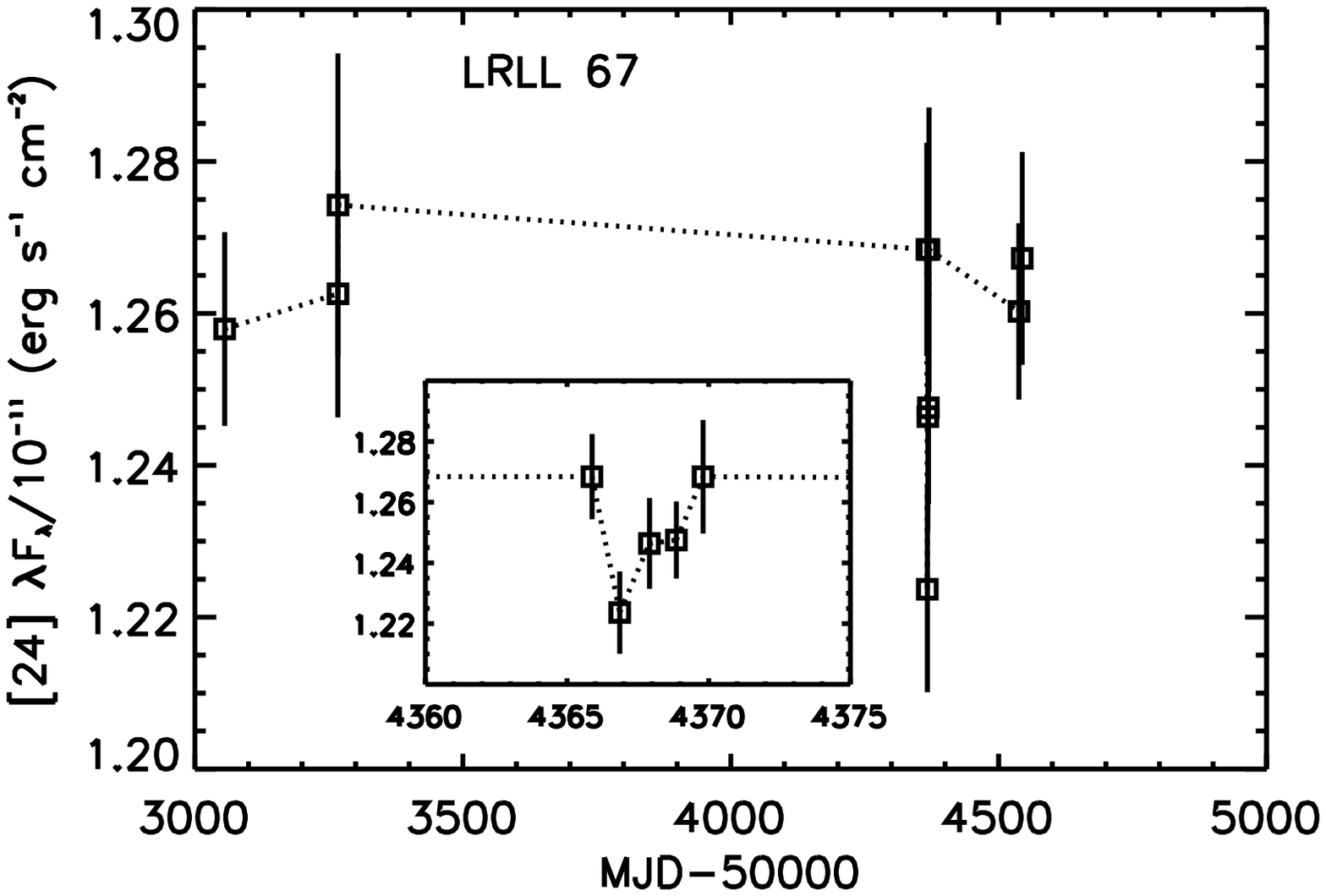}
\includegraphics[scale=.4]{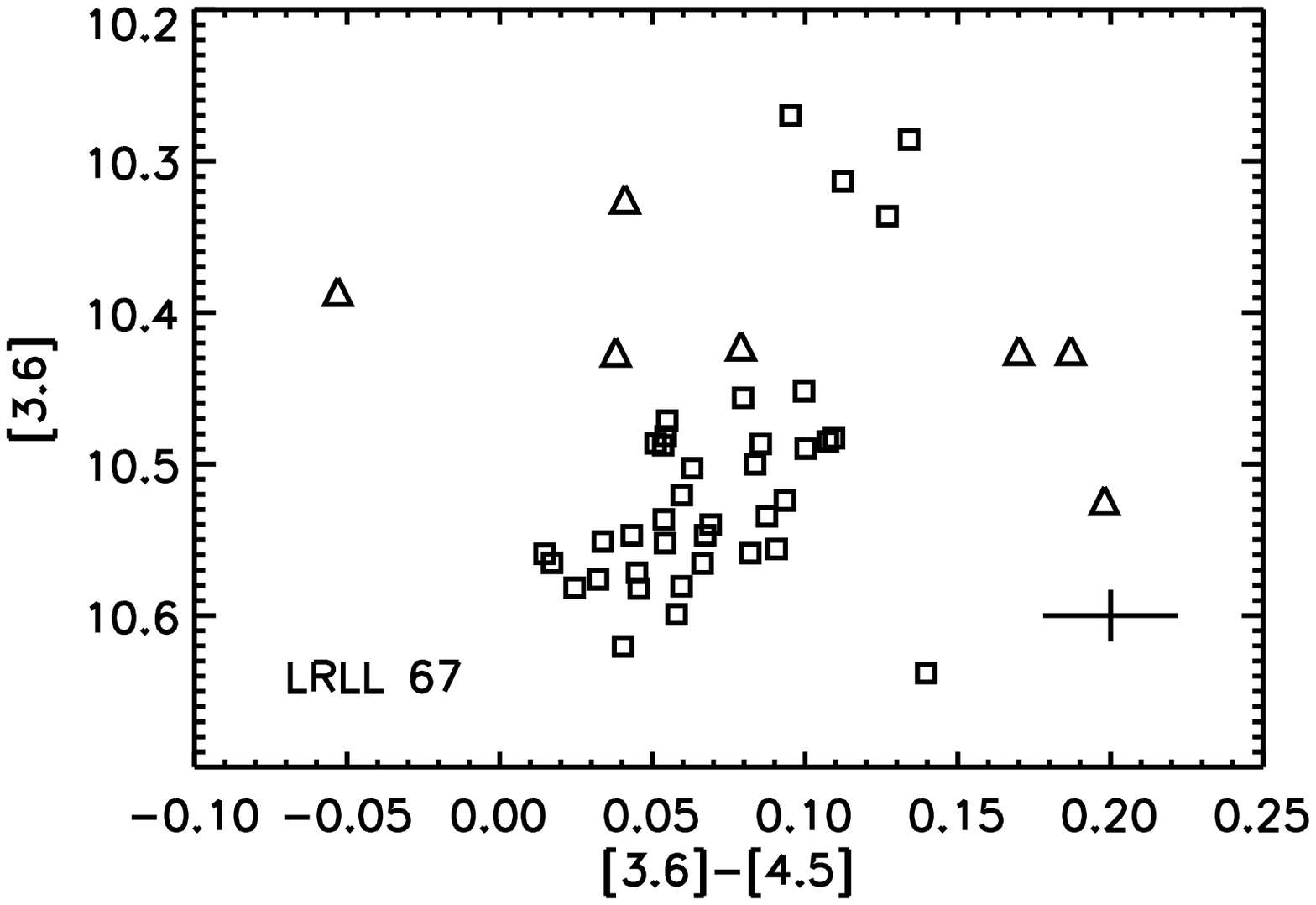}
\caption{Same as Figure~\ref{lrll2_ir2} but for LRLL 67.\label{lrll67_ir2}}
\end{figure*}

\subsubsection{Infrared Variability}
In the IRS spectra, LRLL 67 displays no variations ($<3\%$ at $\lambda<14\micron$ and $<10\%$ at $\lambda\ge14\micron$) (Fig~\ref{lrll67_ir}). The 24$\micron$ photometry display marginal fluctuations at the 2$\sigma$ level, which are consistent with a constant flux, although there is a deviation between the 24\micron\ photometry points and the long-wavelength IRS spectra (Fig~\ref{lrll67_ir2}). The five consecutive days of 3-8$\micron$ photometry show no fluctuations ($<5\%$), but the 3.6 and 4.5$\micron$ monitoring does show large (0.4 mag) variations with no periodicity. Most of the change during the monitoring occurs during a brightening event with a timescale of about a week. Prior to this the infrared flux varied by less than 0.2 mag, which is still significant given the small uncertainties in these data. The measurements are consistent with larger fluctuations at short wavelengths, similar to LRLL 21, possibly involving periods of quiescence. During the 3.6,4.5\micron\ monitoring the [3.6]-[4.5] color gets redder as the source gets brighter (Fig~\ref{lrll67_ir}), in a way that is consistent with a changing flux level from warm dust close to the star, similar to LRLL 21. \citet{esp12} find that a model with no optically thick dust within 10 AU is able to fit the infrared SED. The implications for the variability of the possible lack of optically thick dust close to the star on the variability are discussed in more detail in section~\ref{lrll67}.

\subsubsection{Gas Properties}
\begin{figure}
\hbox{\includegraphics[scale=.5]{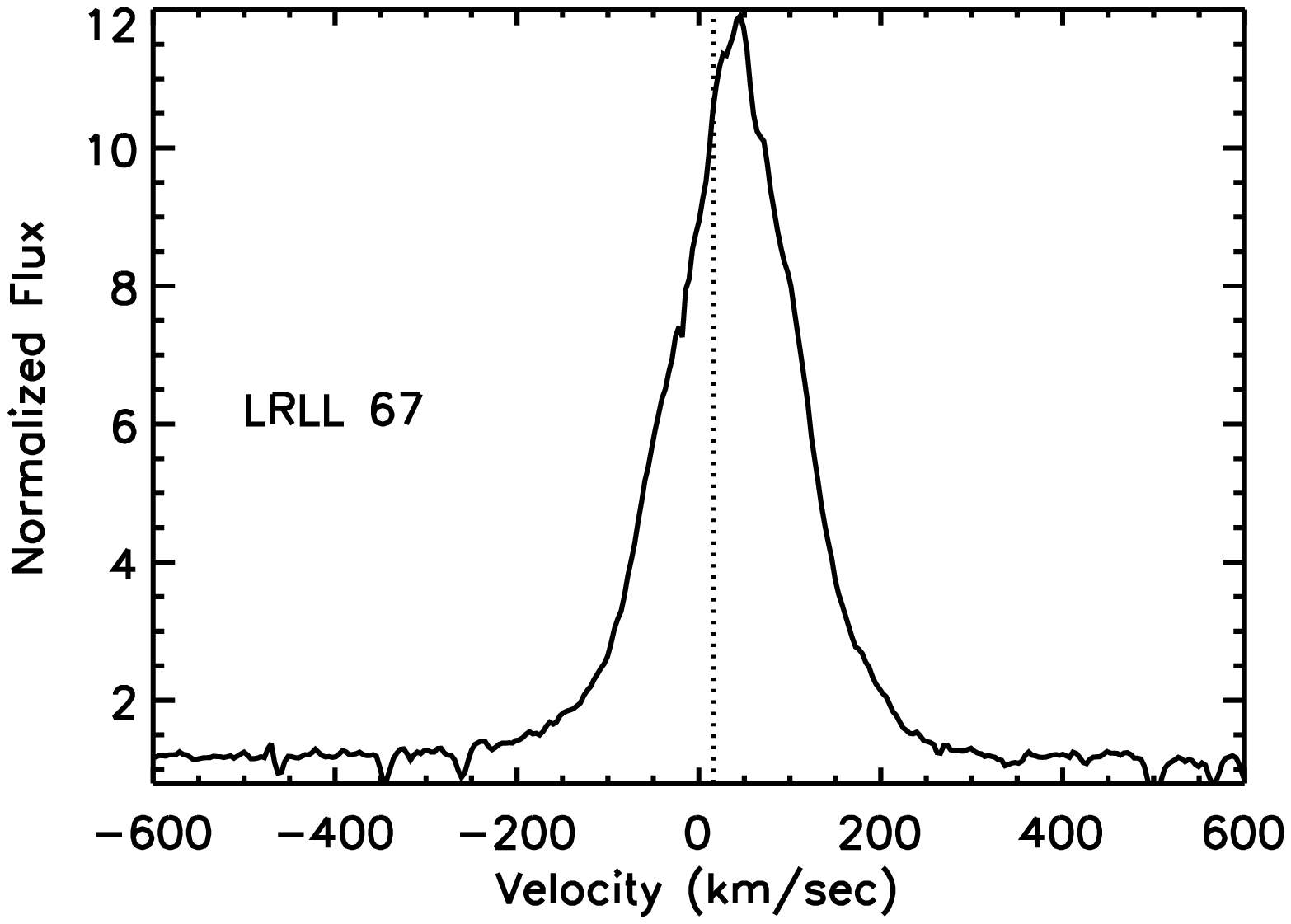}}
\caption{High-resolution optical spectrum of the H$\alpha$ line for LRLL 67 normalized to the continuum. Dashed line is the mean velocity of the cluster.\label{l67_halpha}}
\end{figure}

We have no near-infrared spectra of LRLL 67, but \citet{dah08} measure an H$\alpha$ equivalent width of -28.32\AA\ with an accretion rate of $2\times10^{-9}M_{\odot}$yr$^{-1}$ while \citet{luh03} measure an H$\alpha$ line strength of -35\AA\ and we find EW=-31.4\AA\ (full-width at 10\%\ of 310km/sec) in our high-resolution optical spectrum (Fig~\ref{l67_halpha}). LRLL 67 appears to be actively accreting, with a variable accretion rate, whose accretion luminosity is small (L$_{acc}$/L$_*$=0.06) compared to the stellar luminosity. The accretion rate derived from the width of the H$\alpha$ line \citep{nat04} is $\dot{M}\sim1\times10^{-10}M_{\odot}yr^{-1}$, which confirms that the accretion luminosity is very small. 

\subsection{LRLL 1679}
\subsubsection{Stellar Properties}
LRLL 1679 is the faintest star in our sample and was classified as a M3.5 star with L=0.21L$_{\odot}$ and R$_*$=1.3R$_{\odot}$ by \citet{mue07}. We estimate its mass as $\sim0.3$M$_{\odot}$ using the \citet{sie00} isochrones. Based on five epochs of near-infrared spectra (three in 2008, two in 2009) we find that its extinction is constant at A$_V$=5.8 (Table~\ref{extinction}). There is no mention of periodic optical fluctuations in the literature, although this source may have been too faint for these surveys or outside the field of view (at Dec=31:58:255 it is south of the cluster center).

\begin{figure*}
\center
\includegraphics[scale=.3]{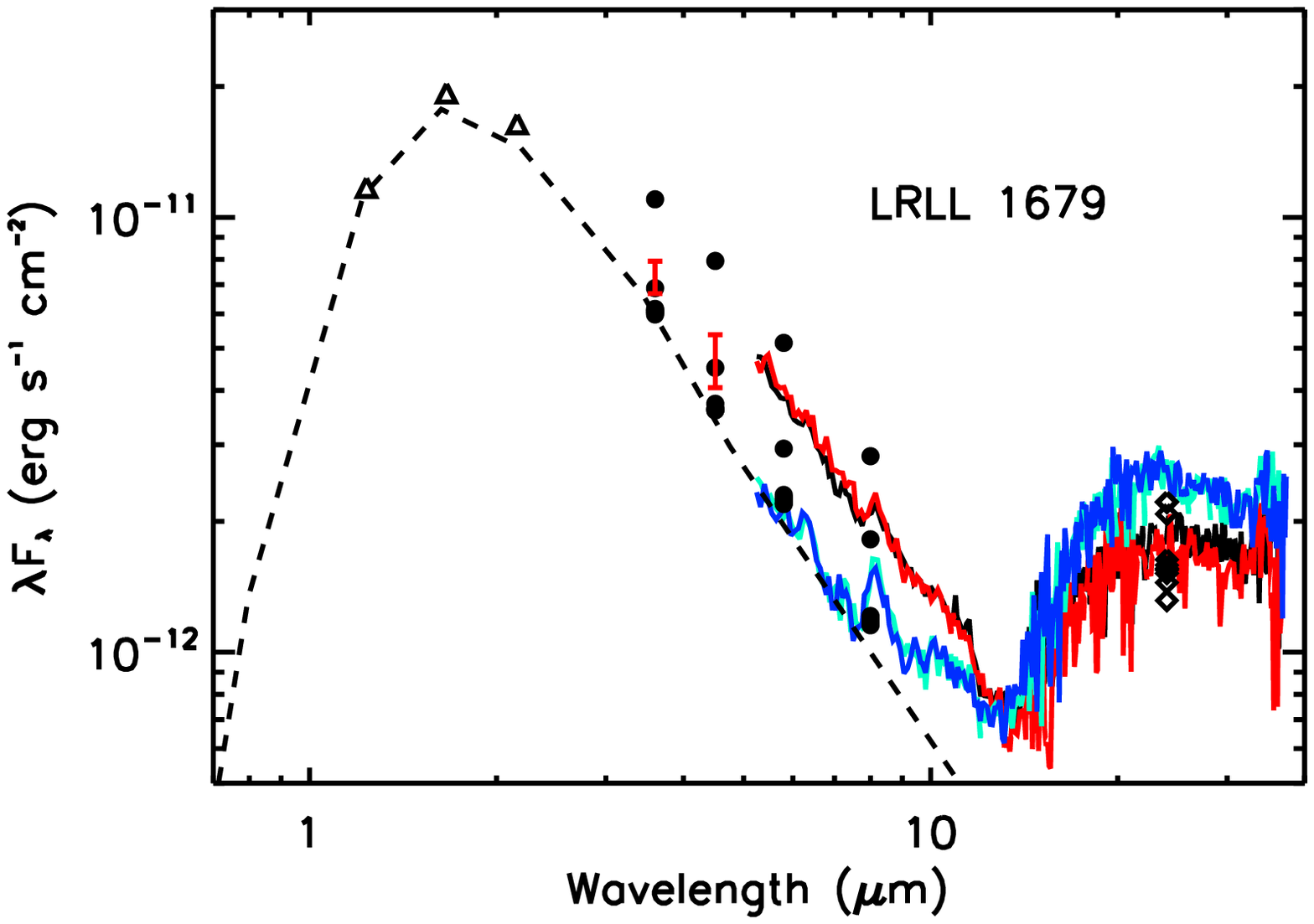}
\includegraphics[scale=.3]{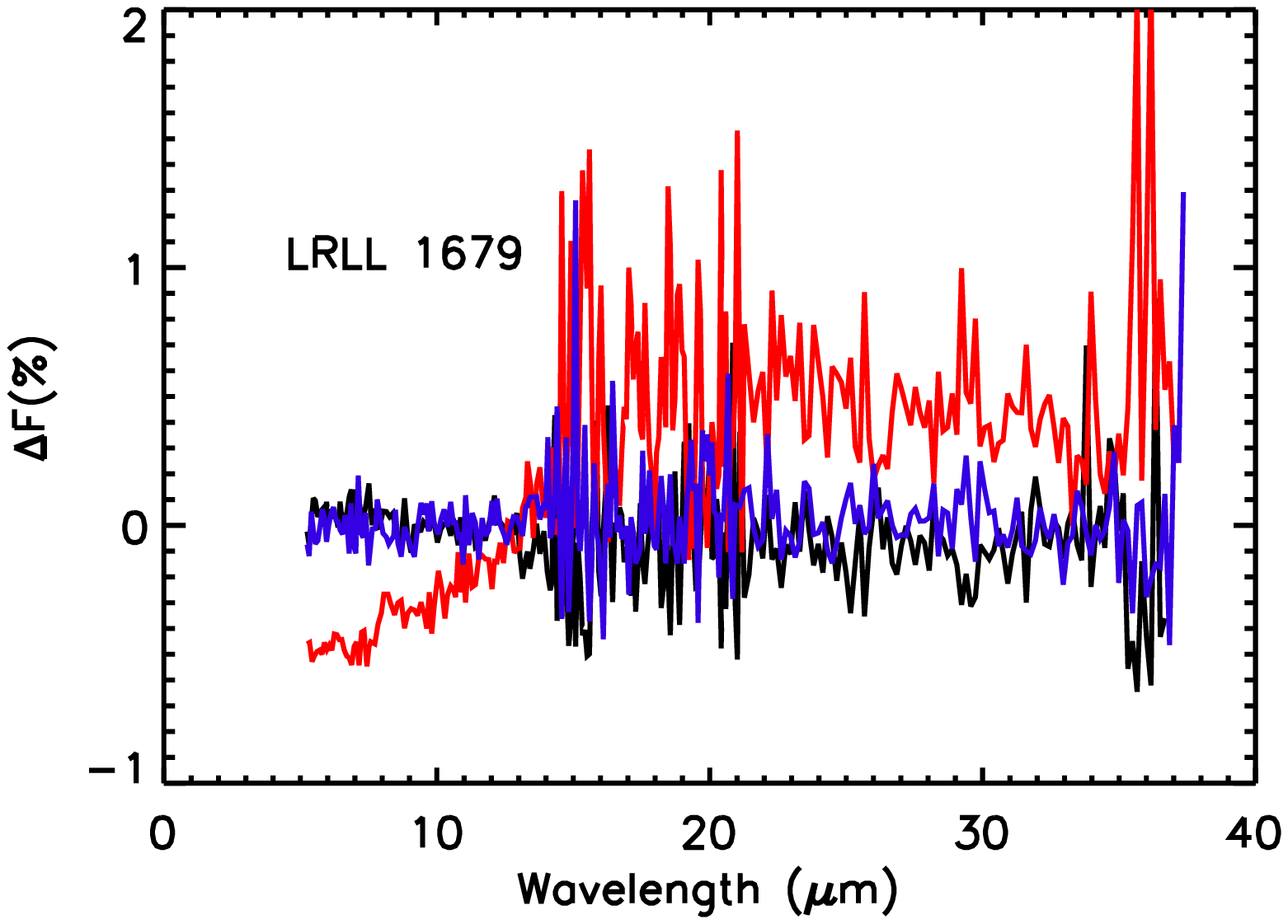}
\includegraphics[scale=.3]{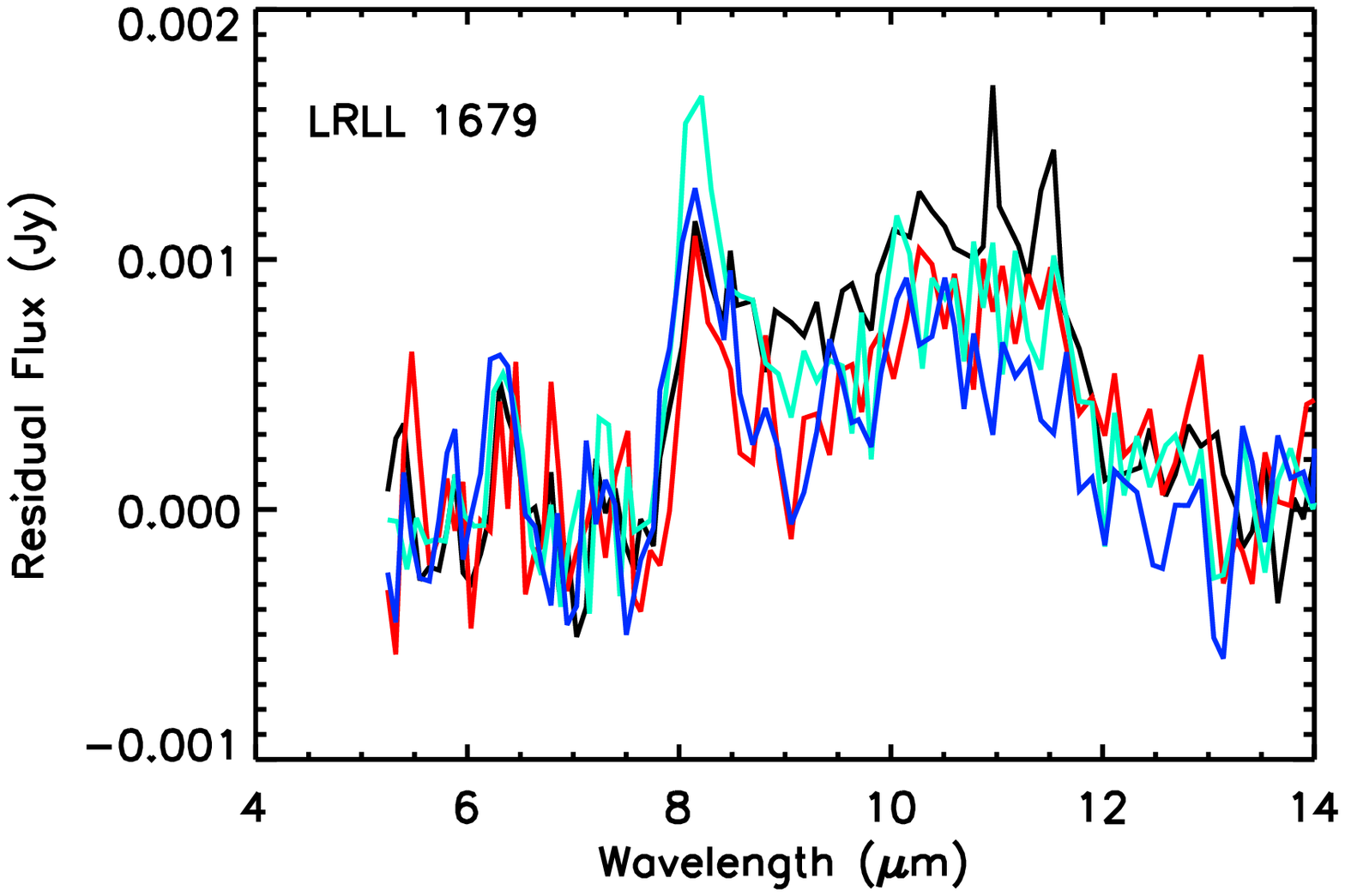}
\caption{Same as Figure~\ref{lrll2_ir} but for LRLL 1679.\label{lrll1679_ir}}
\end{figure*}

\begin{figure*}
\center
\includegraphics[scale=.4]{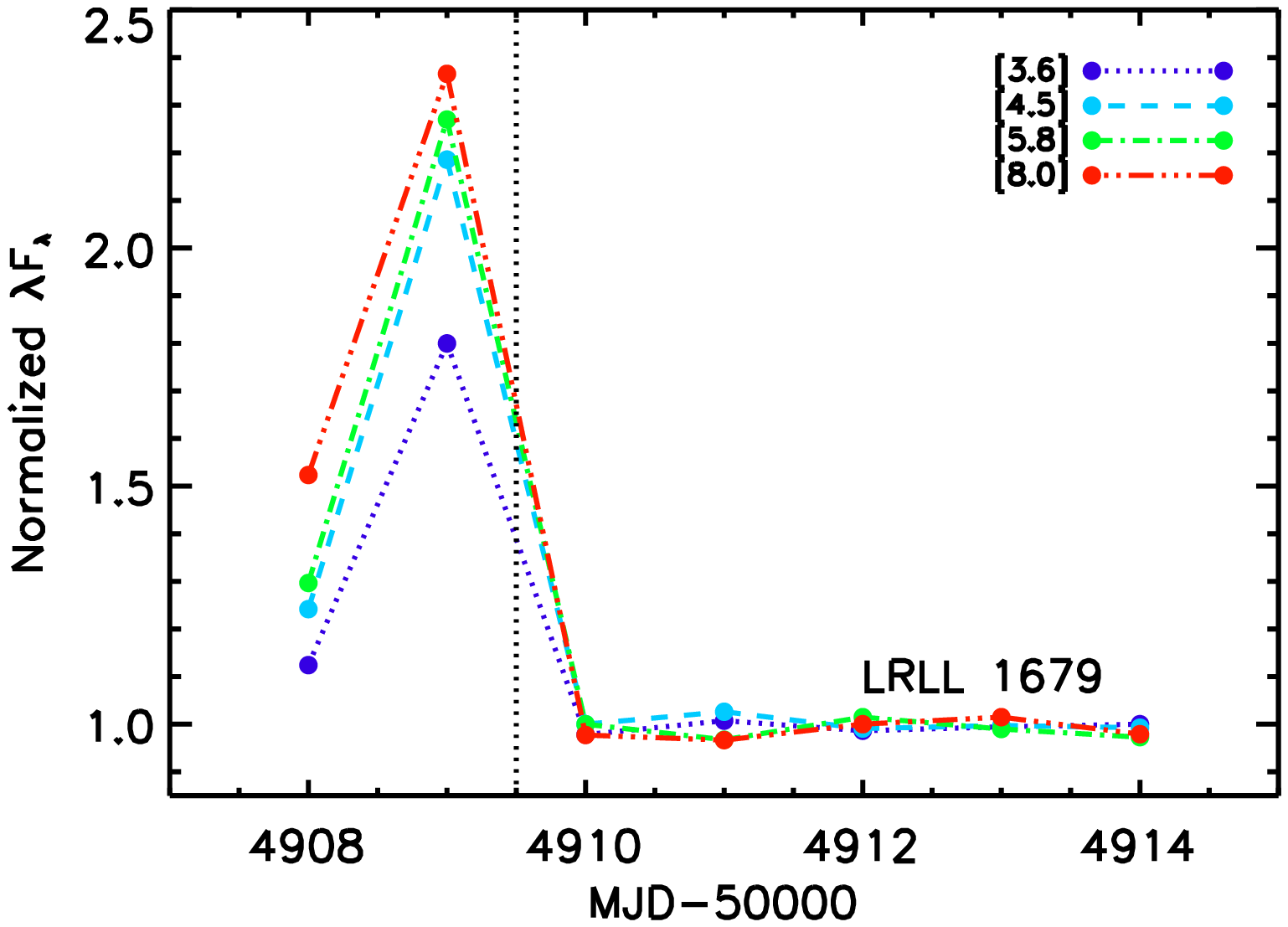}
\includegraphics[scale=.4]{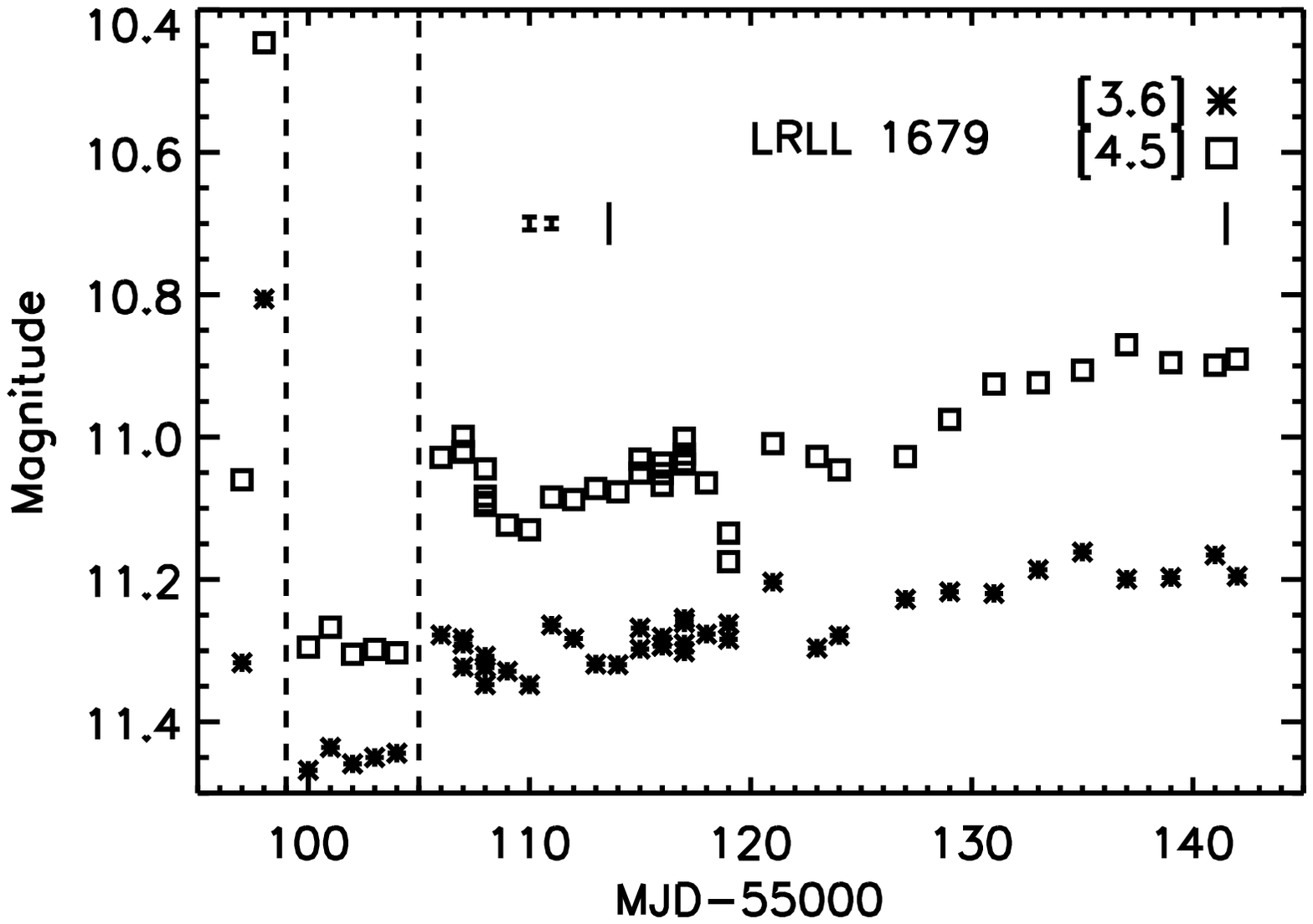}
\includegraphics[scale=.4]{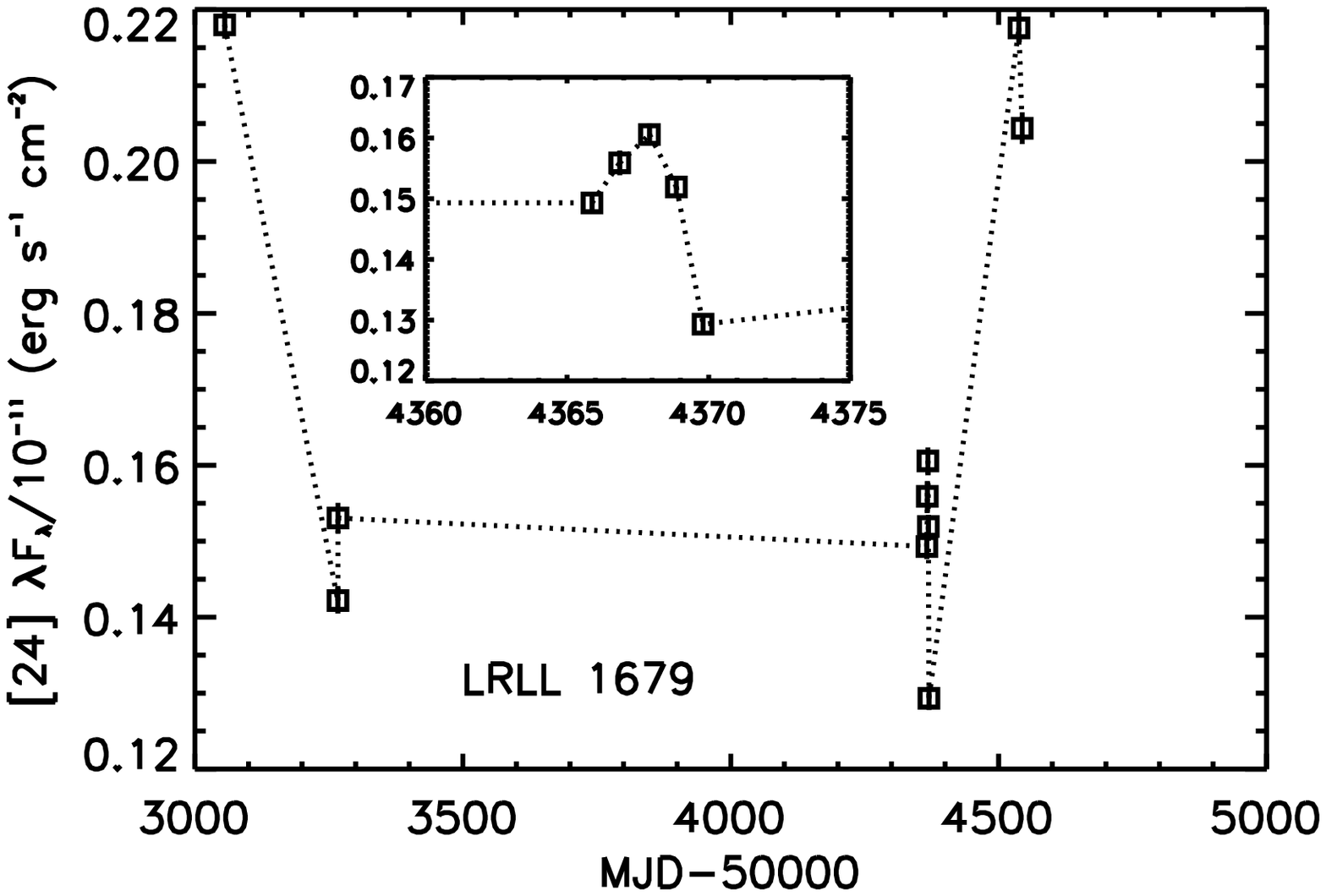}
\includegraphics[scale=.4]{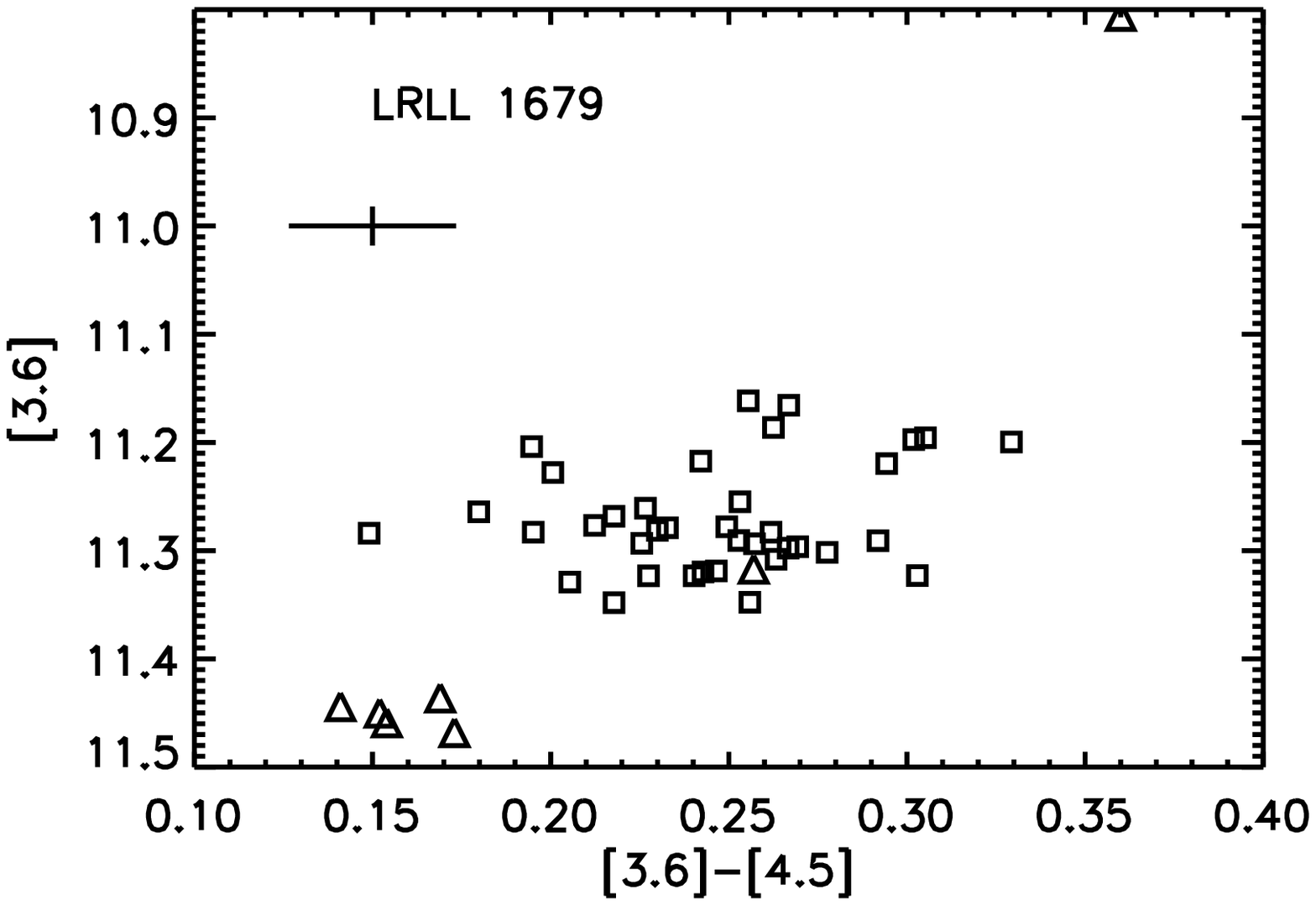}
\caption{Same as Figure~\ref{lrll2_ir2} but for LRLL 1679. \label{lrll1679_ir2}}
\end{figure*}

\begin{figure}
\includegraphics[scale=.5]{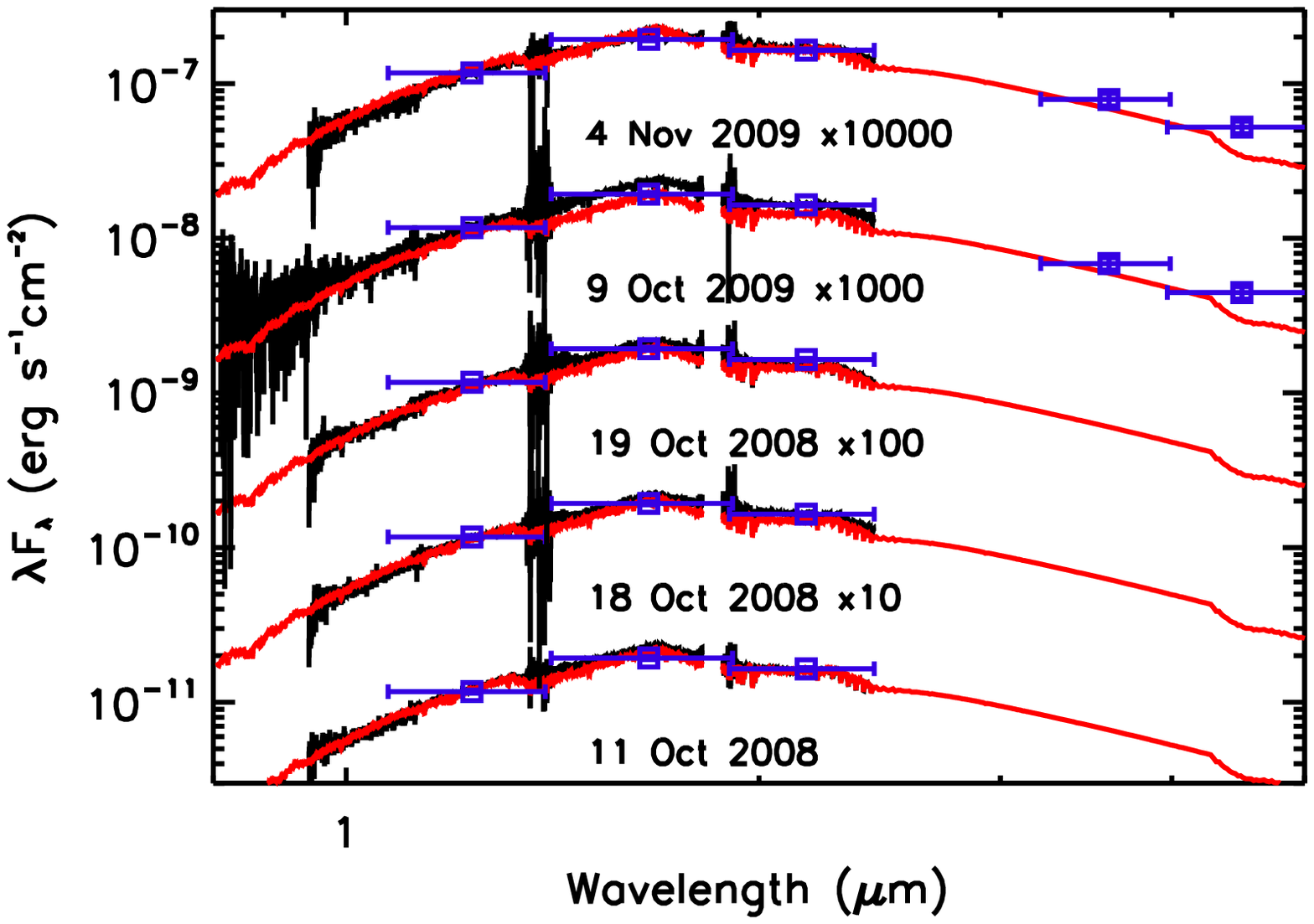}
\caption{Spex spectra (black line) and photometry (blue points) for LRLL 1679. Spectra have been scaled to match 2MASS photometry. A stellar photosphere (LkCa 21 + Kurucz model) shown in red has been reddened by the A$_V$ measured on each night to LRLL 1679 and scaled to the short wavelength flux of the spectra. Very little excess is seen throughout the observations. Spectra have been scaled by a constant factor for clarity. \label{l1679_spex+phot}}
\end{figure}

\subsubsection{Infrared Variability}
In the IRS spectra, LRLL 1679 displays large fluctuations ($\sim50\%$) on timescales of months, but no weekly changes ($<5\%$ for $\lambda<14\micron$ and $<20\%$ for $\lambda>14\micron$) (Fig~\ref{lrll1679_ir}). The SED appears to pivot, similar to LRLL 2, 21, 31 and 58, at $\lambda=13\micron$. The 24$\micron$ photometry displays fluctuations (50\%) that are consistent with those seen in the IRS spectra (Fig~\ref{lrll1679_ir2}). Despite the large fluctuations in the continuum, there is no change in the flux of the silicate feature (although it is the weakest in our sample). The five consecutive days of 3-8$\micron$ photometry show no variation ($<5\%$), consistent with the longer variability timescale. This is also confirmed in the 3.6 and 4.5$\micron$ monitoring, which shows a steady increase in the flux of 0.1 mag over the course of 40 days with no evidence for periodicity. The change in [3.6]-[4.5] color between the 3-8$\micron$ photometry and the 3.6,4.5$\micron$ monitoring, where the source gets redder as it gets brighter, is consistent with optically thick dust in the inner disk getting brighter with time, although there is no evidence of veiling in the 0.8-2.5$\micron$ spectra (Fig~\ref{l1679_spex+phot}). The short-wavelength excess can be very small, with the measurements almost at the level of the photosphere in some epochs, and it is possible that when we obtained our near-infrared spectra the excess was below our detection limit. The non-zero [3.6]-[4.5] color suggests that there is some hot dust close to the star creating a weak excess. As with LRLL 21 and 58 we can estimate the covering fraction in 2009 based on our nearly simultaneous near-infrared spectra and Spitzer photometry. On Oct 9 and Nov 4 we find covering fractions of $\sim2-3\%$ (Table~\ref{ir_excess_table}), which is much smaller than expected for a typical T Tauri star. The small emitting area of the inner disk may be due to either substantial grain growth and settling reducing the scale height of the inner disk, or non-axisymmetric distribution of optically thick material at the inner edge.

\subsubsection{Gas Properties}
\begin{figure}
\hbox{\includegraphics[scale=.5]{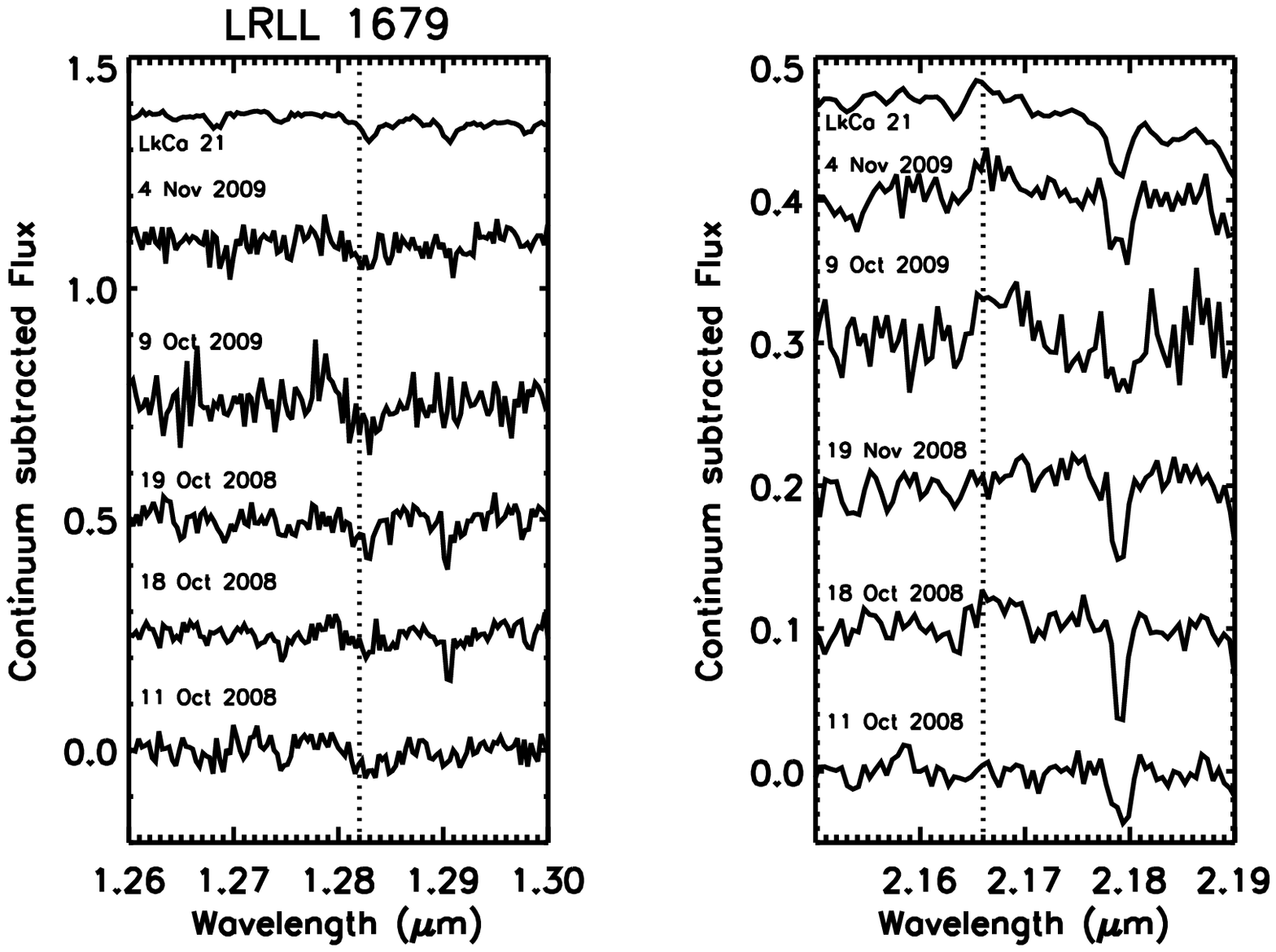}}
\caption{Same as Figure~\ref{l21_lines}, but for LRLL 1679. We do not subtract the photospheric absorption since it is very weak and would only add to the uncertainty in the spectra. There is no evidence for either Pa$\beta$ or Br$\gamma$ emission in any epoch. \label{l1679_lines}}
\end{figure}

\begin{figure}
\includegraphics[scale=.5]{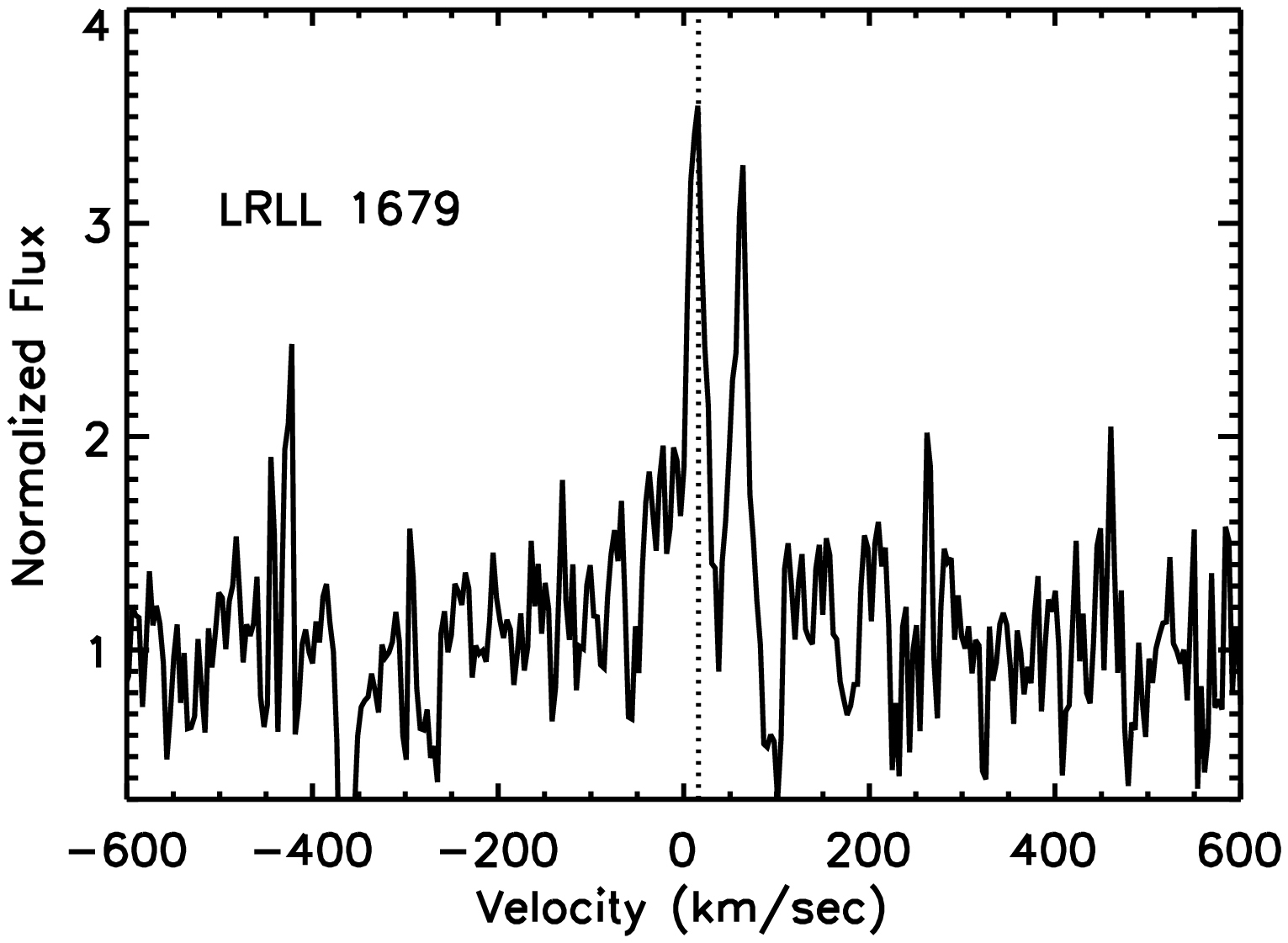}
\caption{Optical spectra of the H$\alpha$ line for LRLL 1679. Dashed line is the mean velocity of the cluster. \label{l1679_halpha}}
\end{figure}

During our five epochs of near-infrared observations, three in 2008 over the course of a week and two in 2009 separated by a month, we find no evidence for Pa$\beta$ or Br$\gamma$ emission (Fig~\ref{l1679_lines}). We can place upper limits on the accretion rate of $\sim10^{-9}M_{\odot}yr^{-1}$ (Table~\ref{accretion}). Our low resolution optical spectrum taken in 2009 during the 3.6,4.5\micron\ monitoring measures an EW of the H$\alpha$ line of -3.3\AA, which is below the accretion boundary for a star of this spectral type \citep{whi03}. The high-resolution H$\alpha$ spectra has an EW of only -1.7\AA\  (full-width at 10\% of 120km/sec) and shows very weak, narrow emission consistent with completely photospheric emission (Fig~\ref{l1679_halpha}). In no epoch do we detect evidence for ongoing accretion in this star. 

\subsection{Summary}
Our sample represents a wide range of stellar masses, from 0.3-3.5M$_{\odot}$. Except for LRLL 21, none of the sample show evidence for large changes in the stellar flux, either due to occultation events or hot spots rotating across the surface of the star. In previous optical monitoring surveys of IC 348 some do show small fluctuations due to cold spots rotating across the surface of the star. In the case of LRLL 21, changes occur from one year to the next, but not on daily or weekly timescales.

The exact size, wavelength dependence and timescale of the infrared fluctuations varies from source to source, but overall all six sources show variability at all infrared wavelengths. Some show very large flux changes, up to 50\%, while others show much smaller variations. Most show fluctuations in which the short wavelength flux decreases when the long wavelength flux increases. Some sources (LRLL 2,31) show a correlation between the silicate flux and the long-wavelength flux, while another source (LRLL 1679) does not, which may reflect differences in the geometry of the dust and the location of the silicate emission zone, in particular whether or not the silicate emission zone can be experience the same shadowing as the outer disk. The timescales range from weeks to months, with no evidence for long year-to-year changes, suggesting that we are not observing the remnants of a single burst or the slow removal/addition of dust to the disk. None of them show strong evidence for periodicity in the warm mission data, which is sensitive to P=2-25 days.  There is some evidence for a dependence of the location of the pivot in the SED on mass with the wavelength of the pivot decreasing with increasing stellar mass, although it is more likely that this pivot wavelength depends more on the exact dust geometry of the system rather than the central mass \citep{esp11}.

In most of the sources there is evidence for variable accretion, which at some epochs dips below our detection limits. The only exception is LRLL 1679 in which there is no evidence of ongoing accretion above our detection limits at any epoch. Overall the accretion luminosity is small relative to the stellar luminosity even when the stars are actively accreting.

\section{Variability Phenomenology}\label{perturbe}

Given all of this information, we seek to address a number of questions: Is the variability in all of these systems caused by similar perturbations to the disk?  Is the type of perturbation common among cluster members? 

\subsection{Is the origin of infrared variability universal?}

\subsubsection{Comparison with LRLL 31}
There is a bewildering array of infrared variability for our six sources. Some show large infrared fluctuations with a clear wavelength dependence in every epoch, some do not. Some (LRLL 58,67,1679) show almost no excess emission for $\lambda<8\micron$ in some epochs, while others (LRLL 21) always have a significant excess. In Table~\ref{var_sum} we summarize the data presented in the previous section regarding the variability of the stellar flux, accretion flux and infrared flux, which we now use to determine if these sources can be included together as one coherent type of object. 

As a starting point we consider the model for the variability of LRLL 31 \citep{fla10,fla11} which relies on a vertical perturbation to the inner disk. \citet{esp11} employ a similar geometry, with a much more detailed model of the dust structure and the radiative transfer, to successfully model the infrared fluctuations of a sample of nearby transition disks, suggesting that this model may be applicable to many sources. The key component is optically thick material close to the dust sublimation radius that experiences some sort of vertical perturbation causing it to rise out of the midplane. The consequences of this model are:
\begin{itemize}
\item An SED that varies at all infrared wavelengths and appears to pivot around a certain wavelength between different epochs. As a vertical perturbation rises out of the inner disk it produces more short wavelength flux while also shadowing the outer disk, causing the long-wavelength flux to decrease. In the case of the LRLL 31, the flux shortward of 8\micron\ would decrease when the long-wavelength flux increased. The pivot point may change depending on the geometry of the disk \citep{esp11}, but it is difficult to produce this wavelength dependence without a variable inner disk scale height \citep{fla10}. This perturbation may or may not be axisymmetric about the star.
\item The infrared flux changes on the dynamical timescale of the inner disk (typically days to weeks). These rapid variations can even be seen at the longest wavelengths, which is dominated by emission from material at a few AU from the star where the dynamical timescales are closer to years and the disk structure cannot substantially change in a matter of days. The long wavelength fluctuations are due to variable shadowing instead of in-situ structural changes.
\item The [3.6]-[4.5] color gets redder as it gets brighter. A change in the disk emission, rather than simply the stellar flux, can produce this effect since the relative contribution of the disk emission, specifically the inner disk which dominates at these wavelengths, is larger at 4.5\micron\ than at 3.6\micron. 
\end{itemize}

Overall each source displays some of the signs of a variable inner disk suggesting that all of these sources are experiencing the same type of variability. LRLL 2 displays a SED that pivots around $\lambda=6\micron$. LRLL 21 has a light curve that is almost identical to LRLL 31; its [3.6]-[4.5] color gets redder as it gets brighter and direct observations of the emission from the inner disk show that it is optically thick and the size of the emitting region gets larger while its temperature stays the same. The 24\micron\ photometry shows that LRLL 21 is variable at this wavelength. LRLL 58 shows a pivot in the SED at $\lambda=8\micron$ when comparing the photometry and the IRS spectra, and the 24\micron\ photometry shows significant fluctuations. It also shows variations on short timescales, consistent with the dynamical timescale of the inner disk (5 days). LRLL 67 shows rapid fluctuations consistent with the dynamical timescale of the inner disk (4 days) and the [3.6]-[4.5] color gets redder as it gets brighter.  LRLL 1679 clearly displays a pivot in its IRS spectra at $\lambda=13\micron$ and there is some evidence that the [3.6]-[4.5] color gets redder as it gets brighter. The small change in optical flux on short timescales in all of these sources suggests that the irradiation of the disk does not change appreciably. If the inner disk radius is set by dust sublimation and thus proportional to the stellar and accretion luminosity, it is not changing in these objects. The rapid infrared fluctuations imply changes in the emitting area of the disk, but since the radius of this dust does not change then the scale height must be variable. That none of them show all of the hallmarks of a variable inner disk may be a sign of differences in the exact geometry of the disk (amount of optically thin dust in the inner disk, location of the outer disk, grain size/composition) rather than differences in the perturbation. In general, the similarities in behavior suggest roughly similar underlying causes for the variability with the possible exception of LRLL 67.

\subsubsection{The case of LRLL 67\label{lrll67}}
One key tenet of this model for the infrared variability is an optically thick inner disk and in most of the transition disks we have strong evidence for such material. LRLL 21 and 31 show direct evidence for optically thick inner disks based on the shape of the 2-5\micron\ excess. The strength of the veiling observed at K band in LRLL 58 is also consistent with strong emission for an optically thick inner disk, rather than the weak emission from an optically thin disk. LRLL 2 and 1679 show a pivot when comparing IRS spectra, while LRLL 58 shows a pivot between the spectra and photometry, which is difficult to explain without shadowing from an optically thick inner disk. For LRLL 67 the evidence for an optically thick inner disk is weaker than in any other star. Its SED is reminiscent of sources such as GM Aur, DM Tau, CS Cha and Coku Tau/4, which have been modeled using only optically thin dust in the inner disk \citep{esp07,dal05,cal05}. We have no near-infrared spectra to look for a sign of hot dust, but the weak 3-8\micron\ excess at all epochs is difficult to explain with optically thick dust. In fact, \citet{esp12} are able to fit the infrared SED without the use of any optically thick dust within 10 AU. If the inner disk is optically thin then changing the scale height will have no effect on the short wavelength flux, since the emission is proportional to the total column density of material and not its exact geometry. The inner disk will also not be able to shadow the outer disk, leading to no variation in the long-wavelength flux.

LRLL 67 has a substantial accretion rate, indicating that some material must be traveling from the massive outer disk and onto the star implying that the inner disk is not completely devoid of material. The measured accretion rate of $2\times10^{-9}M_{\odot}$ yr$^{-1}$ corresponds to a gas surface density of $\sim10$ g cm$^{-2}$ at 1 AU assuming a standard $\alpha$ disk model \citep{dal99}. If the gas to dust mass ratio is 100, then the dust emission should be optically thick. \citet{ric06} find that the outer edge of the dust depleted gap can block some of the inward flowing dust resulting in a very high gas to dust ratio and a much lower dust mass in the inner disk, possibly accounting for the weak dust emission despite the high gas density. Another explanation for the weak excess is if the inner disk were radially optically thick while vertically optically thin. In this case it would be able to shadow the outer disk, while still producing very weak emission. This model has difficulty explaining the variability at 3.6 and 4.5\micron\ with a changing geometry, although if the inner disk were optically thin then a change in its temperature could lead to substantial changes in the infrared flux. This variable heating could be due to an unseen companion on an eccentric orbit \citep{nag10} or changes in the accretion rate, although for LRLL 67 the heating of the disk is dominated by the stellar flux. 

A completely optically thin inner disk would produce no shadowing of the outer disk, and no 'seesaw' behavior in the SED. \citet{esp11} find that the pre-transition disks (i.e. those that have optically thick inner disks) in their sample show the 'seesaw' behavior in the mid-infrared spectra, while the sources without optically thick dust in the inner disk do not. Of our evolved disks, LRLL 67 shows the least amount of evidence for a pivot in the SED or any 'seesaw' behavior in the IRS spectra, leaving open the possibility that the outer disk is not shadowed, although the difference between the 24\micron\ emission and the IRS spectra is difficult to explain without shadowing of the outer disk. This long wavelength variability suggests that some fraction of the inner disk should be optically thick. Observations of the near-infrared excess to determine if it is optically thick, as has been done for LRLL 21 and 31, as well as detailed modeling of the SED are needed to better understand this source.

While the other evolved disks all have strong evidence for optically thick material in the inner disk, they also all have epochs when the short-wavelength excess is very weak, which is difficult to explain with an optically thick disk. For LRLL 21 and 1679 we estimate covering fractions of 0.02-0.04 when the excess is at its weakest (Table~\ref{ir_excess_table}). A perfectly flat disk, which is needed to minimize the flux, would need to be very close to edge on to produce an excess this small. It is also possible that the inner disk is non-axisymmetric, with the majority of it optically thin with only thin trails of optically thick dust. The small covering fraction of these trails would create a very weak excess  \citep{dod11}. LRLL 67 could represent an extreme case with only very thin trails of optically thick material, while the other stars have larger patches of optically thick material.  In addition to vertical perturbations of the inner disk, there may also be variations in the azimuthal coverage of the optically thick material. In some epochs the azimuthal coverage could be very small, producing a minuscule excess at short wavelengths. Further observations and SED modeling of these systems are needed to fully constrain the possibly non-axisymmetric distribution of material in the inner disk.

\subsection{Comparing Transition Disk Variables to Other Variables}
\begin{figure}
\includegraphics[scale=.5]{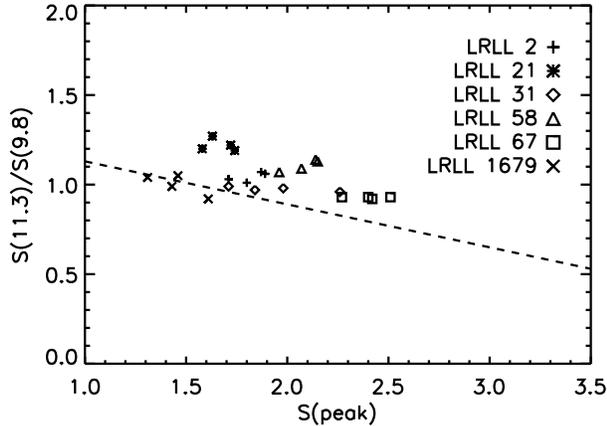}
\caption{Strength versus shape of the 10$\micron$ silicate feature for the evolved disks. Definitions of the axes are from Kessler-Silacci et al. 2006. The dashed line shows a fit to stars from c2d and demonstrates the correlation between strength and shape found for typical stars. The evolved disks tend to lie slightly above this relation, but are not substantially different from other T Tauri stars. \label{silicate_trend}}
\end{figure}

Do these stars represent a unique class of objects, or is the physical mechanism responsible for their variability common to other young stellar objects? We can start to answer this question by comparing the dust grain properties in our evolved disks, as traced by the 10\micron\ silicate emission, to those in a less evolved disk population. The emission can be characterized by the continuum normalized peak flux (S(peak)) and the ratio of the continuum normalized flux on two sides of the feature ($S_{11.3}/S_{9.8}$) as in \citet{kes06}. The wavelengths 11.3 and 9.8\micron\ correspond to the peak of the emission profiles from crystalline and amorphous silicates, allowing us to distinguish processing of the dust grains. In Figure~\ref{silicate_trend} we compare the strength and shape of the silicate feature for each observation of the evolved disks with the trend previously observed by \citet{kes06} and find that the dust properties of the evolved disks studied here are similar to those of other stars. A similar conclusion is found when using the method of \citet{man11} and \citet{fur09}, which apply slightly different indices to trace the strength and shape of the feature, and comparing to the Taurus, Chameleon and Ophiuchus cluster members studied in those papers. Although our sample consists entirely of evolved disks, and by definition the structure of the dust within the disk is different from a typical T Tauri star, there is no clear distinction in their dust properties. The previously mentioned studies have also found that dust properties do not vary strongly between normal and evolved disks. 

\begin{figure*}
\center
\hbox{\includegraphics[scale=.5]{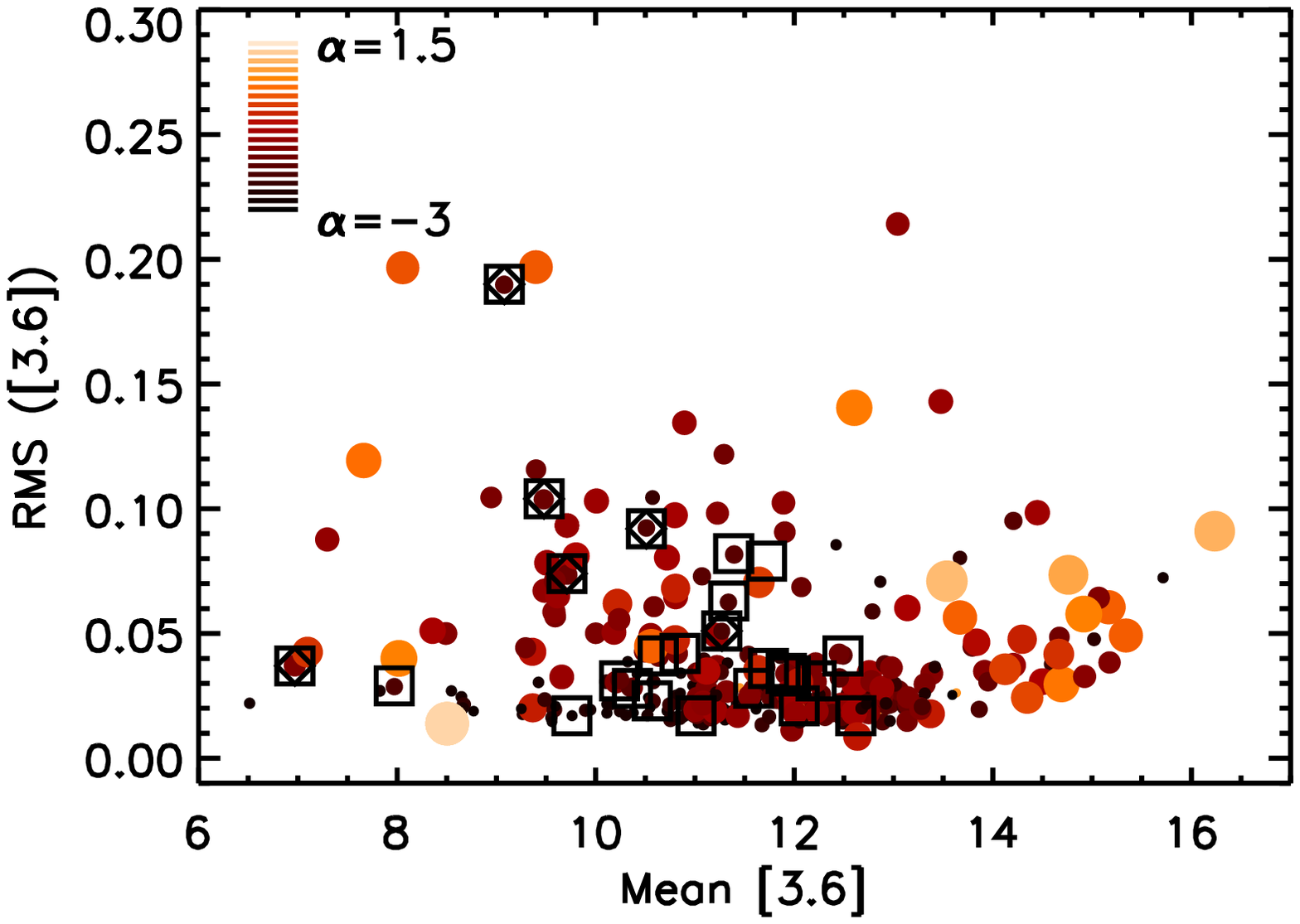}
\includegraphics[scale=.5]{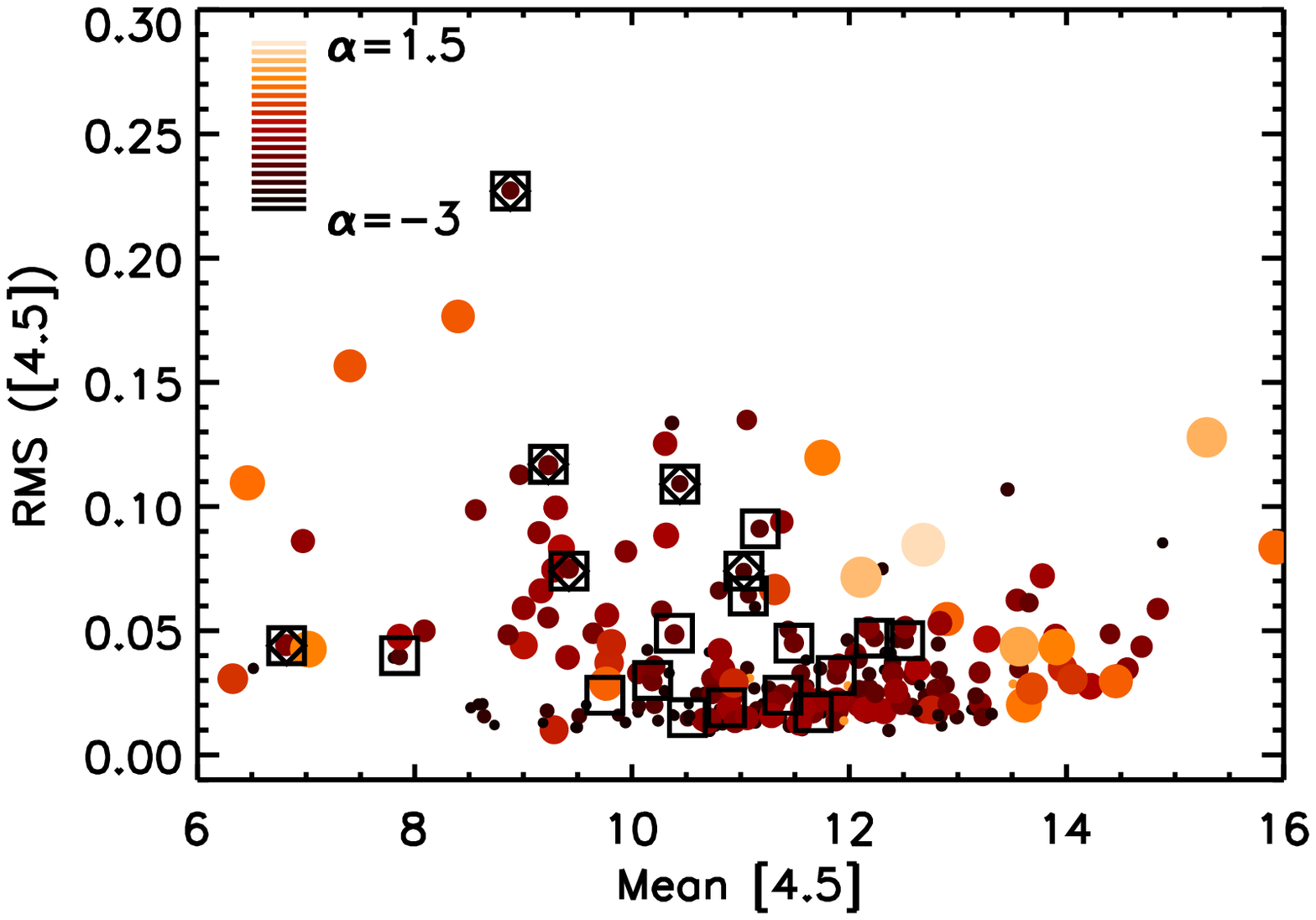}}
\hbox{\includegraphics[scale=.5]{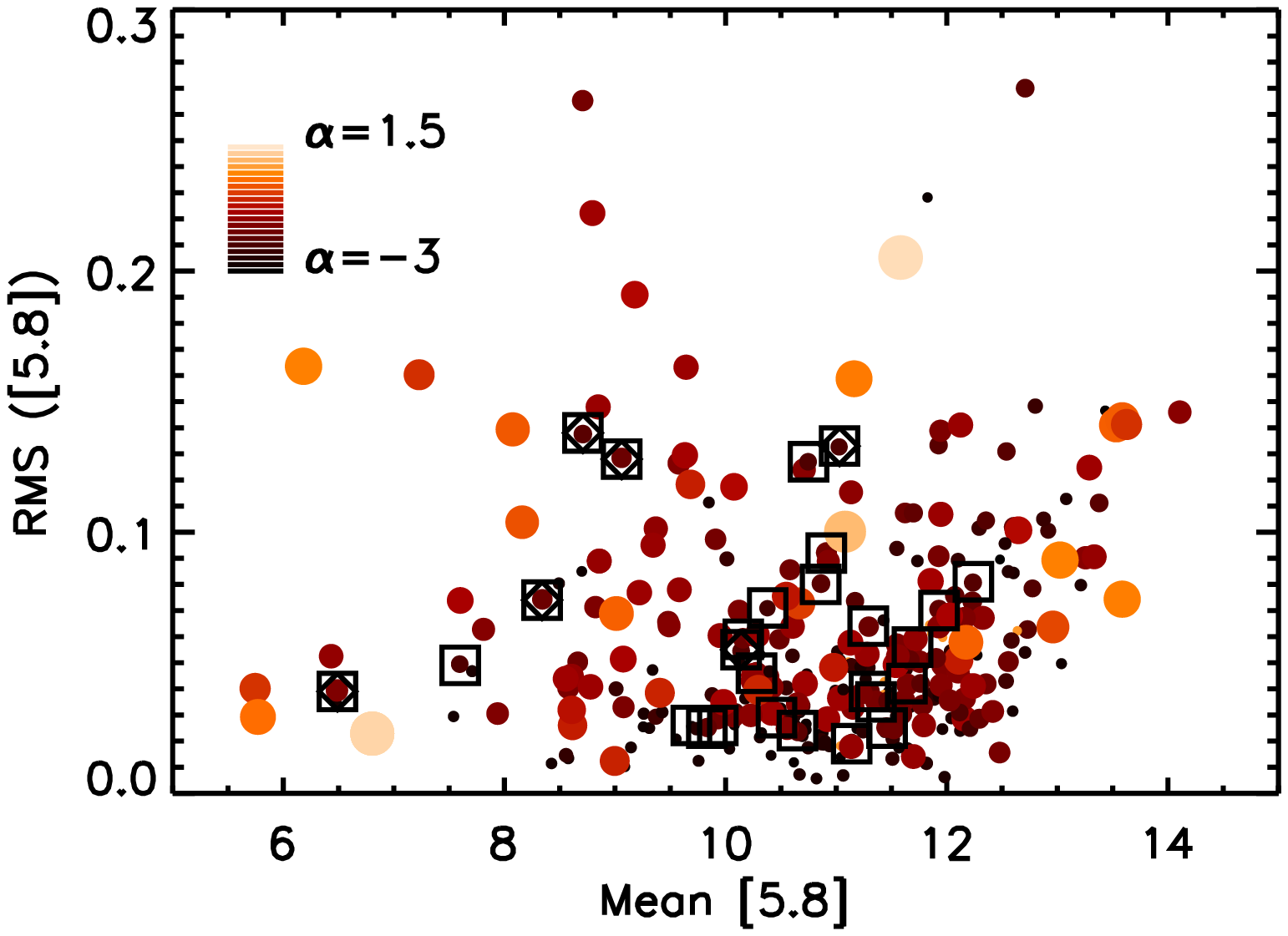}
\includegraphics[scale=.5]{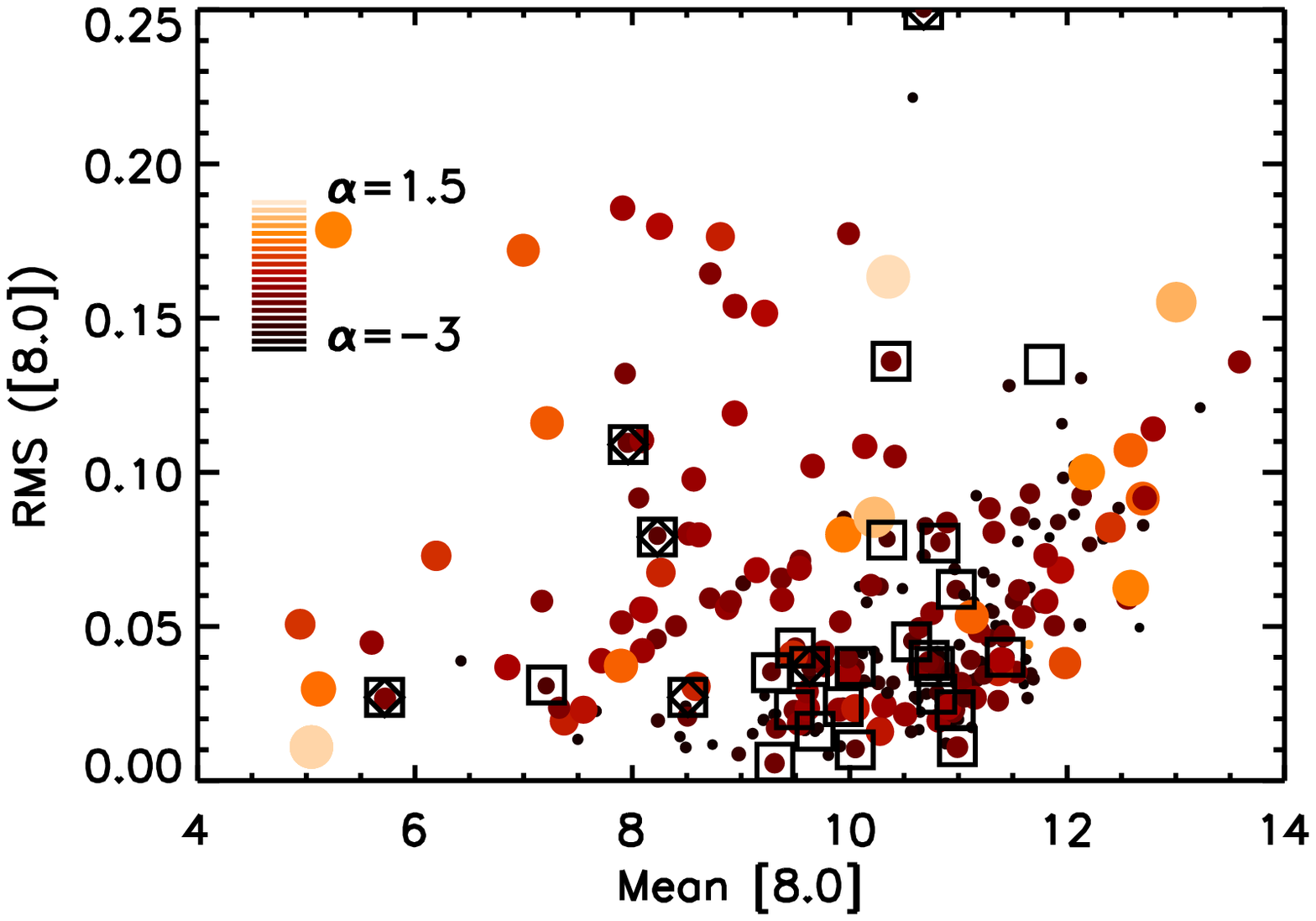}}
\hbox{\includegraphics[scale=.5]{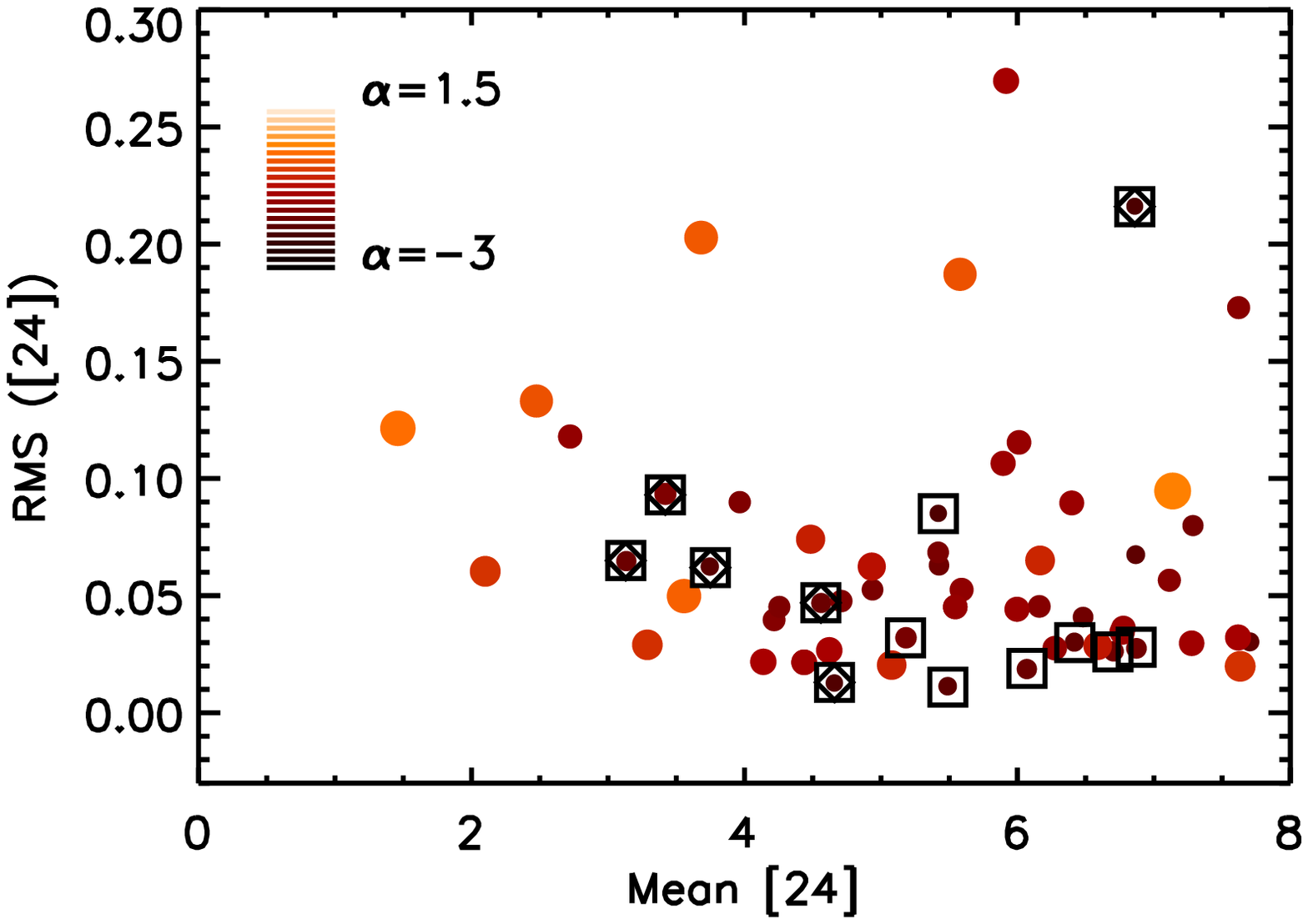}}
\caption{RMS of the fluctuations versus average magnitude for the cluster members. We show the 3.6 and 4.5\micron\ photometry from the warm-mission monitoring as well as the 5.8, 8.0 and 24\micron\ photometry from our cold-mission monitoring. The size and color of each point scales with the power law slope of the SED from 3.6 to 8.0\micron\ (Large, light symbols have rising SEDS while small dark points are stellar photospheres. Small, light points do not have enough information determine the slope of the SED). Transition disks are marked with squares and the evolved disks studied in this paper are marked with diamonds. The size of the fluctuations of the evolved disks is similar to other stars in the sample. \label{rms}}
\end{figure*}

The infrared fluctuations of these evolved disks originally drew us to this sample, and they can be used as a point of comparison with other young stellar objects. We have used our IRAC and MIPS photometry to compare the size and timescale of the infrared variability between the transition disks and the rest of the young stars in IC 348. One way to trace the typical size of the infrared fluctuations is with the rms of the photometry. The rms is defined as \citep{car01}:
\begin{equation}
(rms)^2=\frac{n\Sigma^n_{i=1}w_i(m_i-\bar{m})^2}{(n-1)\Sigma^n_{i=1}w_i}
\end{equation}
where $n$ is the number of observations, $m_i$ is the magnitude on a given epoch, $w_i=1/\sigma_i^2$ and $\sigma_i$ is the photometric noise. In Figure~\ref{rms} we show the rms of the fluctuations at every wavelength for which we have photometry, with the size and color of each point scaled to the slope, $\alpha$, of the IRAC SED, defined as $\lambda F_{\lambda}\propto\lambda^{\alpha}$, since the frequency of infrared variability appears to depend on the strength of the infrared excess \citep{mor11}. For the [3.6] and [4.5] photometry we only include the warm mission photometry since any systematic uncertainties in the photometry may differ between the cold mission and warm mission. For the [5.8] and [8.0] photometry we exclude the c2d photometry because the different observing strategy prevented them from measuring accurate photometry of many of the brightest members. We also mark the transition disks in IC 348 identified by \citet{muz10} for a comparison between evolved disks. The rms for the evolved disks studied here does not stand out from the rest of the cluster members at any wavelength. The number of epochs of data available at each wavelength varies (38 for [3.6] and [4.5], 6 for [5.8] and [8.0] and 10 for [24]) but the conclusion is the same. The fluctuations of our sample are larger than average, but this is mainly because these sources were chosen as the evolved disks that showed the largest fluctuations. 

The similarly sized fluctuations suggests that structural changes in disks, similar to those seen in the evolved disks studied here, are common among young stellar objects. Roughly 60\%\ of the IC 348 cluster members with an infrared excess show variability at 3.6 and 4.5\micron\ larger than 0.04 mag (Flaherty et al in prep.). \citet{mor11} find that 70\%\ of the stars in Orion with an infrared excess are variable at 3.6 and 4.5\micron\ with amplitudes similar to those studied here. The amplitude of the flux variations is very similar between the evolved disks studied here and the other variable disks, which make up a majority of the members with disks. Despite the similarities in rms amplitude it is still possible that the variability mechanisms differ. Further observational evidence is needed to probe whether the normal disks are experiencing vertical perturbations of the inner disk, rather than other structural changes.

There is evidence for varying, non-axisymmetric structure in inner disks around other stars in the literature. \citet{hut94} find a wavelength dependence in the variability of UX Ori and AK Sco that can be explained by changing the scale height of the inner disk \citep{fla10}. \citet{juh07} see infrared variability in SV Cep that they are able to model successfully with a disk whose inner scale height is rapidly varying. None of these objects are characterized by transition disk SEDs.

Other systems also show evidence for non-axisymmetric structure in the inner disk, based on the occasional obscuration of the star by this material. \citet{mor11} find a significant fraction of the variables in Orion display rapid dips in their light curves due to brief stellar occultations by the disk. The COROT satellite monitored the young cluster NGC 2264 for 23 straight days and found that 28\%\ of the young stellar objects exhibit this behavior \citep{ale10}. AA Tau is a well-studied example of a star with periodic occultations due a warp in the inner disk \citep{bou07}. The structure that causes these occultations may be long-lived, or it may rapidly appear and disappear on an orbital timescale. In fact UX Ori, which shows strong evidence for an inner disk with a variable scale height based on its infrared fluctuations, is the canonical member of the UX-Ori class of variables that are characterized by brief drops in their optical flux \citep{her99}, similar to those seen in Orion, NGC 2264 and AA Tau. For this class of objects the disturbances in the inner disk are most likely non-axisymmetric in order to produce periodic occultations. The frequency of observations that are consistent with variable non-axisymmetric inner disk structure suggest that this is a fundamental feature of protoplanetary disks.

\section{What could be causing these variations?}\label{cause}
Our multi-wavelength data tracing more than just the dust emission allows us to put some constraints on the physical cause of the infrared variability. We use these data to eliminate accretion and winds, but companions or magnetic fields are still plausible. 

\subsection{Variable Irradiation of the Disk}
One possible explanation for the observed infrared variability is that the illumination of the disk changes rapidly. As the flux striking the disk increases the temperature of the disk rises. If the inner edge is set by dust sublimation then the location of the inner edge will move outward, increasing its emitting area and hence flux. The increased heating may also raise the temperature of the midplane resulting in an increase in the disk scale height, shadowing the outer disk. The illumination of the disk is dominated by the stellar and accretion luminosity, which we can trace using our near-infrared photometry and our measurements of the Pa$\beta$ and Br$\gamma$ lines as well as with measurements from the literature. The observations of the stellar and accretion flux for these stars are not as dense as the infrared coverage, but we can still look for variability in these properties. Our contemporaneous near-infrared spectra, along with near-infrared and mid-infrared photometry will allow us to look for correlations among the stellar flux, accretion flux and infrared flux although it may be challenging to establish a causal relationship. 

In terms of stellar flux, none of the six evolved disks show evidence for strong fluctuations on timescales similar to that of the infrared variability. LRLL 2, 58, 67 and 1679 display, at the most, small changes due to cold spots rotating across the face of the star. These fluctuations are not large enough to substantially change the temperature of the disk. LRLL 21 and 31 show evidence for year-to-year changes in stellar flux, which is much longer than the timescale for the infrared variability. For both stars we have contemporaneous J band and 3.6,4.5\micron\ photometry in 2009 and we find that the J band flux stays relatively constant despite the large changes in the infrared flux. 

The accretion rate does show large variations in most of these stars, with the only exceptions being LRLL 2, which has large uncertainties on its accretion rate, and LRLL 1679, for which we do not detect any strong ongoing accretion ($<10^{-9}M_{\odot}$ yr$^{-1}$). Despite the accretion rate variability, the average accretion luminosity is small compared to the stellar flux and is not an important source of heating for the disk surface. In none of the evolved disks is the accretion luminosity larger than 10\%\ of the total flux striking the disk. In order to change the scale height by a factor of five, the irradiating luminosity would have to change by a factor of 9  \citep{dul10}, which is much larger than the observed variability. This indicates that even though the accretion luminosity changes significantly, the heating of the disk, and hence its scale height, will not vary at an observable level.

\begin{figure*}
\center
\includegraphics[scale=.3]{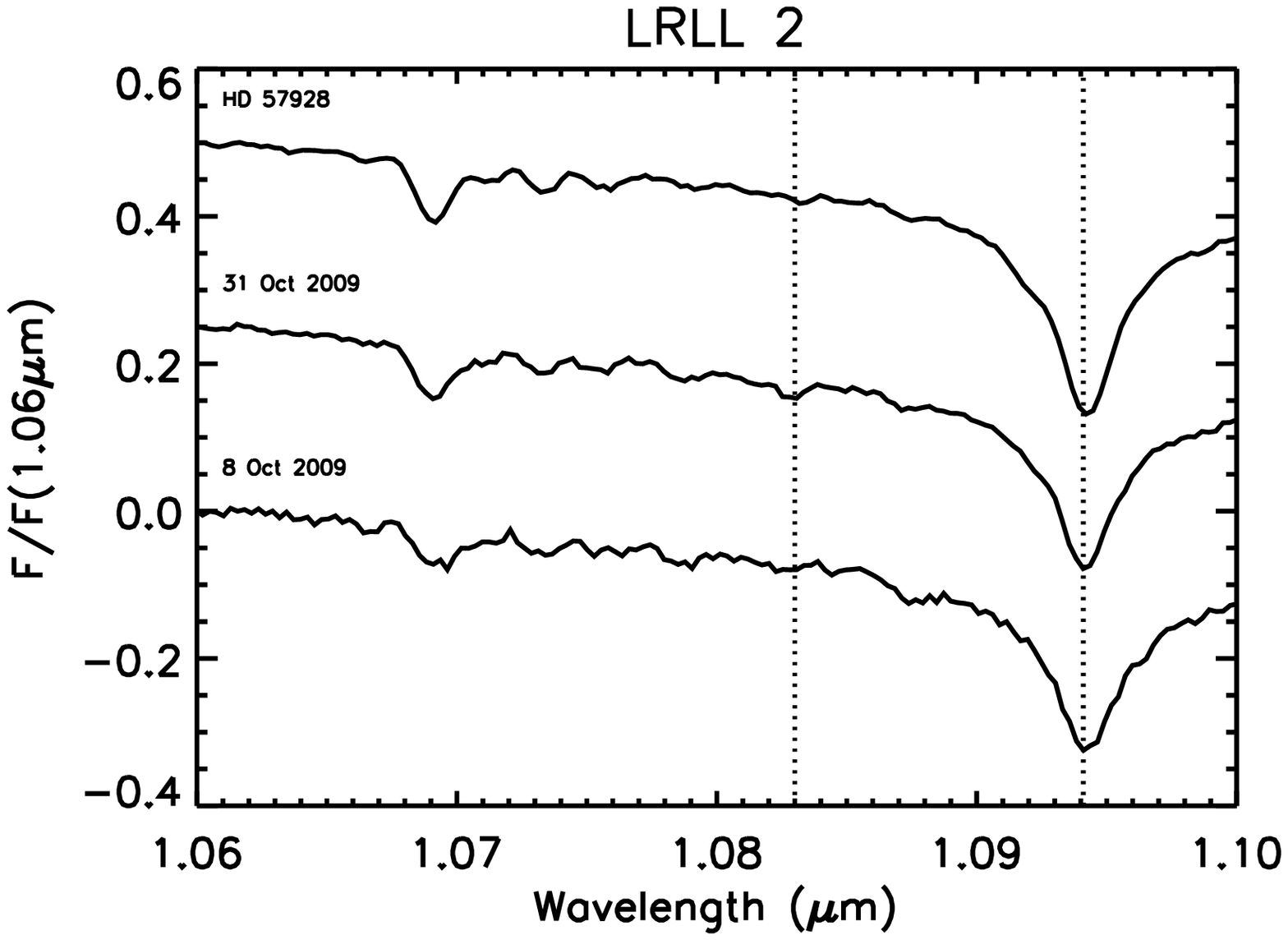}
\includegraphics[scale=.3]{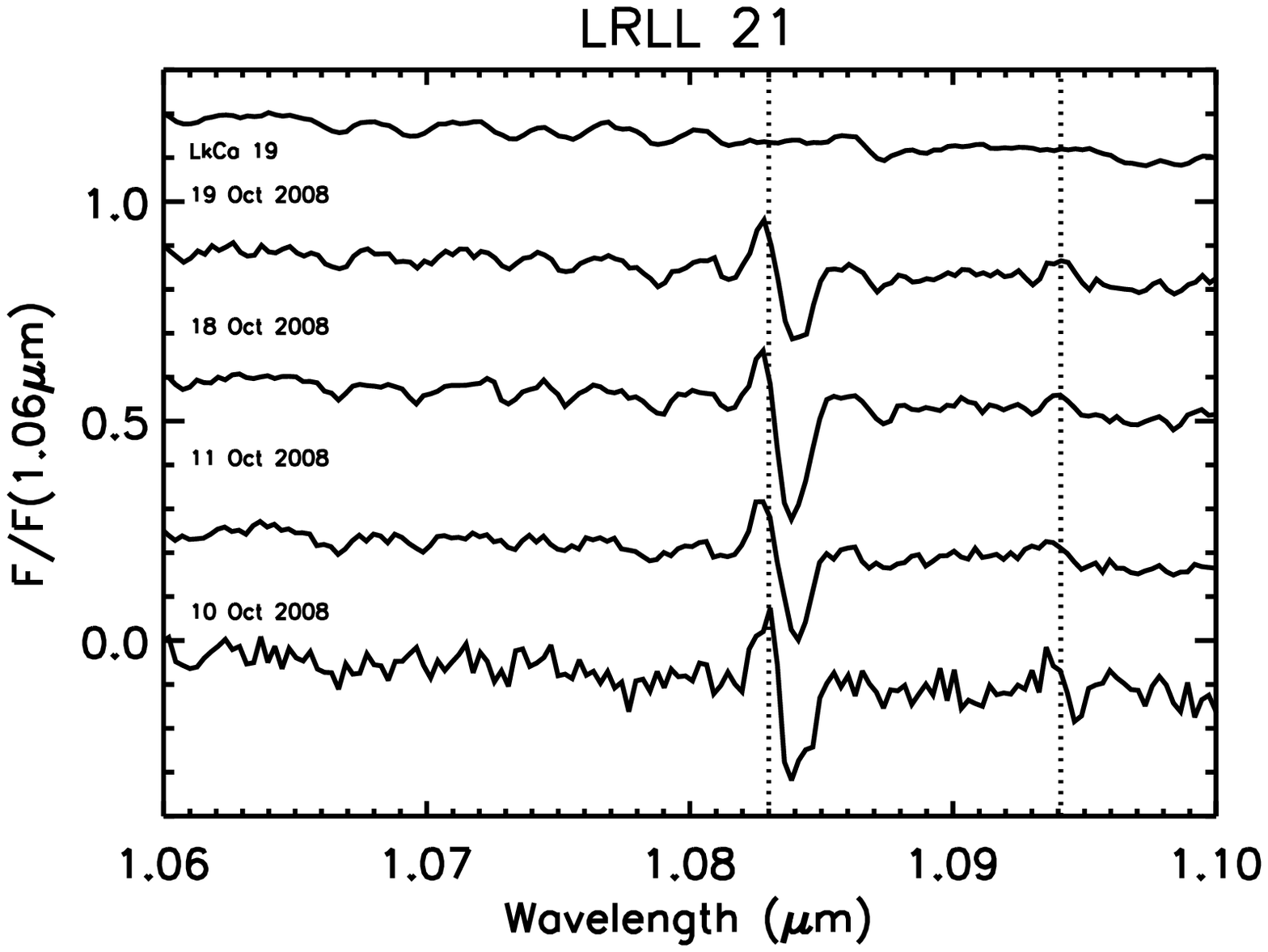}
\includegraphics[scale=.3]{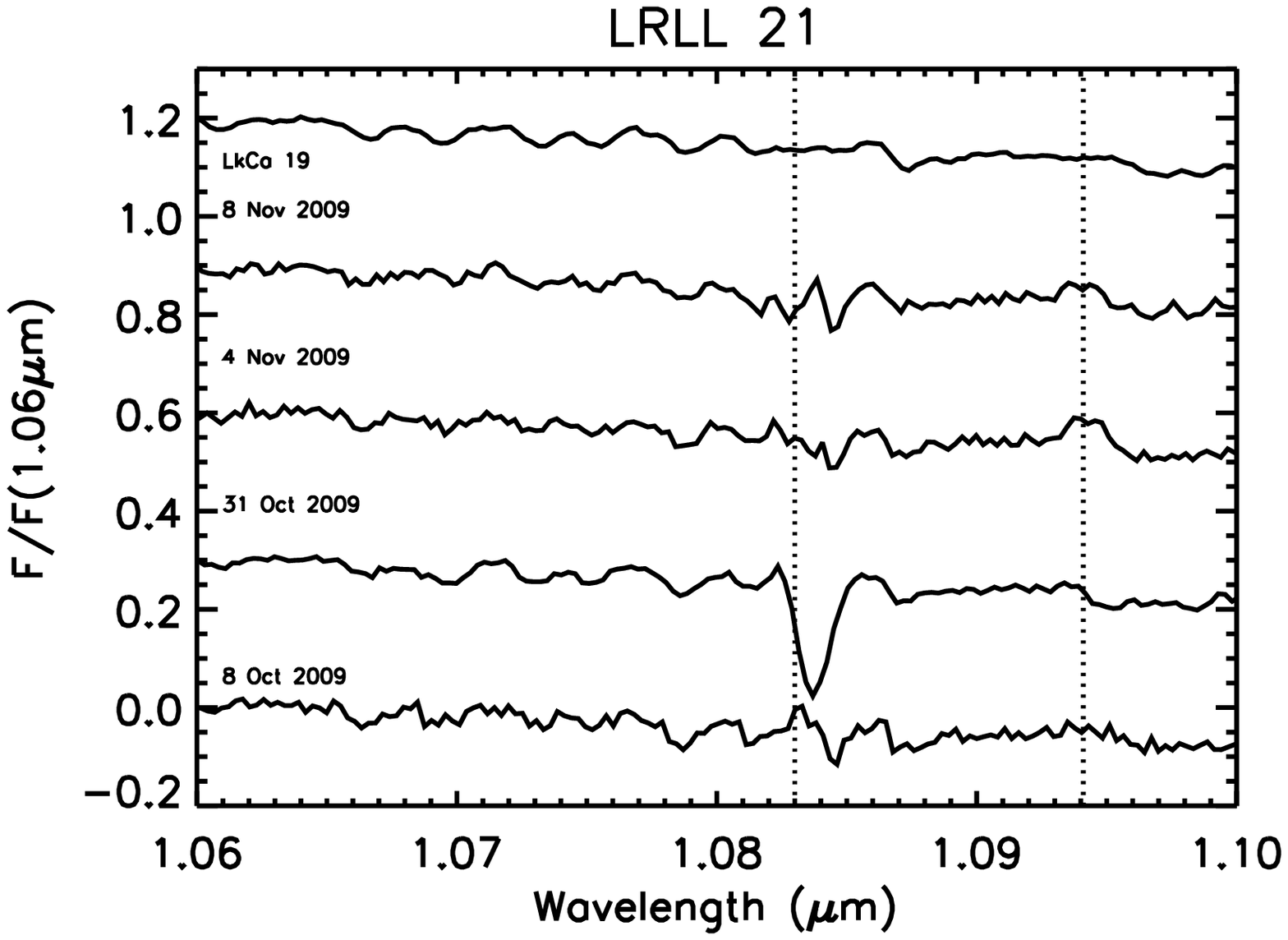}
\includegraphics[scale=.3]{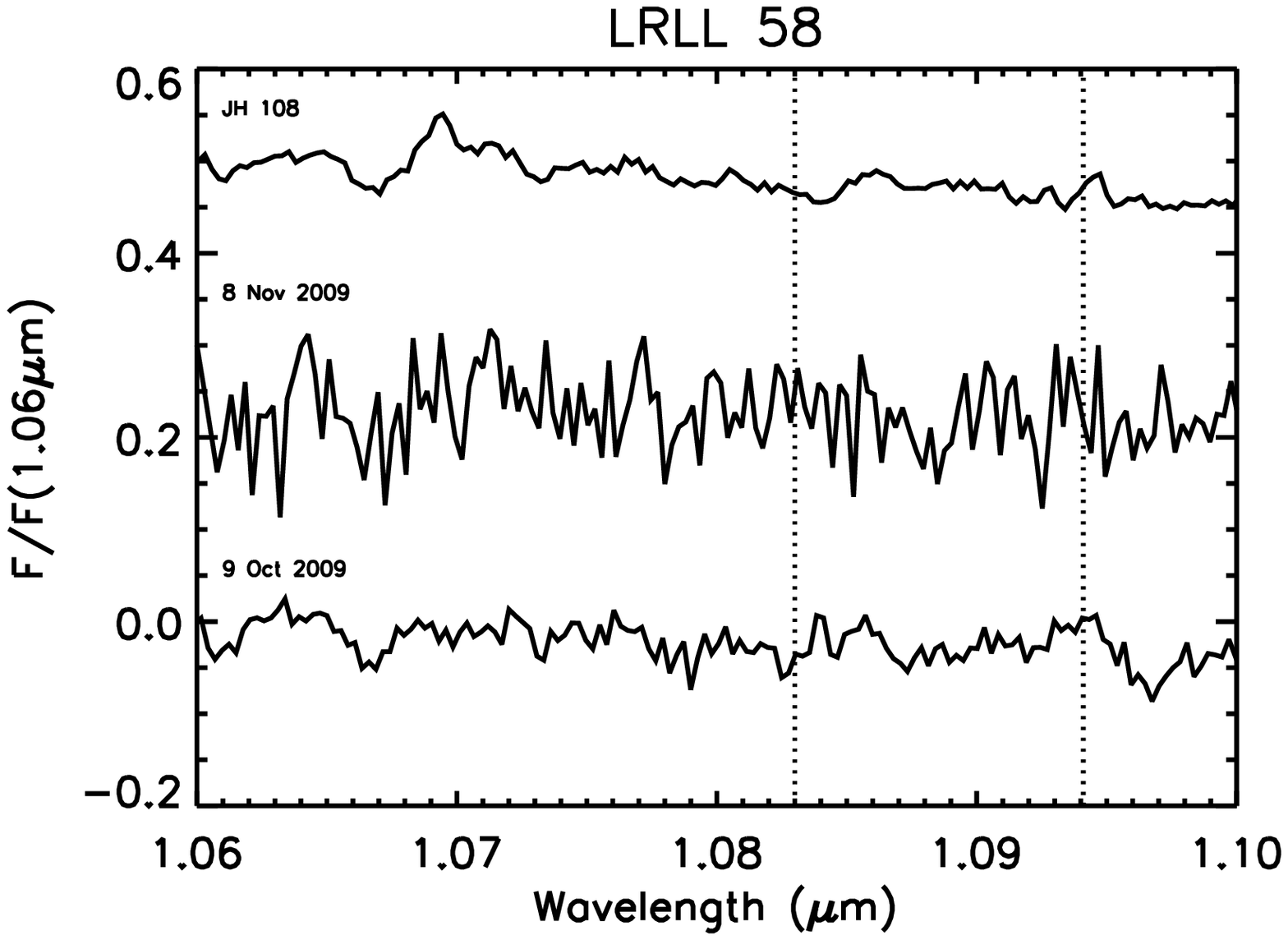}
\includegraphics[scale=.3]{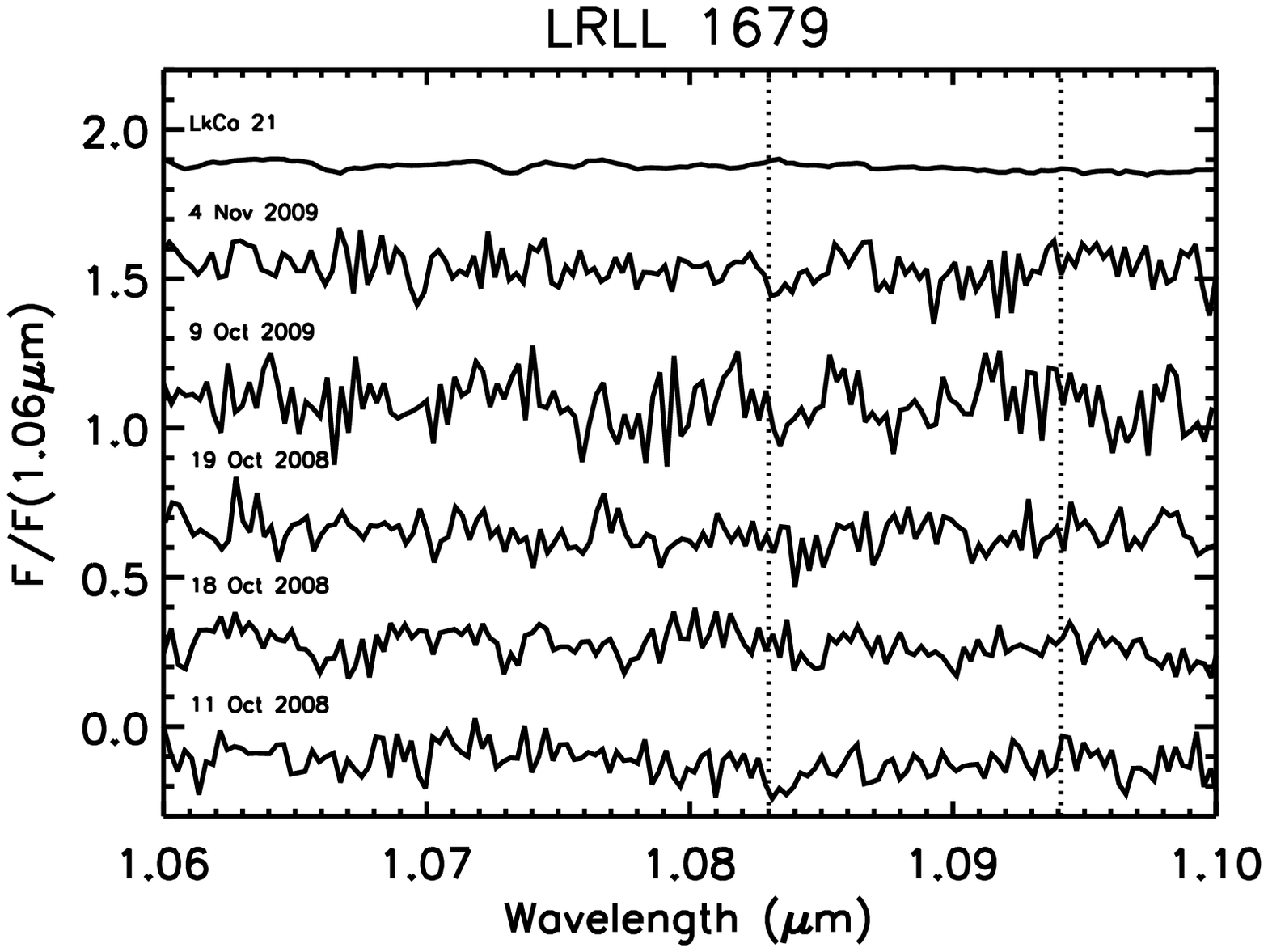}
\caption{HeI ($\lambda=1.083\micron$) and Pa$\gamma$($\lambda=1.0941\micron$) lines in the evolved disks. LRLL 21 is the only source that shows strong absorption or emission in the HeI line, and this line is highly variable.\label{heI}}
\end{figure*}

For LRLL 21 and 31 we have enough contemporaneous data in 2009 to look for any correlation between the accretion rate and infrared flux. Both of these stars show similar infrared light curves, with a sharp drop during the first week of observations followed by a slow increase over the next four weeks. We were able to obtain Pa$\beta$ and Br$\gamma$ measurements during the IR minimum as well as a few weeks later after the infrared flux had increased for both stars. In LRLL 31, the Pa$\beta$ line shows almost no emission during the infrared minimum while it has strong emission after the infrared flux has increased, suggesting a strong correlation between the gas and dust behavior. In LRLL 21, the Pa$\beta$ line shows absorption during the infrared minimum (Oct 8) as well as 3 weeks later (Oct 31) after the infrared flux has increased by 0.7 mag. One week later (Nov 4) the Pa$\beta$ line is in emission even though the infrared flux has only increased by 0.1 mag. It is possible that unresolved absorption on Oct 31 makes the line appear very weak. The HeI line, which traces the inflow and outflow of gas near the star, shows very strong red-shifted absorption due to the accretion in this epoch, but not during any other epoch (Fig~\ref{heI}). There is also evidence for red-shifted absorption in the Pa$\gamma$ line. This suggests that there may be additional absorption that suppresses the Pa$\beta$ emission making it appear very weak. If this is the case then the Pa$\beta$ flux may be correlated with the infrared flux as in LRLL 31. Even if the accretion rate and infrared flux are correlated it is unlikely that the accretion variability is the source of the infrared fluctuations. As discussed earlier, the accretion luminosity is small compared to the stellar flux heating the disk. It is also unlikely that a variable accretion flow through the disk leads to the variable disk emission because the timescale for an over-dense clump of dust and gas to move through the disk is the viscous timescale, which is on the order of hundreds of years at the location of the inner disk edge for the objects.

\subsection{Winds}
A highly variable wind, carrying both gas and dust out of the disk, could lead to variable infrared emission \citep{kon00,vin07}.  As material is lifted out of the disk its emitting area increases and it may shadow the outer disk, creating the observed infrared variability. It is also possible that the wind is asymmetric about the midplane, likely due to a misaligned stellar magnetic field \citep{lov10}. If the mass flow in the wind is larger on one side of the disk than the other, then the asymmetric loss of angular momentum will lead to warping of the inner disk. Both of these models require the wind to be launched outside of the dust sublimation radius, which is more typical of a disk wind \citep{kon00} than an X-wind \citep{shu94}. As discussed below, the co-rotation radius, which is approximately the radius at which an X-wind is launched, lies within the sublimation radius for three of our evolved disks (LRLL 2, 21 and 31) preventing this type of wind from directly affecting the dust. We can trace the wind using the blue-shifted absorption/emission of the HeI (1.08\micron) line \citep{edw06}. When a clump of material is lifted out of the disk midplane by a wind \citep{suz10} the clump needs to be optically thick to stellar photons to shadow the outer disk and produce the observed wavelength dependence of the variability. This sets a lower limit on the column density of material in the clump of $A_V\sim1$, which corresponds to $\log(N_H)\sim21$ \citep{dra03}. A wind with this column density that arose from $\sim$0.5AU and was able to remove material in a week would have an outflow rate of $\dot{M_w}\sim10^{-9}M_{\odot}$ yr$^{-1}$. This is comparable to the accretion rates detected in the evolved disks, and should produce detectable HeI emission \citep{kwa11}. In LRLL 2, 58 and 1679 we see no evidence for absorption or emission in the HeI line (Fig~\ref{heI}). Only LRLL 21 and 31 show blue-shifted emission in some epochs, but not nearly at the strength relative to the continuum expected for such a high column density of material. This suggests that any wind that is present in the system is not dense enough to shadow the outer disk.

An asymmetric wind does not require the density in the wind to be as high, since it is the warped disk that is doing the shadowing of the outer disk, rather than the wind itself. With this model we would expect the strongest blue-shifted absorption/emission, corresponding to the strongest wind, to occur during the epochs when the infrared flux, and hence the warping of the disk, is the largest. For LRLL 21 in 2009, when we have contemporaneous measurements of HeI and infrared flux, there is no sign of a correlation between HeI and the infrared flux (Fig~\ref{heI}). In LRLL 31, in 2009, the strongest blue-shifted absorption occurs during the minimum of the infrared light curve, which is the opposite of the expected correlation \citep{fla11}. This suggests that an asymmetric wind is not the source of the observed infrared variability.

\subsection{Companion}
\begin{figure*}
\center
\includegraphics[scale=.3]{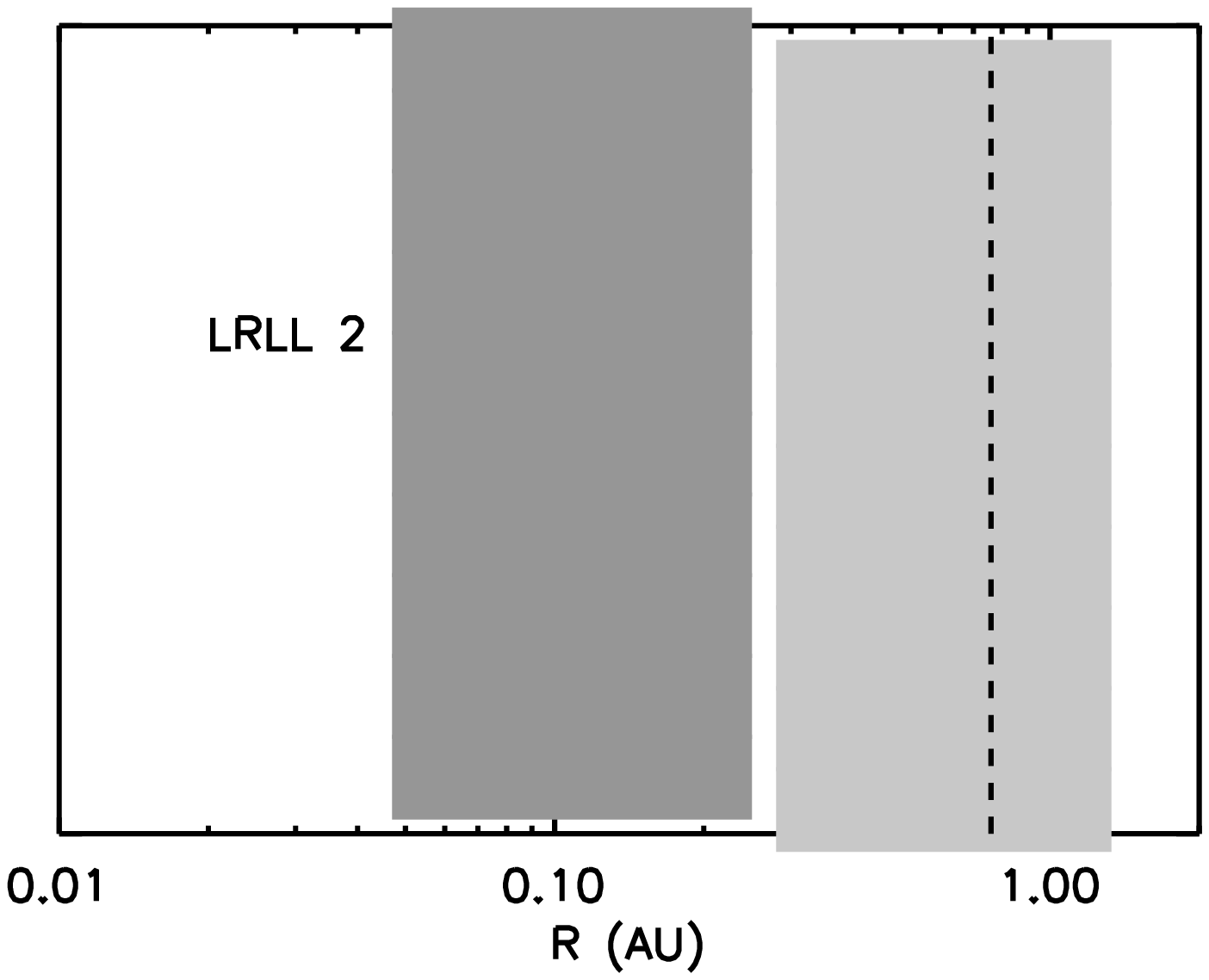}
\includegraphics[scale=.3]{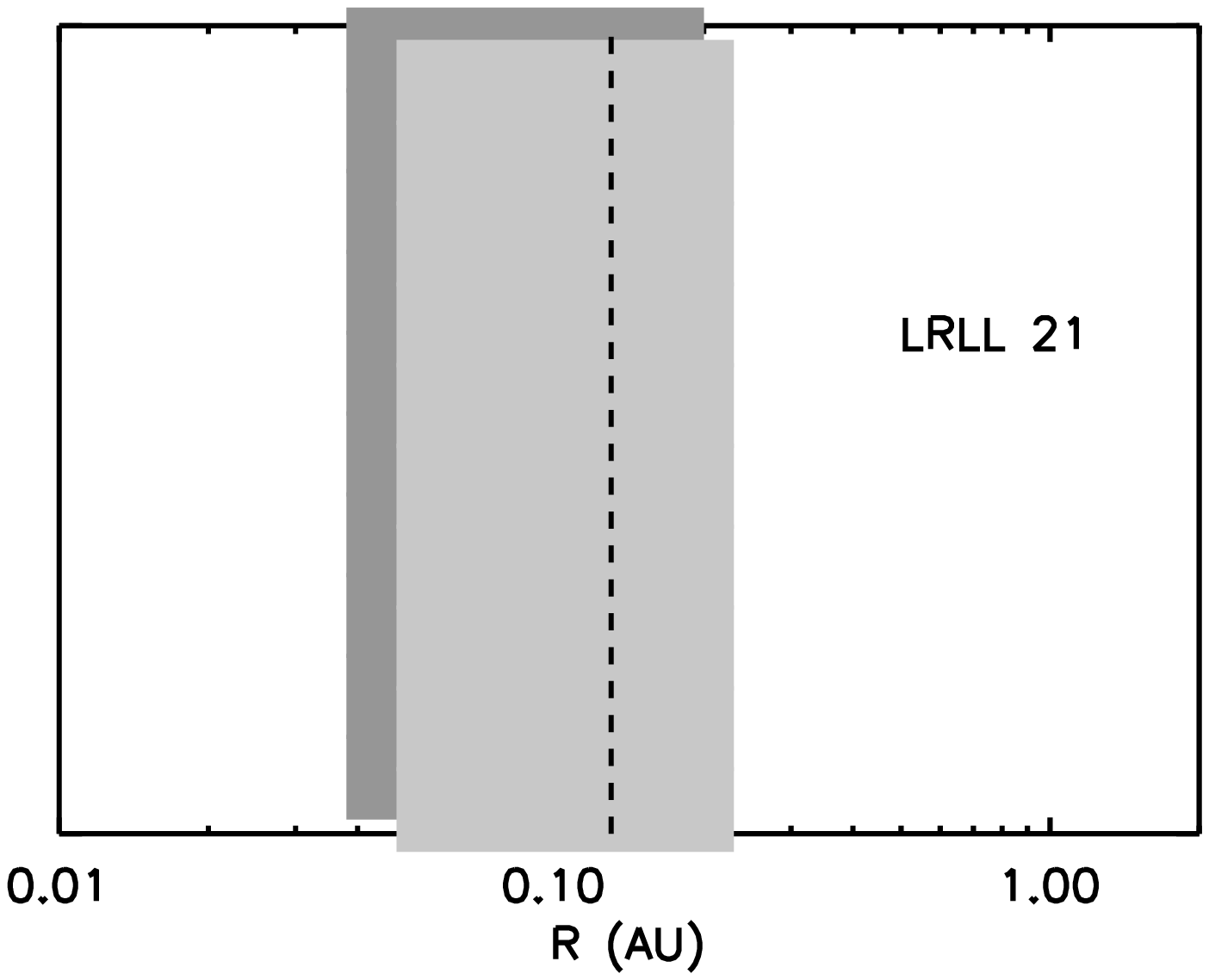}
\includegraphics[scale=.3]{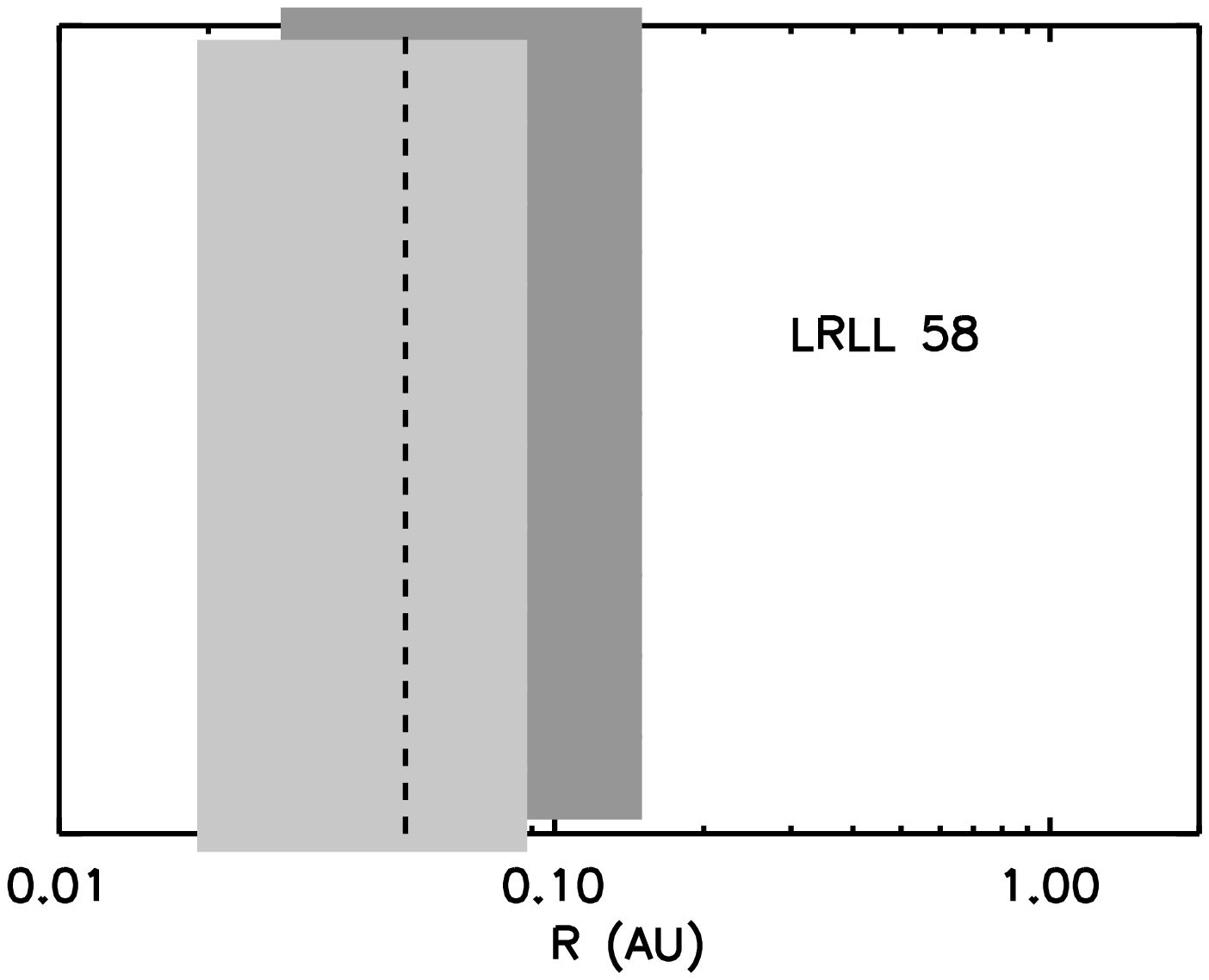}
\includegraphics[scale=.3]{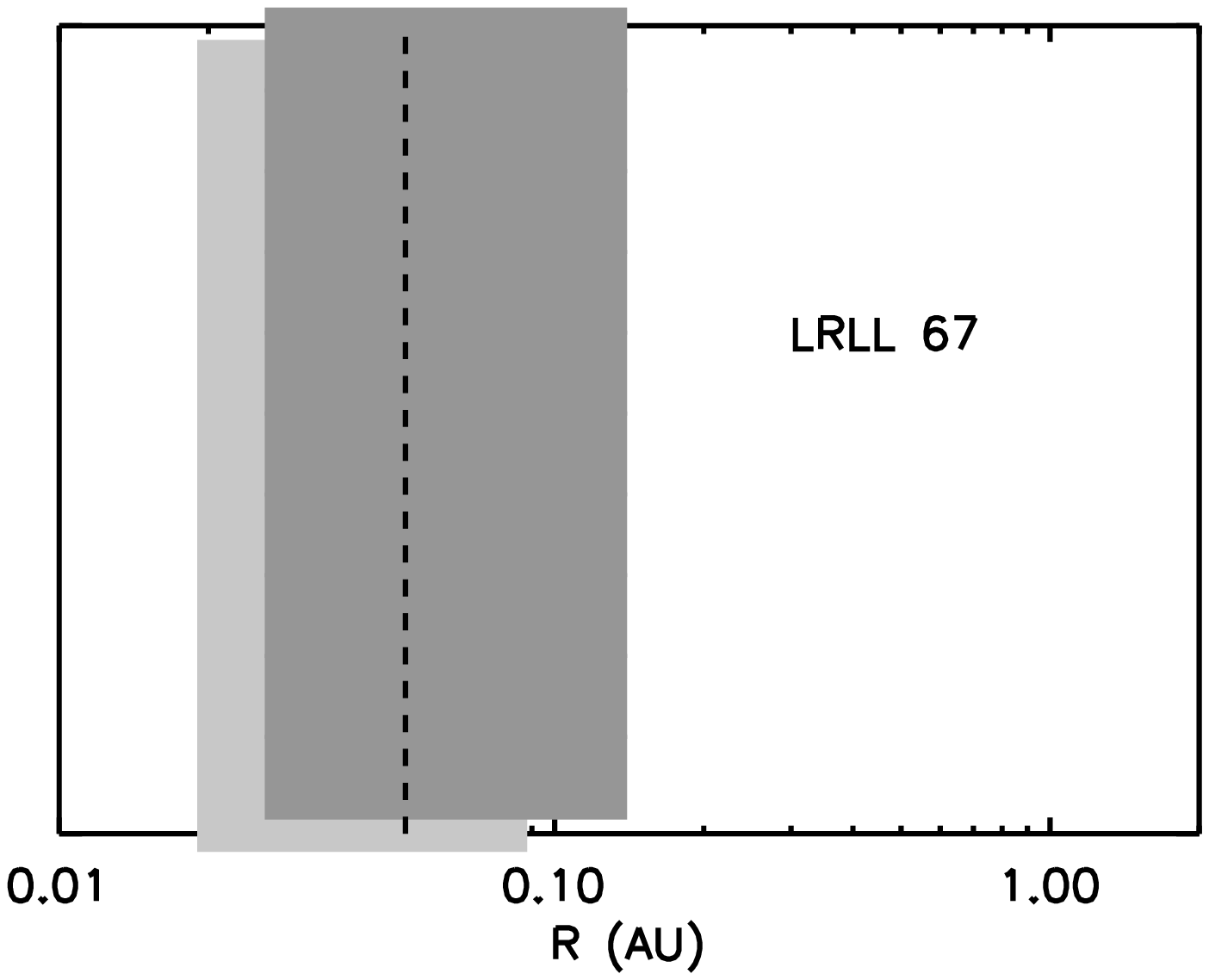}
\includegraphics[scale=.3]{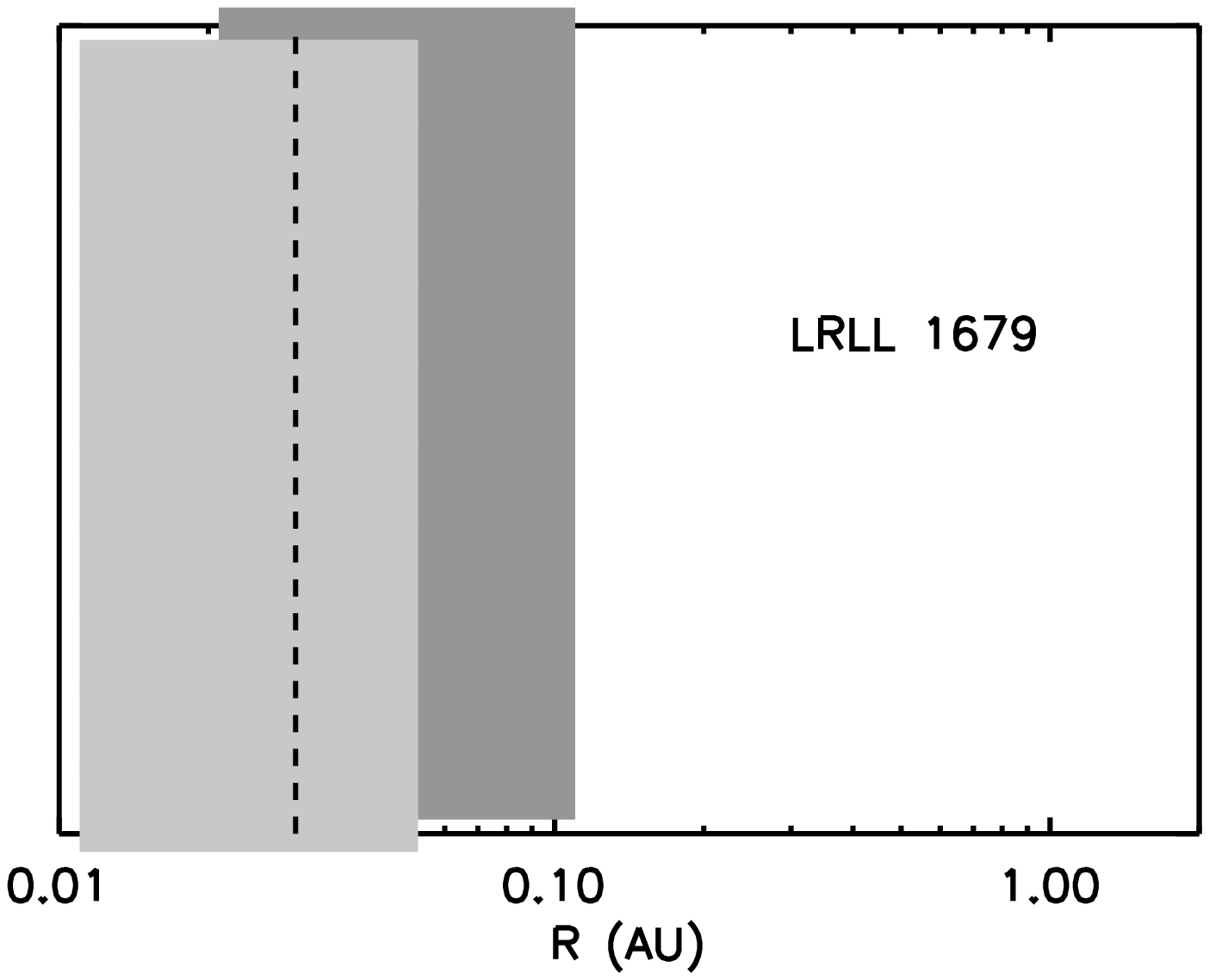}
\caption{Areas near the star that are excluded from having a companion. The dark grey boxes have been excluded based on the lack of periodicity in the 3.6 and 4.5\micron\ monitoring while the light grey boxes have been excluded based on the presence of hot dust in the inner disk (the dust destruction radius is marked with a dashed line).\label{companion}}
\end{figure*}

A companion on an orbit that is misaligned with the disk will lead to a warp whose height varies as the companion drags material out of the midplane \citep{fra09}. If a companion were coplanar with the disk, then it could lead to variable heating of the disk \citep{nag10} or variable mass flow through the disk in the gap surrounding the planet \citep{art96}. The fluctuating mass flow through a gap could lead to variable shock heating of the inner disk which would then change the scale height of the inner disk. The perturbation from such an object would occur on a timescale equal to the companion's orbital period, which could be as little as a few days. There is evidence for radial velocity variations in the optical spectra of LRLL 21 and 31, which are consistent with the presence of a massive companion. We can put some constraints on the possible location of a companion based on two observations: (1) the existence of hot dust, responsible for the 2-5\micron\ excess and (2) the lack of periodicity in the infrared for P=2-25 days. The first constraint prevents a companion from being located within $0.3a<r<1.7a$ where $a$ is the location of the inner dust, or else the inner disk would be removed \citep{art96}. It is possible the the inner disk has a finite radial extent, in which case the companion is excluded from $0.3a_0<r<1.7a_1$ where $a_0$ and $a_1$ are the inner and outer edges of the inner disk. We assume that the inner disk is infinitesimally thin in order to place the most conservative constraints on the location of a companion. We estimate the location of the dust destruction radii assuming T$_{inner}$=1500 K and ISM like grains \citep{dul10}. Here we have assumed that every system has optically thick dust at the dust sublimation radius. For LRLL 21 and 31 we have direct measurements of this excess, while in LRLL 2, 58 and 1679 we have indirect evidence based on the strength of the K-band veiling and the 'seesaw' behavior in the IRS spectra. As discussed earlier, the evidence for optically thick dust close to LRLL 67 is weak, although we include it here for completeness. We use ISM-like grains, but note that a change in the grain sizes and composition can lead to substantially different estimates of the dust sublimation radius \citet{esp12}. The large spectral type range for our sample results in a large range of inner disk radii (Table~\ref{inner_radii}), from 0.76AU for LRLL 2 down to 0.03 AU for LRLL 1679. The second constraint excludes companions from orbits with periods between 2 and 25 days, assuming that a companion would perturb the material on every orbit. In Figure~\ref{companion} we indicate the radii for which a companion is excluded in these evolved disks. For LRLL 2 we can place the tightest constraints, and exclude a companion from 0.05-1.3AU. 

Two important, and related, caveats to the above analysis: (1) dust can extend closer to the companion if it is in streamers rather than a uniform wall and (2) the effects of a companion on the disk may not be periodic. Simulations of the clearing of a disk by a companion find that small streamers of material can extend from the undisturbed outer disk to the companion \citep{art96}. The limits used above for the location of the disk refer to the undisturbed outer disk and streamers can fill in some of the space that has been mostly cleared out by the gravitational influence of the companion. If these streamers are optically thick then they could create the small excess that is observed in some of our sources, due to the very small covering fraction of this material. The models also show that the structure of these streamers may vary stochastically, with no strongly periodic component. This runs contrary to our assumption that a companion would perturb the disk on every orbit. The above analysis is still applicable to a companion on an inclined orbit, but breaks down when dealing with optically thick streamers of material that can arise from a companion being embedded within the disk. More direct tracers are needed to rule out or confirm the perturbation of the disk by a companion.

\subsection{Magnetic Fields}
Magnetic fields play an important role in channeling material from the disk onto the star \citep{bou07b} and possibly for driving the turbulence within the disk that leads to the inward flow of material \citep{bal91}. Many theoretical models predict that the structure of the magnetic field in these young systems should not be constant and these fluctuations may be the source of the vertical perturbations of the inner disk. Magnetic fields in the turbulent disk may occasionally become buoyant, lifting dust and gas off the disk on very short timescales \citep{tur10,hir11}. The stellar magnetic field, and where it intersects the disk, may also be important for variability. As material from the disk flows inward it eventually is loaded onto the stellar magnetic field lines, is lifted out of the midplane, and free-falls onto the star. A highly dynamic interface between the stellar magnetic field and the disk may lead to rapid changes in disk structure. The stellar magnetic field may expand and contract on weekly timescales, leading to a warping of the disk as the field expands \citep{goo99}. If the stellar magnetic field is misaligned with the disk, then this could also cause a warp \citep{lai08}. Simulations also show that the inflow of material along field lines is unstable and very patchy \citep{rom09}. If dust is entrained in this patchy inflow then it could occasionally create a large short-wavelength excess as it is lifted out of the midplane as well as shadow the outer disk.

These last two models (tilted magnetic field and patchy inflow) require dust to extend inwards toward the point at which material is loaded onto the magnetic field lines. This radius is typically taken as near the co-rotation radius, where the periods of the stellar and disk rotation are equal, since outside of this radii the angular momentum of the gas and dust is large enough that it would be flung out of the system rather than accreted onto the star. The innermost radius of the dust is set by the point at which the dust is hot enough to sublimate, which depends on the luminosity of the central star, as well as the dust composition \citep{dul10}. Table~\ref{inner_radii} lists estimates of the co-rotation radii, calculated using either measured rotation periods or the observed range of rotation periods for pre-main sequence stars in IC 348 \citep{cie06}, as well as the dust sublimation radii, calculated earlier. For the low-luminosity sources (LRLL 58, 67, 1679) dust can extend close enough to the star to be loaded onto the stellar magnetic field lines and lifted out of the midplane. For the high-luminosity sources the dust disk is truncated at much larger radii, even if we assume that the disk is entirely composed of very large ($>10\micron$) grains, which can extend closer to the star than small (0.1\micron) grains. We include estimates of dust sublimation radius using very large grains in order to put a lower limit on its location, despite the fact that the large silicate emission features in many of our IRS spectra are inconsistent with an inner disk dominated by large grains. 

Lifting dust out of the midplane would occasionally lead to higher extinction and decreased stellar flux coincident with changes in the infrared flux as our line of sight passes through one of these dense clumps. This would be more prevalent the closer the dust approached the star. In none of our stars do we see evidence for an increase in extinction or a decrease in stellar flux associated with the infrared flux variations. It is unlikely that for all of these six systems we have been lucky enough that our line of sight never intersects the dusty clump, although the paucity of measurements of the extinction of stellar flux makes it difficult for us to say that the stars are never occulted. For LRLL 21 and 31, where we have the densest coverage of the extinction contemporaneous with the 2009 infrared monitoring, we do not see any correlation between these properties. This suggests that the coverage of the sky by the clump as seen from the star is small.  Additional monitoring for extinction events associated with material being lifted out of the midplane would help to further constrain this theory.

\subsection{X-ray Flares}
Young stellar objects are known to exhibit stronger average X-ray emission as well as stronger X-ray flares than a main sequence star of similar mass \citep{fei07}. X-ray flares last for up to 2 days and can substantially heat the disk \citep{wol05}. Recently \citet{ke12} showed that flares similar in strength to the most powerful outbursts seen in Orion may even be responsible for the change in inner disk scale height. A large X-ray flare will dramatically increase the temperature and ionization of the dust and gas. The highly ionized dust will then be accelerated either inward or outward by the stellar magnetic field and this acceleration will lead to a change in the scale height of the disk. \citet{ke12} show that given a typical X-ray flare, amid assumptions about the disk structure, they can reproduce the 'seesaw' behavior on timescales of days to weeks that we observe here. Since the response of the disk to the X-ray flare is nearly instantaneous and lasts for days, contemporaneous monitoring of the X-ray and infrared flux should be able to constrain this theory. And because the strength of the stellar magnetic field drops strongly with radius, those sources with dust close to the co-rotation radius should exhibit the largest fluctuations, assuming that different sources have similar X-ray flares. In our small sample we have objects that range from having dust very close to the co-rotation radius to objects with dust very far from the co-rotation radius and do not find any clear trend in the strength of the variability between these sources.

\section{Conclusion}
We have studied the mid-infrared variability of six evolved disk systems in the young stellar cluster IC 348.
\begin{itemize}
\item We find strong infrared variability at all wavelengths in every source. These fluctuations are typically rapid (days to weeks) and often show a wavelength dependence where the short-wavelength flux increases as the long-wavelength flux decreases.
\item The most likely source of this infrared variability is a vertical perturbation of the inner disk, based on the wavelength dependence and rapid timescales of the variability, as well as direct measurements of the emission from the inner disk. 
\item The strength and timescale of the infrared variability in these evolved disks is not unlike that seen in the rest of the IC 348 cluster members with disks, of which roughly 60\%\ are measurably variable. This result suggests that these types of perturbations are common.
\item It is unlikely that variable heating from fluctuations in the stellar flux or accretion rate are responsible for the infrared variability. We do not observe strong stellar flux changes on the same timescale as the infrared variability, and the accretion luminosity is not a significant component of the flux striking the disk.
\item We can also exclude the influence of strong winds and companions at certain orbital radii within the disk. 
\item Possible explanations for the behavior are either companions at permitted orbital radii, effects from the system magnetic field, or the influence of very large X-ray flares. However these models predictions that are not yet confirmed - in the first case of periodicities, in the second of the variable extinction of the star and in the third of correlated X-ray and infrared flux.
\end{itemize}

The size and timescale of the infrared flux variations seen in the six evolved disks studied here are very similar to the other variable disks in IC 348, which make up roughly half of the cluster members with disks. This suggests that many young stellar objects undergo similar perturbations to the structure of their inner disks. Further observations are needed to determine if the physical causes of the variability are similar between these two samples, but the similarity in the size of the fluctuations suggest that variability could be a useful tool for studying the inner disks around many young stellar objects.

\acknowledgements
We would like to thank Brandon Kelly for the helpful discussions about time-variable statistics. Our results are partly based on observations obtained at the Centro Astrono\'omico Hispano Alem\'an (CAHA) at Calar Alto, operated jointly by the Max-Planck Institut f\"ur Astronomie and the Instituto de Astrof\'{\i}sica de Andaluc\'{\i}a (CSIC). The observations were supported by OPTICON. OPTICON has received research funding from European Community's Sixth Framework Programme under contract number RII3-CT-001566. This work is based in part on observations made with the Spitzer Space Telescope, which is operated by the Jet Propulsion Laboratory, California Institute of Technology under a contract with NASA. Our results are partly based on observations obtained at the Infrared Telescope Facility, which is operated by the University of Hawaii under Cooperative Agreement no. NNX-08AE38A with the National Aeronautics and Space Administration, Science Mission Directorate, Planetary Astronomy Program. This work was also supported by contract 1255094 from Caltech/JPL to the University of Arizona.

\begin{deluxetable}{ccccc}
%\tablewidth{0pt}
\tabletypesize{\scriptsize}
\tablecaption{Observing Log\label{obs_log}}
\tablehead{\colhead{Date}&\colhead{MJD}&\colhead{Wavelength}&\colhead{Resolution}&\colhead{Instrument}}
\startdata
\cutinhead{LRLL 2}
08 Oct 2007 & 54381.9 & 5-40\micron & $\sim$600 & IRS spectra\\
15 Oct 2007 & 54388.9 & 5-40\micron & $\sim$600 & IRS spectra\\
24 Mar 2008 & 54549.2 & 5-40\micron & $\sim$600 & IRS spectra\\
 31 Mar 2008 & 54556.7 & 5-40\micron & $\sim$600 & IRS spectra\\
08 Oct 2009\tablenotemark{a} & 55112.6 & 0.8-2.5\micron & $\sim$800 & Spex\\
31 Oct 2009\tablenotemark{a} & 55137.3 & 0.8-2.5\micron & $\sim$800 & Spex\\
\cutinhead{LRLL 21}
09 Oct 2007 & 54382.3 & 5-40\micron & $\sim$600 & IRS spectra\\
15 Oct 2007 & 54388.8 & 5-40\micron & $\sim$600 & IRS spectra\\
24 Feb 2008 & 54520.3 & 5-40\micron & $\sim$600 & IRS spectra\\
29 Feb 2008 & 54525.1 & 6460-6650\AA & $\sim$ 35000 & Hectochelle\\
01 Mar 2008 & 54526.9 & 5-40\micron & $\sim$600 & IRS spectra\\
10 Oct 2008 & 54750.5 & 0.8-5\micron & $\sim$1500 & Spex\\
11 Oct 2008 & 54751.5 & 0.8-2.5\micron & $\sim$ 800 & Spex\\
18 Oct 2008 & 54757.4 & 0.8-5\micron & $\sim$1500 & Spex\\
19 Oct 2008 & 54758.4 & 0.8-5\micron & $\sim$1500 & Spex\\
08 Oct 2009\tablenotemark{a} & 55112.6 & 0.8-5\micron & $\sim$ 800 & Spex\\
12 Oct 2009\tablenotemark{a} & 55116.1 & 6000-9000\AA & $\sim$ 3500 & CAFOS\\
31 Oct 2009\tablenotemark{a} & 55137.3 & 0.8-5\micron & $\sim$ 800 & Spex\\
04 Nov 2009\tablenotemark{a} & 55141.5 & 0.8-5\micron & $\sim$ 800 & Spex\\
08 Nov 2009 & 55144.5 & 0.8-2.5\micron & $\sim$ 800 & Spex\\
\cutinhead{LRLL 58}
08 Oct 2007 & 54381.9 & 5-40\micron & $\sim$600 & IRS spectra\\
15 Oct 2007 & 54388.9 & 5-40\micron & $\sim$600 & IRS spectra\\
29 Feb 2008 & 54525.1 & 6460-6650\AA & $\sim$ 35000 & Hectochelle\\
24 Mar 2008 & 54549.2 & 5-40\micron & $\sim$600 & IRS spectra\\
31 Mar 2008 & 54556.8 & 5-40\micron & $\sim$600 & IRS spectra\\
09 Oct 2009\tablenotemark{a} & 55113.6 & 0.8-2.5\micron & $\sim$800 & Spex\\
13 Oct 2009\tablenotemark{a} & 55117.1 & 6000-9000\AA & $\sim$3500 & CAFOS\\
08 Nov 2009 & 55144.5 & 0.8-2.5\micron & $\sim$800 & Spex\\
\cutinhead{LRLL 67}
09 Oct 2007 & 54382.4 & 5-40\micron & $\sim$600 & IRS spectra\\
15 Oct 2007 & 54388.9 & 5-40\micron & $\sim$600 & IRS spectra\\
24 Feb 2008 & 54520.3 & 5-40\micron & $\sim$600 & IRS spectra\\
29 Feb 2008 & 54525.1 & 6460-6650\AA & $\sim$ 35000 & Hectochelle\\
03 Mar 2008 & 54527.0 & 5-40\micron & $\sim$600 & IRS spectra\\
\cutinhead{LRLL 1679}
08 Oct 2007 & 54381.9 & 5-40\micron & $\sim$600 & IRS spectra\\
15 Oct 2007 & 54388.9 & 5-40\micron & $\sim$600 & IRS spectra\\
29 Feb 2008 & 54525.1 & 6460-6650\AA & $\sim$ 35000 & Hectochelle\\
24 Mar 2008 & 54549.1 & 5-40\micron & $\sim$600 & IRS spectra\\
31 Mar 2008 & 54556.8 & 5-40\micron & $\sim$600 & IRS spectra\\
11 Oct 2008 & 54751.5 & 0.8-2.5\micron & $\sim$ 800 & Spex\\
18 Oct 2008 & 54757.4 & 0.8-2.5\micron & $\sim$1500 & Spex\\
19 Oct 2008 & 54758.4 & 0.8-2.5\micron & $\sim$1500 & Spex\\
09 Oct 2009\tablenotemark{a} & 55113.6 & 0.8-2.5\micron & $\sim$800 & Spex\\
13 Oct 2009\tablenotemark{a} & 55117.1 & 6000-9000\AA & $\sim$3500 & CAFOS\\
04 Nov 2009\tablenotemark{a} & 55141.5 & 0.8-2.5\micron & $\sim$ 800 & Spex\\
\cutinhead{All Stars}
21 Feb 2004 & 53056.6 & 24\micron & photometry & MIPS\\
19 Sep 2004 & 53267.9 & 24\micron & photometry & MIPS\\
23 Sep 2007 & 54366.9 & 24\micron & photometry & MIPS\\
24 Sep 2007 & 54367.9 & 24\micron & photometry & MIPS\\
25 Sep 2007 & 54368.9 & 24\micron & photometry & MIPS\\
26 Sep 2007 & 54369.9 & 24\micron & photometry & MIPS\\
27 Sep 2007 & 54370.8 & 24\micron & photometry & MIPS\\
12 Mar 2008 & 54538.7 & 24\micron & photometry & MIPS\\
19 Mar 2008 & 54544.8 & 24\micron & photometry & MIPS\\
11 Feb 2004 & 53046 & 3-8\micron & photometry & IRAC cold-mission\\
08 Sep 2004 & 53257 & 3-8\micron & photometry & IRAC cold-mission\\
19 Mar 2009 & 54910 & 3-8\micron & photometry & IRAC cold-mission\\
20 Mar 2009 & 54911 & 3-8\micron & photometry & IRAC cold-mission\\
21 Mar 2009 & 54912 & 3-8\micron & photometry & IRAC cold-mission\\
22 Mar 2009 & 54913 & 3-8\micron & photometry & IRAC cold-mission\\
23 Mar 2009 & 54914 & 3-8\micron & photometry & IRAC cold-mission\\
03 Oct-07 Nov 2009 & 55107-55142 & 3.6,4.5\micron & photometry & IRAC warm-mission\\
\enddata
\tablenotetext{a}{Data were taken during the IRAC warm-mission monitoring campaign in fall 2009.}
\end{deluxetable}

\begin{deluxetable}{cccccc}
%\tablewidth{0pt}
\tablecaption{MIPS photometry\label{mips_phot}}
\tablehead{\colhead{MJD}&\colhead{LRLL 2}&\colhead{LRLL 21}&\colhead{LRLL 58}&\colhead{LRLL 67}&\colhead{LRLL 1679}}
\startdata
53055.56 & 3.288 $\pm$ 0.015 & 3.651 $\pm$ 0.021 & 4.686 $\pm$ 0.015 & 4.658 $\pm$ 0.011 & 6.561 $\pm$ 0.009\\
53266.87 & 3.611 $\pm$ 0.019 & 3.866 $\pm$ 0.014 & 4.602 $\pm$ 0.015 & 4.654 $\pm$ 0.014 & 7.025 $\pm$ 0.013\\
53267.09 & 3.587 $\pm$ 0.017 & 3.872 $\pm$ 0.013 & 4.557 $\pm$ 0.010 & 4.644 $\pm$ 0.017 & 6.945 $\pm$ 0.014\\
54365.88 & 3.413 $\pm$ 0.013 & 3.717 $\pm$ 0.011 & 4.531 $\pm$ 0.011 & 4.649 $\pm$ 0.012 & 6.972 $\pm$ 0.012\\
54366.86 & 3.407 $\pm$ 0.020 & 3.753 $\pm$ 0.011 & 4.528 $\pm$ 0.014 & 4.688 $\pm$ 0.012 & 6.925 $\pm$ 0.014\\
54367.91 & 3.409 $\pm$ 0.019 & 3.741 $\pm$ 0.012 & 4.520 $\pm$ 0.012 & 4.668 $\pm$ 0.013 & 6.893 $\pm$ 0.012\\
54368.86 & 3.402 $\pm$ 0.013 & 3.719 $\pm$ 0.010 & 4.526 $\pm$ 0.013 & 4.667 $\pm$ 0.011 & 6.953 $\pm$ 0.011\\
54369.80 & 3.386 $\pm$ 0.016 & 3.778 $\pm$ 0.007 & 4.546 $\pm$ 0.017 & 4.649 $\pm$ 0.016 & 7.128 $\pm$ 0.013\\
54537.66 & 3.367 $\pm$ 0.013 & 3.631 $\pm$ 0.020 & 4.581 $\pm$ 0.016 & 4.656 $\pm$ 0.010 & 6.563 $\pm$ 0.011\\
54543.78 & 3.337 $\pm$ 0.015 & 3.730 $\pm$ 0.016 & 4.546 $\pm$ 0.014 & 4.650 $\pm$ 0.012 & 6.631 $\pm$ 0.011\\
\enddata
\end{deluxetable}

\begin{deluxetable}{ccccc}
%\tablewidth{0pt}
\tabletypesize{\scriptsize}
\tablecaption{IRAC Photometry\label{irac_cm_phot}}
\tablehead{\colhead{Date}&\colhead{[3.6]}&\colhead{[4.5]}&\colhead{[5.8]}&\colhead{[8.0]}}
\startdata
\cutinhead{LRLL 2}
11 Feb 2004 & 6.971 $\pm$ 0.002 & 6.797 $\pm$ 0.002 & 6.555 $\pm$ 0.002 & 5.761 $\pm$ 0.004\\
08 Sep 2004 & - & - & - & -\\
19 Mar 2009 & 7.015 $\pm$ 0.002 & 6.905 $\pm$ 0.003 & 6.455 $\pm$ 0.002 & 5.724 $\pm$ 0.005 \\
20 Mar 2009 & 6.982 $\pm$ 0.002 & 6.791 $\pm$ 0.002 & 6.454 $\pm$ 0.002 & 5.721 $\pm$ 0.005 \\
21 Mar 2009 & 6.995 $\pm$ 0.002 & 6.748 $\pm$ 0.002 & 6.479 $\pm$ 0.002 & 5.687 $\pm$ 0.004 \\
22 Mar 2009 & 6.948 $\pm$ 0.002 & 6.796 $\pm$ 0.002 & 6.468 $\pm$ 0.002 & 5.713 $\pm$ 0.003 \\
23 Mar 2009 & 6.984 $\pm$ 0.002 & 6.800 $\pm$ 0.002 & 6.500 $\pm$ 0.002 & 5.742 $\pm$ 0.004 \\
\cutinhead{LRLL 21}
11 Feb 2004 & 9.086 $\pm$ 0.003 & 8.912 $\pm$ 0.003 & 8.619 $\pm$ 0.006 & 8.130 $\pm$ 0.02 \\
08 Sep 2004 & 9.069 $\pm$ 0.004 & 8.657 $\pm$ 0.003 & 8.413 $\pm$ 0.006 & 8.014 $\pm$ 0.02 \\
19 Mar 2009 & 9.221 $\pm$ 0.003 & 9.025 $\pm$ 0.003 & 8.862 $\pm$ 0.007 & 8.306 $\pm$ 0.03 \\
20 Mar 2009 & 9.212 $\pm$ 0.003 & 9.033 $\pm$ 0.003 & 8.806 $\pm$ 0.006 & 8.313 $\pm$ 0.03 \\
21 Mar 2009 & 9.119 $\pm$ 0.003 & 8.961 $\pm$ 0.003 & 8.787 $\pm$ 0.006 & 8.270 $\pm$ 0.03 \\
22 Mar 2009 & 9.063 $\pm$ 0.003 & 8.885 $\pm$ 0.003 & 8.669 $\pm$ 0.005 & 8.193 $\pm$ 0.02 \\
23 Mar 2009 & 8.934 $\pm$ 0.003 & 8.746 $\pm$ 0.003 & 8.511 $\pm$ 0.005 & 8.168 $\pm$ 0.02 \\
\cutinhead{LRLL 58}
11 Feb 2004 & 9.789 $\pm$ 0.004 & 9.588 $\pm$ 0.004 & 9.330 $\pm$ 0.015 & 8.543 $\pm$ 0.05 \\
08 Sep 2004 & 10.104 $\pm$ 0.004 & 9.959 $\pm$ 0.004 & 9.637 $\pm$ 0.017 & 8.557 $\pm$ 0.05 \\
19 Mar 2009 & 9.605 $\pm$ 0.004 & 9.372 $\pm$ 0.003 & 9.059 $\pm$ 0.014 & 8.513 $\pm$ 0.04 \\
20 Mar 2009 & 9.619 $\pm$ 0.003 & 9.347 $\pm$ 0.003 & 9.043 $\pm$ 0.012 & 8.493 $\pm$ 0.04 \\
21 Mar 2009 & 9.703 $\pm$ 0.004 & 9.305 $\pm$ 0.003 & 9.011 $\pm$ 0.014 & 8.506 $\pm$ 0.05 \\
22 Mar 2009 & 9.556 $\pm$ 0.004 & 9.254 $\pm$ 0.004 & 8.973 $\pm$ 0.012 & 8.484 $\pm$ 0.05 \\
23 Mar 2009 & 9.515 $\pm$ 0.003 & 9.245 $\pm$ 0.004 & 8.939 $\pm$ 0.012 & 8.494 $\pm$ 0.05 \\
\cutinhead{LRLL 67}
11 Feb 2004 & 10.425 $\pm$ 0.003 & 10.255 $\pm$ 0.003 & 10.063 $\pm$ 0.007 & 9.575 $\pm$ 0.010 \\
08 Sep 2004 & 10.425 $\pm$ 0.003 & 10.238 $\pm$ 0.004 & 9.986 $\pm$ 0.007 & 9.509 $\pm$ 0.011 \\
19 Mar 2009 & 10.422 $\pm$ 0.003 & 10.343 $\pm$ 0.003 & 10.225 $\pm$ 0.008 & 9.675 $\pm$ 0.010 \\
20 Mar 2009 & 10.325 $\pm$ 0.002 & 10.284 $\pm$ 0.003 & 10.129 $\pm$ 0.008 & 9.633 $\pm$ 0.010 \\
21 Mar 2009 & 10.524 $\pm$ 0.003 & 10.325 $\pm$ 0.003 & 10.149 $\pm$ 0.008 & 9.644 $\pm$ 0.010 \\
22 Mar 2009 & 10.386 $\pm$ 0.003 & 10.439 $\pm$ 0.004 & 10.111 $\pm$ 0.008 & 9.605 $\pm$ 0.009 \\
23 Mar 2009 & 10.426 $\pm$ 0.003 & 10.388 $\pm$ 0.004 & 10.163 $\pm$ 0.008 & 9.658 $\pm$ 0.012 \\
\cutinhead{LRLL 1679}
11 Feb 2004 & 11.317 $\pm$ 0.004 & 11.060 $\pm$ 0.004 & 10.781 $\pm$ 0.011 & 10.287 $\pm$ 0.010 \\
08 Sep 2004 & 10.806 $\pm$ 0.003 & 10.446 $\pm$ 0.003 & 10.173 $\pm$ 0.008 & 9.809 $\pm$ 0.007 \\
19 Mar 2009 & 11.468 $\pm$ 0.004 & 11.295 $\pm$ 0.005 & 11.063 $\pm$ 0.012 & 10.769 $\pm$ 0.014 \\
20 Mar 2009 & 11.436 $\pm$ 0.004 & 11.267 $\pm$ 0.005 & 11.099 $\pm$ 0.013 & 10.781 $\pm$ 0.014 \\
21 Mar 2009 & 11.459 $\pm$ 0.004 & 11.305 $\pm$ 0.005 & 11.047 $\pm$ 0.012 & 10.744 $\pm$ 0.015 \\
22 Mar 2009 & 11.449 $\pm$ 0.004 & 11.298 $\pm$ 0.005 & 11.074 $\pm$ 0.013 & 10.728 $\pm$ 0.016 \\
23 Mar 2009 & 11.444 $\pm$ 0.004 & 11.303 $\pm$ 0.005 & 11.093 $\pm$ 0.013 & 10.769 $\pm$ 0.015 \\
\enddata
\end{deluxetable}

\begin{deluxetable}{cccccccccccc}
%\tablewidth{0pt}
\tablecolumns{11}
\tabletypesize{\scriptsize}
\tablecaption{IRAC Warm Mission Photometry\label{irac_wm_phot}}
\tablehead{\colhead{ }&\multicolumn{2}{c}{LRLL 2}&\multicolumn{2}{c}{LRLL 21}&\multicolumn{2}{c}{LRLL 58}&\multicolumn{2}{c}{LRLL 67}&\multicolumn{2}{c}{LRLL 1679}\\ \colhead{MJD-55000}&\colhead{[3.6]}&\colhead{[4.5]}&\colhead{[3.6]}&\colhead{[4.5]}&\colhead{[3.6]}&\colhead{[4.5]}&\colhead{[3.6]}&\colhead{[4.5]}&\colhead{[3.6]}&\colhead{[4.5]}}
\startdata
106.91 & 7.015 & 6.862 & 8.996 & 8.809 & 9.741 & 9.489 & 10.487 & 10.401 & 11.278 & 11.029\\
107.30 & 7.009 & 6.863 & 8.979 & 8.804 & 9.778 & 9.496 & 10.500 & 10.417 & 11.323 & 11.020\\
107.57 & 7.012 & 6.833 & 9.027 & 8.805 & 9.759 & 9.469 & 10.503 & 10.439 & 11.291 & 10.999\\
107.81 & 6.993 & 6.870 & 9.006 & 8.809 & 9.746 & 9.431 & 10.521 & 10.461 & 11.283 & 11.021\\
108.13 & 7.017 & 6.863 & 8.991 & 8.805 & 9.711 & 9.435 & 10.559 & 10.477 & 11.308 & 11.045\\
108.38 & 6.959 & 6.838 & 8.971 & 8.809 & 9.724 & 9.410 & 10.556 & 10.465 & 11.348 & 11.092\\
108.58 & 6.991 & 6.833 & 9.054 & 8.808 & 9.754 & 9.423 & 10.540 & 10.471 & 11.323 & 11.083\\
108.81 & 6.988 & 6.802 & 9.037 & 8.822 & 9.749 & 9.452 & 10.524 & 10.431 & 11.323 & 11.096\\
109.58 & 6.939 & 6.833 & 9.197 & 8.977 & 9.853 & 9.536 & 10.534 & 10.447 & 11.329 & 11.124\\
110.50 & 6.988 & 6.823 & 9.137 & 8.977 & 9.735 & 9.447 & 10.452 & 10.352 & 11.348 & 11.130\\
111.07 & 7.024 & 6.849 & 9.201 & 8.995 & 9.704 & 9.398 & 10.485 & 10.378 & 11.264 & 11.084\\
112.43 & 7.008 & 6.848 & 9.209 & 9.060 & 9.638 & 9.325 & 10.471 & 10.416 & 11.283 & 11.088\\
113.07 & 6.999 & 6.799 & 9.223 & 9.082 & 9.655 & 9.355 & 10.483 & 10.374 & 11.319 & 11.072\\
114.73 & 6.967 & 6.802 & 9.288 & 9.185 & 9.715 & 9.395 & 10.486 & 10.435 & 11.319 & 11.077\\
115.37 & 6.943 & 6.799 & 9.325 & 9.164 & 9.691 & 9.409 & 10.488 & 10.434 & 11.298 & 11.031\\
115.97 & 6.954 & 6.792 & 9.309 & 9.158 & 9.684 & 9.386 & 10.489 & 10.389 & 11.268 & 11.050\\
116.32 & 6.905 & 6.766 & 9.294 & 9.125 & 9.620 & 9.366 & 10.482 & 10.427 & 11.281 & 11.051\\
116.62 & 6.920 & 6.751 & 9.289 & 9.129 & 9.635 & 9.316 & 10.537 & 10.483 & 11.294 & 11.036\\
116.72 & 6.972 & 6.812 & 9.290 & 9.138 & 9.644 & 9.344 & 10.565 & 10.499 & 11.293 & 11.068\\
117.13 & 6.954 & 6.746 & 9.269 & 9.102 & 9.649 & 9.357 & 10.552 & 10.498 & 11.291 & 11.038\\
117.21 & 6.969 & 6.771 & 9.275 & 9.107 & 9.633 & 9.338 & 10.547 & 10.479 & 11.261 & 11.034\\
117.46 & 6.945 & 6.782 & 9.273 & 9.086 & 9.659 & 9.371 & 10.547 & 10.504 & 11.255 & 11.002\\
117.69 & 6.939 & 6.757 & 9.215 & 9.076 & 9.665 & 9.391 & 10.551 & 10.517 & 11.301 & 11.024\\
118.60 & 6.909 & 6.771 & 9.146 & 9.010 & 9.742 & 9.425 & 10.576 & 10.544 & 11.277 & 11.064\\
119.25 & 6.955 & 6.784 & 9.167 & 9.008 & 9.725 & 9.442 & 10.565 & 10.548 & 11.263 & 11.175\\
119.93 & 6.944 & 6.777 & 9.151 & 8.963 & 9.739 & 9.452 & 10.582 & 10.537 & 11.284 & 11.135\\
121.21 & 6.949 & 6.789 & 9.129 & 8.912 & 9.788 & 9.495 & 10.582 & 10.557 & 11.204 & 11.009\\
123.07 & 6.921 & 6.775 & 9.054 & 8.792 & 9.856 & 9.591 & 10.572 & 10.529 & 11.296 & 11.027\\
124.96 & 6.924 & 6.784 & 9.018 & 8.800 & 9.928 & 9.615 & 10.620 & 10.580 & 11.279 & 11.046\\
127.68 & 6.945 & 6.794 & 8.966 & 8.748 & 9.766 & 9.487 & 10.581 & 10.521 & 11.228 & 11.027\\
129.81 & 6.940 & 6.797 & 9.044 & 8.746 & 9.758 & 9.504 & 10.559 & 10.545 & 11.217 & 10.975\\
131.27 & 6.938 & 6.792 & 8.910 & 8.632 & 9.825 & 9.529 & 10.599 & 10.541 & 11.219 & 10.925\\
133.38 & 6.954 & 6.832 & 8.887 & 8.683 & 9.753 & 9.456 & 10.638 & 10.498 & 11.186 & 10.924\\
135.61 & 6.975 & 6.841 & 8.875 & 8.608 & 9.649 & 9.338 & 10.456 & 10.377 & 11.162 & 10.906\\
137.52 & 7.004 & 6.867 & 8.649 & 8.379 & 9.594 & 9.329 & 10.314 & 10.201 & 11.199 & 10.870\\
139.07 & 7.029 & 6.879 & 8.634 & 8.364 & 9.645 & 9.337 & 10.336 & 10.209 & 11.197 & 10.895\\
141.52 & 7.030 & 6.915 & 8.776 & 8.485 & 9.612 & 9.300 & 10.269 & 10.175 & 11.166 & 10.899\\
142.11 & 7.038 & 6.925 & 8.715 & 8.478 & 9.643 & 9.314 & 10.286 & 10.152 & 11.196 & 10.890\\
\enddata
\end{deluxetable}

\begin{deluxetable}{cc}
%\tablewidth{0pt}
\tabletypesize{\scriptsize}
\tablecaption{Extinction\label{extinction}}
\tablehead{\colhead{Date}&\colhead{A$_V$}}
\startdata
\cutinhead{LRLL 2}
08 Oct 2009 & 2.8\\
31 Oct 2009 & 3.0\\
\cutinhead{LRLL 21}
10 Oct 2008 & 4.1\\
11 Oct 2008 & 4.4\\
18 Oct 2008 & 4.2\\
19 Oct 2008 & 4.6\\
08 Oct 2009 & 3.9\\
31 Oct 2009 & 4.1\\
04 Nov 2009 & 4.2\\
08 Nov 2009 & 4.3\\
\cutinhead{LRLL 58}
09 Oct 2009 & 3.4\\
08 Nov 2009 & 3.4\\
\cutinhead{LRLL 1679}
11 Oct 2008 & 5.8\\
18 Oct 2008 & 5.8\\
19 Oct 2008 & 5.8\\
09 Oct 2009 & 5.8\\
04 Nov 2009 & 5.8\\
\enddata
\end{deluxetable}

\begin{deluxetable}{cccc}
%\tablewidth{0pt}
\tabletypesize{\scriptsize}
\tablecaption{Emission Lines\label{line_strength_table}}
\tablehead{\colhead{Date}&\colhead{Pa$\beta$ EW ($\AA$)}&\colhead{Br$\gamma$ EW ($\AA$)}&\colhead{Corrected Br$\gamma$ EW ($\AA$)}}
\startdata
\cutinhead{LRLL 2}
9 Oct 2009 & -0.46 $\pm$ 0.04 & -3.45 $\pm$ 0.27 & -1.85 $\pm$ 0.17\\
31 Oct 2009 & -3.40 $\pm$ 0.03 & -5.23 $\pm$ 0.17 & -3.16 $\pm$ 0.10\\
\cutinhead{LRLL 21}
10 Oct 2008 & 0.63 $\pm$ 0.35 & -0.01 $\pm$ 0.27 & -0.07 $\pm$ 0.26\\
11 Oct 2008 & 0.96 $\pm$ 0.07 & 0.32 $\pm$ 0.05 & 0.27 $\pm$ 0.04\\
18 Oct 2008 & -1.21 $\pm$ 0.05 & 0.16 $\pm$ 0.04 & 0.09 $\pm$ 0.03\\
19 Oct 2008 & -1.25 $\pm$ 0.08 & 0.07 $\pm$ 0.05 & 0.02 $\pm$ 0.04\\
8 Oct 2009 & 0.83 $\pm$ 0.06 & -0.29 $\pm$ 0.05 & -0.33 $\pm$ 0.05\\
31 Oct 2009 & 1.90 $\pm$ 0.07 & -0.14 $\pm$ 0.04 & -0.22 $\pm$ 0.03\\
4 Nov 2009 & -0.70 $\pm$ 0.07 & -1.02 $\pm$ 0.08 & -1.11 $\pm$ 0.08\\
8 Nov 2009 & -0.56 $\pm$ 0.06 & -0.88 $\pm$0.09 & -0.94 $\pm$ 0.09\\
\cutinhead{LRLL 58}
9 Oct 2009 & -0.96 $\pm$ 0.09 & -0.66 $\pm$ 0.09 & -0.86 $\pm$ 0.09\\
8 Nov 2009 & -1.95 $\pm$ 0.51 & -1.63 $\pm$ 0.45 & -1.84 $\pm$ 0.47\\
\cutinhead{LRLL 1679}
11 Oct 2008 & 0.58  $\pm$ 0.22 & 0.12  $\pm$ 0.07 & -\\
18 Oct 2008 & 0.23  $\pm$ 0.24 & -0.66  $\pm$ 0.20 & -\\
19 Oct 2008 & 1.03  $\pm$ 0.33 & -0.13  $\pm$ 0.15 & -\\
9 Oct 2009 & 0.89  $\pm$ 0.49 & -0.51  $\pm$ 0.42 & -\\
4 Nov 2009 & 0.78  $\pm$ 0.31 & -0.17  $\pm$ 0.13 & -\\

\enddata
\end{deluxetable}

\begin{deluxetable}{cccccc}
%\tablewidth{0pt}
\tabletypesize{\scriptsize}
\tablecaption{Accretion Rates\label{accretion}}
\tablehead{\colhead{Date}&\colhead{Flux ($10^{-14}$ erg s$^{-1}$ cm$^{-2}$)}&\colhead{$\log$(L/L$_{\odot}$)}&\colhead{$\log$(L$_{acc}$/L$_{\odot}$)}&\colhead{L$_{acc}$/L$_*$}&\colhead{$\dot{M}$ ($10^{-8} M_{\odot}yr^{-1}$)}}
\startdata
\cutinhead{LRLL 2 Pa$\beta$}
9 Oct 2009 & 25.2 $\pm$ 6.8 & -3.09 $\pm$ 0.12 & -0.37 $\pm$ 0.14 & 0.003 & 3.14 $\pm$ 0.44\\
31 Oct 2009 & 185 $\pm$ 29 & -2.23 $\pm$ 0.07 & 0.61 $\pm$ 0.08 & 0.03 & 29.9 $\pm$ 2.4\\
\cutinhead{LRLL 2 Br$\gamma$}
9 Oct 2009 & 29.2 $\pm$ 3.6 & -3.03 $\pm$ 0.05 & 0.61 $\pm$ 0.06 & 0.03 & 29.9 $\pm$ 1.8\\
31 Oct 2009 & 49.8 $\pm$ 5.2 & -2.80 $\pm$ 0.04 & 0.90 $\pm$ 0.05 & 0.06 & 58.4 $\pm$ 2.9\\
\cutinhead{LRLL 58 Pa$\beta$}
9 Oct 2009 & 1.35 $\pm$ 0.24 & -4.37 $\pm$ 0.08 & -1.83 $\pm$ 0.09 & 0.02 & 0.23 $\pm$ 0.02\\
8 Nov 2009 & 2.81 $\pm$ 0.56 & -4.05 $\pm$ 0.09 & -1.47 $\pm$ 0.10 & 0.05 & 0.52 $\pm$ 0.05\\
\cutinhead{LRLL 58 Br$\gamma$}
9 Oct 2009 & 0.74 $\pm$ 0.09 & -4.63 $\pm$ 0.06 & -1.40 $\pm$ 0.08 & 0.06 & 0.62 $\pm$ 0.05\\
8 Nov 2009 & 1.57 $\pm$ 0.23 & -4.30 $\pm$ 0.06 & -0.99 $\pm$ 0.08 & 0.14 & 1.58 $\pm$ 0.13\\
\cutinhead{LRLL 21 Pa$\beta$}
10 Oct 2008 & $<3.52$ & $<-3.95$ & $<-1.35$ & $<0.02$ & $<0.26$\\
11 Oct 2008 & $<0.77$ & $<-4.61$ & $<-2.11$ & $<0.003$ & $<0.05$\\
18 Oct 2008 & 4.68 $\pm$ 0.74 & -3.83 $\pm$ 0.07 & -1.22 $\pm$ 0.08 & 0.02 & 0.388 $\pm$ 0.031\\
19 Oct 2009 & 5.33 $\pm$ 0.86 & -3.77 $\pm$ 0.07 & -1.15 $\pm$ 0.08 & 0.02 & 0.455 $\pm$ 0.036\\
8 Oct 2009 & $<0.79$ & $<-4.60$ & $<-2.09$ & $<0.002$ & $<0.05$\\
31 Oct 2009 & $<0.95$ & $<-4.52$ & $<-2.01$ & $<0.003$ & $<0.07$\\
4 Nov 2009 & 3.36 $\pm$ 0.57 & -3.97 $\pm$ 0.07 & -1.38 $\pm$ 0.09 & 0.01 & 0.268 $\pm$ 0.024\\
8 Nov 2009 & 2.75 $\pm$ 0.49 & -4.06 $\pm$ 0.08 & -1.48 $\pm$ 0.10 & 0.01 & 0.212 $\pm$ 0.021\\
\cutinhead{LRLL 21 Br$\gamma$}
10 Oct 2008 & $<1.14$ & $<-4.34$ & $<-1.04$ & $<0.003$ & $<0.53$\\
11 Oct 2008 & $<0.22$ & $<-5.15$ & $<-2.06$ & $<0.002$ & $<0.05$\\
18 Oct 2008 & $<0.19$ & $<-5.23$ & $<-2.16$ & $<0.003$ & $<0.05$\\
19 Oct 2008 & $<0.26$ & $<-5.09$ & $<-1.98$ & $<0.04$ & $<0.07$\\
8 Oct 2009 & 0.663 $\pm$ 0.164 & -4.68 $\pm$ 0.11 & -1.47 $\pm$ 0.14 & 0.01 & 0.218 $\pm$ 0.031\\
31 Oct 2009 & 0.501 $\pm$ 0.103 & -4.79 $\pm$ 0.09 & -1.61 $\pm$ 0.11 & 0.007 & 0.158 $\pm$ 0.017\\
4 Nov 2009 & 2.78 $\pm$ 0.31 & -4.05 $\pm$ 0.05 & -0.67 $\pm$ 0.06 & 0.06 & 1.38 $\pm$ 0.08\\
8 Nov 2008 & 2.30 $\pm$ 0.29 & -4.13 $\pm$ 0.05 & -0.77 $\pm$ 0.06 & 0.05 & 1.09 $\pm$ 0.07\\
\cutinhead{LRLL 1679 Pa$\beta$}
11 Oct 2008 & $<0.35$ & $<-4.96$ & $<-2.49$ & $<0.02$ & $<0.08$\\
18 Oct 2008 & $<0.38$ & $<-4.92$ & $<-2.46$ & $<0.02$ & $<0.08$\\
19 Oct 2008 & $<0.52$ & $<-4.78$ & $<-2.30$ & $<0.02$ & $<0.12$\\
9 Oct 2009 & $<0.78$ & $<-4.61$ & $<-2.10$ & $<0.04$ & $<0.18$\\
4 Nov 2009 & $<0.49$ & $<-4.81$ & $<-2.33$ & $<0.07$ & $<0.11$\\
\cutinhead{LRLL 1679 Br$\gamma$}
11 Oct 2008 & $<0.06$ & $<-5.70$ & $<-2.75$ & $<0.008$ & $<0.04$\\
18 Oct 2008 & $<0.18$ & $<-5.25$ & $<-2.19$ & $<0.03$ & $<0.15$\\
19 Oct 2008 & $<0.13$ & $<-5.37$ & $<-2.34$ & $<0.02$ & $<0.11$\\
9 Oct 2009 & $<0.38$ & $<-5.92$ & $<-3.03$ & $<0.004$ & $<0.02$\\
4 Nov 2009 & $<0.12$ & $<-5.43$ & $<-2.41$ & $<0.02$ & $<0.09$\\
\enddata
\end{deluxetable}

\begin{deluxetable}{cccc}
%\tablewidth{0pt}
\tablecaption{LRLL 21 NIR Photometry\label{nir_phot}}
\tablehead{\colhead{Date}&\colhead{J}&\colhead{H}&\colhead{K}}
\startdata
2MASS & 11.02 & 9.99 & 9.47\\
10 Oct 2008 & 11.31 & 10.17 & 9.71\\
11 Oct 2008 & 11.30 & 10.21 & 9.73\\
19 Oct 2008 & 11.19 & 10.14 & 9.60\\
8 Oct 2009 & 10.96 & 9.98 & 9.59\\
31 Oct 2009 & 10.99 & 10.05 & 9.52\\
4 Nov 2009 & 10.95 & 9.91 & 9.39\\
8 Nov 2009 & 10.96 & 9.94 & 9.43\\
\enddata
\end{deluxetable}

\begin{deluxetable}{cc}
%\tablewidth{0pt}
\tablecaption{LRLL 21 Stellar Luminosity\label{lstar}}
\tablehead{\colhead{Date}&\colhead{L$_*$ (L$_{\odot}$)}}
\startdata
10 Oct 2008 & 2.52 pm 0.39\\
11 Oct 2008 & 2.75 pm 0.43\\
19 Oct 2008 & 3.20 pm 0.50\\
08 Oct 2009 & 3.30 pm 0.52\\
31 Oct 2009 & 3.38 pm 0.53\\
04 Nov 2009 & 3.60 pm 0.56\\
08 Nov 2009 & 3.66 pm 0.57\\
\enddata
\end{deluxetable}

\begin{deluxetable}{cccc}
%\tablewidth{0pt}
\tablecaption{Radial and Rotational Velocity\label{velocity}}
\tablehead{\colhead{ID}&\colhead{v$_r$}&\colhead{v$\sin$i}&\colhead{Reference}}
\startdata
LRLL 21 & 17.01 & 17.84 & 1\\
 & -15.7\tablenotemark{a} & 21.9\tablenotemark{a} & 2\\
 & 66.8\tablenotemark{a} & $<$15\tablenotemark{a} & 2\\
LRLL 58 & 14.2 & $<$15 & 2\\
LRLL 67 & 15.0 & $<$15 & 2\\
 & 15.09 & 13.6 & 1\\
 & 16.9 & $<$11 & 3\\
\enddata
\tablenotetext{a}{In our high-resolution spectra of LRLL 21 we find two peaks in the cross-correlation function and report the radial and rotational velocity of each component.}
\tablecomments{References: (1) \citet{dah08} (2) This paper (3) \citet{nor06}.}
\end{deluxetable}

\begin{deluxetable}{cccccc}
%\tablewidth{0pt}
\tablecaption{Infrared Excess\label{ir_excess_table}}
\tablehead{\colhead{Date}&\colhead{r$_J$}&\colhead{r$_H$}&\colhead{r$_K$}&\colhead{T$_{dust}$}&\colhead{Covering Fraction}}
\tablecolumns{6}
\startdata
\cutinhead{LRLL 21}
10 Oct 2008 & 0.13 & 0.19 & 0.26 & 1900 & 0.06\\
11 Oct 2008 & 0.10 & 0.22 & 0.35 & - & -\\
18 Oct 2008 & 0.01 & 0.17 & 0.36 & 2020 & 0.08\\
19 Oct 2008 & 0.01 & 0.17 & 0.37 & 2180 & 0.08\\
08 Oct 2009 & 0.02 & 0.09 & 0.19 & - & $\sim0.04$\tablenotemark{a}\\
31 Oct 2009 & 0.06 & 0.29 & 0.56 & 1860 & 0.13\\
04 Nov 2009 & 0.07 & 0.29 & 0.54 & 1700 & 0.13\\
08 Nov 2009 & 0.15 & 0.29 & 0.46 & - & -\\
\cutinhead{LRLL 58\tablenotemark{b}}
09 Oct 2009 & 0.12 & 0.28 & 0.46 & - & $\sim0.13$\tablenotemark{a}\\
08 Nov 2009 & 0.18 & 0.32 & 0.48 & - & -\\
\cutinhead{LRLL 1679}
11 Oct 2008 & -0.02 & -0.004 & 0.01 & - & -\\
18 Oct 2008 & -0.09 & -0.07 & -0.04 & - & -\\
19 Oct 2008 & -0.16 & -0.11 & -0.05 & - & -\\
09 Oct 2009 & -0.11 & -0.02 & 0.07 & - & $\sim0.02$\tablenotemark{a}\\
04 Nov 2009 & -0.04 & 0.02 & 0.09 & - & $\sim0.03$\tablenotemark{a}\\
\enddata
\tablenotetext{a}{For days when 2-5\micron\ spectra were not available, but we have Spitzer 3.6 and 4.5\micron\ photometry, we estimate a blackbody fit to the excess assuming T=1900K, a temperature that was found to fit those days that do have the 2-5\micron\ spectra.}
\tablenotetext{b}{For LRLL 58 we use the standard JH 108 to measure the veiling even though it shows emission near the Br$\gamma$ line, which may indicate accretion and possibly a disk. The line strengths fit better than an M1 giant or dwarf standard. The measured veiling is only a lower limit, although the lack of infrared excess for JH 108 indicates that within our uncertainties we are measuring the true veiling}
\tablecomments{Uncertainties on the veiling measurements are $\sim$0.1. No veiling is measured for LRLL 2 since the only photospheric absorption lines are the hydrogen series, which will be contaminated from emission by the accretion flow.}
\end{deluxetable}

\begin{deluxetable}{ccccccccc}
%\tablewidth{0pt}
%\rotate
\tabletypesize{\scriptsize}
\tablecolumns{9}
\tablecaption{Variability Summary\label{var_sum}}
\tablehead{\colhead{Star}&\colhead{Stellar Flux}&\colhead{}&\multicolumn{3}{c}{Accretion Flux}&\colhead{}&\multicolumn{2}{c}{Infrared Variability}\\
\cline{2-2} \cline{4-6} \cline{8-9}\\
\colhead{(Spectral Type)}&\colhead{Variable?\tablenotemark{a}}&\colhead{}&\colhead{\# of Epochs}&\colhead{Variable?}&\colhead{Rate\tablenotemark{b} ($10^{-8}M_{\odot}yr^{-1}$)}&\colhead{}&\colhead{Strength}&\colhead{Pivot Point}}
\startdata
\cline{1-9}\\
LRLL 2 & no & & 2 & no?\tablenotemark{c} & $\sim$30  & & 20\% in IRS & 6\micron\\
A2 & & & & & & & 25\% at [24] & \\
 & & & & & & & 0.1 mag at [3.6],[4.5] & \\
 \cline{1-9}\\
LRLL 21 & P: 2.5 days & & 8 & yes & $<0.05-0.46$ & & 40\% for $\lambda<14\micron$ in IRS & 8\micron\\
K0 & & & & & & &  $<8\%$ for $\lambda>14\micron$ in IRS & \\
& NP: yearly  & & & & & & 20\% at [24] & \\
& & & & & & & 0.5 mag at [3.6],[4.5] & \\
\cline{1-9}\\
LRLL 31 & P: 3.4 days & & 9 & yes & 0.25-1.62 & & 60\% in IRS & 8\micron\\
G6 &  &  & & & & & 30\% at [24] & \\
& & & & & & & 0.3 mag at [3.6],[4.5] & \\
\cline{1-9}\\
LRLL 58 & P: 7.4 days & & 2 & yes &  0.23-0.52 & & $<$5\% for $\lambda<14\micron$ in IRS  & 8\micron\\
M1.25 & & & & & & & $<20\%$ for $\lambda>14\micron$ in IRS & \\
 & & & & & & & 14\% at [24] & \\
& & & & & & & 0.3 mag at [3.6],[4.5] & \\
\cline{1-9}\\
LRLL 67 & no & & 3\tablenotemark{d} & yes\tablenotemark{d} & $\sim$0.2\tablenotemark{d} & & $<3\%$ for $\lambda<14\micron$ in IRS& none detected\\
M0.75 & & & & & & & $<20\%$ for $\lambda>14\micron$ in IRS  & \\
 & & & & & & & 40\% at [24]\tablenotemark{e} & \\
& & & & & & & 0.4 mag at [3.6],[4.5] & \\
\cline{1-9}\\
LRLL 1679 & unknown & & 5 & undetected & $<0.18$ & & 50\% in IRS & 13\micron\\
M3.5 & & & & & & & 50\% at [24] & \\
& & & & & & & 0.1 mag at [3.6],[4.5] & \\
\enddata
\tablenotetext{a}{P=Periodic fluctuations. NP=fluctuations detected with no known period. For variable sources we include the period or the typical timescale of the non-periodic variability.}
\tablenotetext{b}{Derived from our observations of the Pa$\beta$ line.}
\tablenotetext{c}{The large systematic uncertainties in characterizing the continuum around the Pa$\beta$ line prevent us from determining if the accretion rate is variable even though the derived accretion rates in the two epochs are substantially different.}
\tablenotetext{d}{We have no Pa$\beta$ spectra for LRLL 67, which prevents us from measuring an accretion rate. The accretion rate is taken from \citet{dah08} and the presence of variability is based on the changing H$\alpha$ EW.}
\tablenotetext{e}{The 24\micron\ variability of LRLL 67 is only detected at the $\sim$2$\sigma$ level}

\end{deluxetable}

\begin{deluxetable}{cccc}
%\tablewidth{0pt}
\tablecaption{Strength and shape of silicate feature\label{silicate_table}}
\tablehead{\colhead{Epoch}&\colhead{EW(10$\micron$)}&\colhead{$S(peak)$}&\colhead{$S(11.3)/S(9.8)$}}
\startdata
\cutinhead{LRLL 2}
1 & 2.98 & 0.97 & 1.00\\
2 & 2.71 & 0.91 & 1.06\\
3 & 2.97 & 0.96 & 1.11\\
4 & 2.96 & 0.96 & 1.13\\
\cutinhead{LRLL 21}
1 & 2.19 & 0.76 & 1.39\\
2 & 2.13 & 0.80 & 1.52\\
3 & 1.87 & 0.65 & 1.53\\
4 & 1.89 & 0.68 & 1.81\\
\cutinhead{LRLL 31}
1 & 2.37 & 1.98 & 0.98\\
2 & 3.31 & 2.26 & 0.96\\
3 & 2.37 & 1.84 & 0.97\\
4 & 2.37 & 1.71 & 0.99\\
\cutinhead{LRLL 58}
1 & 2.51 & 0.79 & 1.13\\
2 & 2.76 & 0.87 & 1.18\\
3 & 3.07 & 1.03 & 1.27\\
4 & 3.21 & 1.05 & 1.24\\
\cutinhead{LRLL 67}
1 & 3.73 & 1.62 & 0.87\\
2 & 3.51 & 1.49 & 0.87\\
3 & 3.92 & 1.72 & 0.87\\
4 & 3.77 & 1.63 & 0.86\\
\cutinhead{LRLL 1679}
1 & 1.68 & 0.69 & 0.93\\
2 & 1.33 & 0.47 & 1.16\\
3 & 1.74 & 0.69 & 0.70\\
4 & 1.19 & 0.53 & 1.21\\
\enddata
\tablecomments{These parameters have been derived based on a dereddened silicate feature. The extinction is assumed to be constant between each epoch and the spectra are dereddened using the \citet{mcc09} extinction law}
\end{deluxetable}

\begin{deluxetable}{ccc}
%\tablewidth{0pt}
\tablecaption{Inner radii of disk\label{inner_radii}}
\tablehead{\colhead{ID}&\colhead{Co-rotation Radius (AU)}&\colhead{Sublimation Radius (AU)}}
\startdata
LRLL 2 & (0.03-0.18) & 0.76 ($>$0.41)\\
LRLL 21 & 0.04 & 0.13 ($>$0.07)\\
LRLL 31 & 0.05 & 0.13 ($>$0.07)\\
LRLL 58 & 0.07 & 0.05 ($>$0.03)\\
LRLL 67 & (0.02-0.10) & 0.04 ($>$0.02)\\
LRLL 1679 & (0.01-0.08) & 0.03 ($>$0.02)\\
\enddata
\tablecomments{Co-rotation radii are calculated based on observed rotational periods estimated from optical light curves. When no observed period is available we estimate the co-rotation radius for periods of 1 and 15 days, which span the typical periods observed for pre-main sequence stars in IC 348 \citep{cie06}. The lower limit on the sublimation radius, included in parenthesis, assumes very large grains ($>10\micron$) at 1500K.}
\end{deluxetable}

\end{document}